\documentclass[showpacs,twocolumn,aps,prx,superscriptaddress,10pt]{revtex4-1}

\usepackage{amsmath,amssymb,amsfonts,bm,graphicx,mathtools,braket}
\usepackage[caption=false]{subfig}

\usepackage{hyperref}
\hypersetup{%
	pdftitle={Josephson}, %
	pdfauthor={Zhiyuan Sun},%
	pdfpagemode={UseNone},%
	pdfstartview={FitH},%
	breaklinks=true,%
	citecolor=blue,%
	colorlinks=true,%
	linkcolor=blue,%
	urlcolor=blue
}

\newcommand{\unit}[1]{\,\mathrm{#1}} 
\newcommand{\equa}[1]{Eq.~\eqref{#1}} 
\newcommand{\fig}[1]{Fig.~\ref{#1}}

\newcommand{\rom}[1]{\uppercase\expandafter{\romannumeral #1\relax}}

\graphicspath{{./}}

\begin{document}

\title{Second order Josephson effect in excitonic insulators}

\author{Zhiyuan Sun}
\affiliation{Department of Physics, Columbia University, 538 West 120th Street, New York, New York 10027}

\author{Tatsuya Kaneko}
\affiliation{Department of Physics, Columbia University, 538 West 120th Street, New York, New York 10027}

\author{Denis Gole\v{z}}
\affiliation{Center for Computational Quantum Physics, Flatiron Institute, 162 5th Avenue, New York, NY, 10010}

\affiliation{Faculty of Mathematics and Physics, University of Ljubljana, Jadranska 19, SI-1000 Ljubljana, Slovenia}

\affiliation{Jo\v{z}ef Stefan Institute, Jamova 39, SI-1000, Ljubljana, Slovenia}

\author{Andrew J. Millis}
\affiliation{Department of Physics, Columbia University, 538 West 120th Street, New York, New York 10027}
\affiliation{Center for Computational Quantum Physics, Flatiron Institute, 162 5th Avenue, New York, NY, 10010}

\begin{abstract}
We show that in electron-hole bilayers with excitonic order arising from conduction and valence bands formed by atomic orbitals that  have different parities, nonzero interlayer tunneling leads to a second order Josephson effect. This means the interlayer  electrical current is related to the phase of the excitonic order parameter  as $J = J_c \sin2\theta$ instead of $J = J_c \sin \theta$, and that the system has two degenerate ground states at $\theta=0, \pi$ that can be switched by an interlayer voltage pulse.
When generalized to a three dimensional stack of alternating electron-hole  planes or a two dimensional stack of chains, AC Josephson effect implies that electric field pulses perpendicular to the layers and chains can steer the order parameter phase between the two degenerate ground states, making these devices  ultrafast memories.
The order parameter steering also applies to the excitonic insulator candidate Ta$_2$NiSe$_5$. 
\\
\end{abstract}

\maketitle

Excitonic condensation \cite{Mott1961, Keldysh1965, Jerome1967,Halperin.1968, Keldysh1968} has been experimentally realized in electron-hole bilayers (EHB) \cite{Butov.1994,Butov2002,Du2017,Li2017,Burg.2018, Wang.2019,Ma.2021strongly, Eisenstein2014,Fogler2014a,Liu.2017} where electrons in one layer pair with holes in the other layer to form excitons that condense into a single macroscopic state. 
In 1976, Kulik and Shevchenko \cite{Kulik.1976,Shevchenko1977} (see also Refs. \cite{Lozovik.1997,Wen.1992,Shevchenko1994}) noted that nonzero interlayer tunneling endowes the EHB with a Josephson effect similar to that in superconductors. This effect was observed in 2000 by Spielman \textit{et al.} in quantum hall bilayers \cite{Spielman2000,Spielman2001} and explained in detail in Refs.~\cite{Fogler2001a,Stern2001,Joglekar.2001,Balents.2001}. 

\begin{figure}
	\includegraphics[width=0.85 \linewidth]{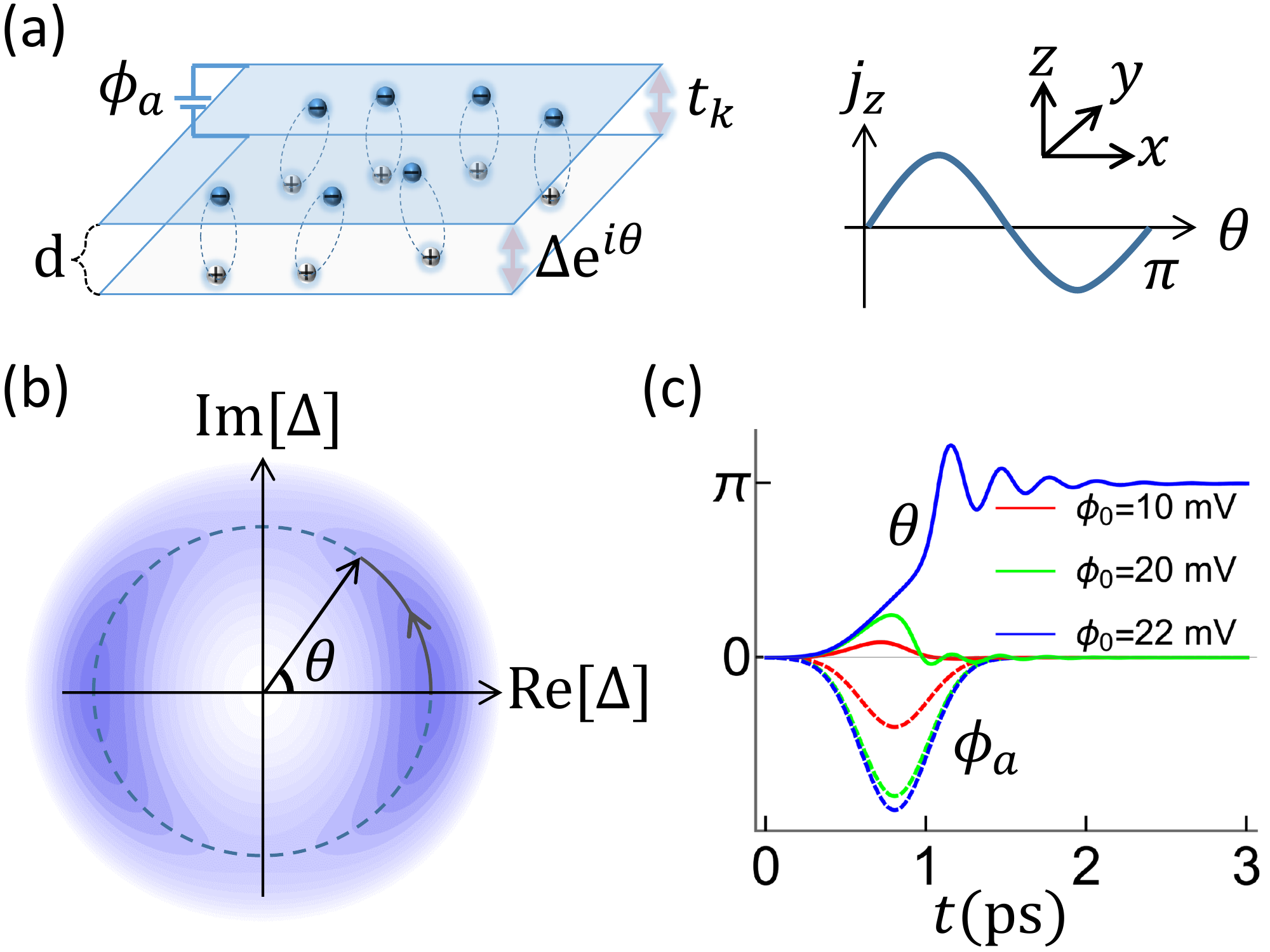}
	\caption{(a) Schematics of the electron-hole bilayer showing electrons ($-$) and holes ($+$) and the interlayer current-phase relation in the excitonic insulating phase. (b) False color representation of the free energy on the plane of complex order parameter  where lower energy appears bluer. (c) Solid curve: time dependence of order parameter phase after a voltage pulse $\phi_a=-\phi_0 e^{-(t-t_0)^2/T_0^2}$ (dashed curve) computed from \equa{eqn:EOM} with $\Delta_{\text{p}}=14 \unit{meV}$, $T_0=0.3 \unit{ps}$,  $\gamma=0.3 \Delta_{\text{p}}$, $D=2$ and $C=1$.}
	\label{fig:bilayer}
\end{figure}

\begin{figure}
	\includegraphics[width= \linewidth]{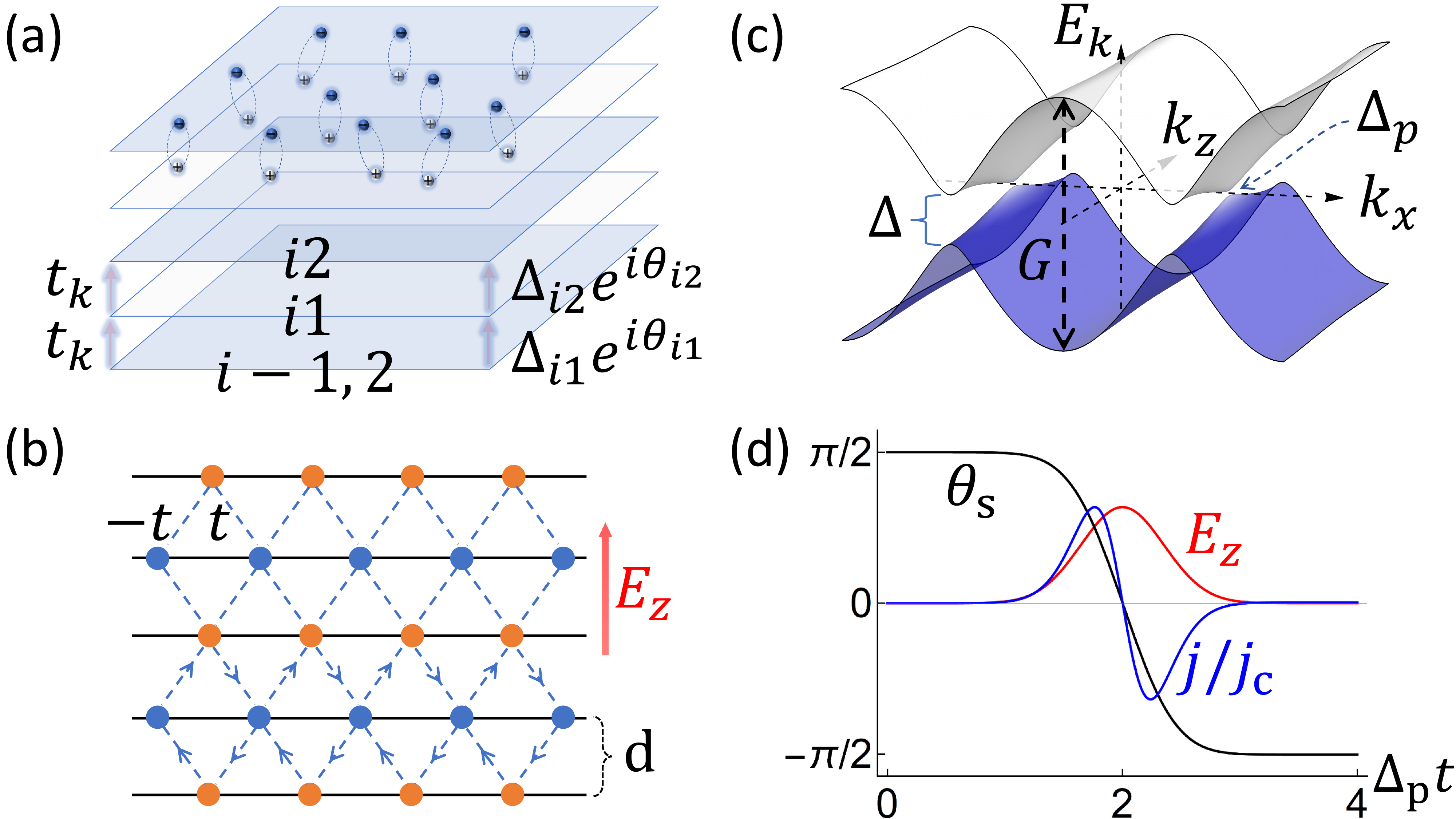}
	\caption{(a) Schematic of the 3D stack of alternating electron (blue) and hole (unshaded) planes with pairing order parameters labeled. (b) Schematic of the 2D stack of alternating electron and hole chains. The orange and blue dots represent atomic orbitals forming the conduction and valence bands. Their different parities lead to   asymmetric  inter chain hoping $t/-t$ \cite{kaneko2020bulk}. Arrows represent the spontaneous circulating currents. (c) The ground state band dispersion of the 2D stack. (d) The order parameter phase dynamics (black curve) and the Josephson current (blue curve) induced by an electric field pulse  $E_z(t)=E_{\text{max}} e^{-(t-t_0)^2/T_0^2}$ (red curve) implied by \equa{eqn:steering}, with $\Delta=10\Delta_{\text{p}}$, $E_{\text{max}}=3.55 \Delta_{\text{p}}/d$ and $T_0=0.5/\Delta_{\text{p}}$.}
	\label{fig:stacks}
\end{figure}

If the electron and hole bands are formed by atomic orbitals that transform differently under crystal symmetries, the intrinsic tunneling (hybridization) vanishes at high symmetry points of the Brillouin zone and is very small nearby, such that the excitonic insulator (EI) transition breaks a discrete symmetry \cite{Halperin.1968, Halperin.1968_2, Portengen1996, Mazza2020,kaneko2020bulk,Lenk.2020}.  In this paper, we show that if the orbitals lie at different spatial locations as shown in \fig{fig:bilayer}, a difference of symmetries (e.g. $p$ and $d$ orbitals) implies that the ordered state sustains a \emph{second order} Josephson effect as the tunneling has to create or annihilate two excitons  each time. A similar effect is already well known in carefully designed superconducting Josephson junctions \cite{Golubov.2004} (e.g., a 45$^\circ$ junction between d-wave superconductors or a junction between s and d-wave superconductors  \cite{Tanaka.1994,Yip.1995,Huck.1997, Zagoskin_1997,Ilichev.1999,Ilichev.2001,Asano.2001, zeng2021phasefluctuation}). We show that it naturally occurs in EIs, which leads to symmetry breaking degenerate ground states that are easily distinguishable and switchable. In an isolated EHB the two ground states break parity and have opposite in-plane electrical polarization. In three dimensional (3D) stacks of coupled planes or two dimensional (2D) stacks of coupled chains (\fig{fig:stacks}), the two EI states break time reversal  symmetry with opposite anomalous hall conductivity \cite{Sun.2009_quadratic_touching, Ren.2021}, and potentially form topologically nontrivial states.  In all cases the excitonic order parameter may be `steered' by applied interlayer or interchain electric fields via the AC Josephson effect, enabling controlled switching of degenerate ground states.
This order parameter steering applies as well to the EI candidate Ta$_2$NiSe$_5$~\cite{Kaneko2013, Lu2017, Werdehausen2018, Sugimoto2018, Mazza2020,Ning.2020, Kim:2021wf, volkov2021failed,andrich2020imaging}. 

\emph{The electron-hole bilayer} shown in \fig{fig:bilayer}(a)  consists of two planes  labelled $1$ and $2$ with two-component electron creation operator  $\psi^\dagger=(\psi_1^\dagger , \, \psi_2^\dagger)$ from the two bands. The  Hamiltonian  is 
\begin{align}
H_{\text{EHB}} =& \sum_k \psi_k^\dagger
\begin{pmatrix}
\xi_1(k+A_1)  + \phi_1  & e^{id A_z} t_{k+A} \\
e^{-id A_z} t^\ast_{k+A} & \xi_2(k+A_2)  + \phi_2 
\end{pmatrix}
\psi_k 
\notag\\
&  + 
\int {dr dr^\prime} V(r-r^\prime) \rho(r) \rho(r^\prime)
\label{eqn:hamiltonian}
\end{align}
where $\psi_{k}=\int dr  e^{i k r} \psi(r)$, $\rho(r)=\psi^\dagger(r) \psi(r)$ is the density, $\xi_{1,2}(k)$ is the kinetic energy describing in-plane motion with $\xi_1$ dispersing upwards from a minimum $-G/2$ and $\xi_2$ dispersing downwards from a maximum $G/2$ at the same momentum $k=0$, both  isotropic.  
$(\phi_i,\, A_i)$ is the electromagnetic (EM) potential at layer $i$, $A=(A_1+A_2)/2$  is the average in-plane component of the vector potential, $A_z$ is the average out of plane component and we have set $e=c=\hbar=1$. 
We assume the Hamiltonian is invariant under time reversal $\hat{T}$ and in-plane inversion defined as $\hat{P}: r \rightarrow -r, (\psi_1, \, \psi_2)_r \rightarrow (\psi_1, \, -\psi_2)_{-r} $ where $r=(x,y)$,  implying $\xi_{1,2}(k)=\xi_{1,2}(-k)$ and that the intrinsic interlayer tunneling satisfies  $t_{-k}=t^\ast_k=-t_k$. Thus one can write $t_k=i\Delta_{\text{p}} f_k$ where $f_k$ is odd under $k \rightarrow -k$, $\Delta_{\text{p}}>0$ is real and the subscript `p' denotes the $k$-odd nature. We distinguish the Bardeen-Cooper-Schrieffer (BCS) case ($G>0$) where the two bands cross at a Fermi momentum $k_{\text{F}}$ with Fermi velocity $v_{\text{F}}$, and the Bose-Einstein condensation (BEC) case ($G<0$) where they don't overlap. While all the equations and qualitative conclusions hold for both cases, the quantitative coefficients are presented for the analytically tractable \emph{BCS} weak coupling case ($\Delta \ll G$), unless otherwise specified. Without loss of generality, we set $f_k=c_{\text{f}} \sin k_x$ where $c_{\text{f}}$ is chosen such that $|f_{k_{\text{F}}}|=1$.

To study the excitonic order we write the model as a path integral and decompose the interaction in the electron-hole pairing channel: $Z=\int D[\psi,\Delta_k, A] e^{\int d\tau dr L_0(\psi,\Delta_k, A)}$ where $\Delta_k$ is the Hubbard-Stratonovich field. The excitonic state appears as a saddle point with the order parameter $\Delta_k= \sum_{k^\prime} V_{k-k^\prime} \langle \psi_{2k^\prime}^\dagger \psi_{1k^\prime}   \rangle $ where $V_q$ is the Fourier transform of $V(r)$. 
For physically reasonable interactions, the energetically favored  order parameter $\Delta e^{i\theta}$ has $s$-wave symmetry  \cite{sun2020topological} so the $k$ dependence may be neglected.
The quasiparticle properties are described by replacing the term $e^{id A_z} t_{k+A} $ in \equa{eqn:hamiltonian}   by $\Delta e^{i\theta}+e^{id A_z} t_{k+A} $ \cite{2note}.  There is always an odd parity phonon \cite{Halperin.1968_2, Kaneko2013, Kaneko.2015, Golez.2020, Murakami.2020}  (e.g., shear motion between the two layers) that couples linearly to $\Delta$ but may be integrated out.

Integrating out the fermions, phonons and the order parameter amplitude  fluctuations one obtains a low energy effective Lagrangian for the order parameter phase: 
\begin{align}
L=& \frac{1}{2}\nu \Bigg[- (\partial_t \theta + \phi_a)^2  + v_g^2(\nabla \theta-A_a)^2
\notag \\
& - \frac{1}{D}\Delta_{\text{p}}^2 \cos (2(\theta-A_z d))
\Bigg]
\label{eqn:L_bilayer}
\,
\end{align}
where $(\phi_a, A_a)=(\phi_1-\phi_2, A_1-A_2)/2$ is the layer-antisymmetric component of the EM field \cite{Sun.2020_BaSh}.  The last term  arises from expanding $L$ to second order in $t_k$ (assumed small relative to $\Delta$ or temperature), observing that  terms linear in $t$ vanish (see Ref.~\cite{SI}
\nocite{Altland.2010, Sun.2020_collective_modes, Herbut.2007,Tinkham,Li.2014,Dai:2014,Rodin.2014_phosphorene,Tong:2017uh, Basov2016,Sun.2015} 
Sec.~\ref{SI:bilayer}). An inversion even $t_k$ would change this  term to $\propto t \cos \theta$, giving rise to the usual Josephson effect \cite{Kulik.1976,Lozovik.1997,Fogler2001a,Stern2001,Joglekar.2001}.
The z-dipole density is $\rho_a=\delta L/\delta (\partial_t \theta)=-\nu  (\partial_t \theta + \phi_a)$ and \equa{eqn:L_bilayer} should be supplemented by the electric field energy $\sum_q \phi_a(q)^2/(2V_{\text{eff}}(q))$ representing the dipole-dipole interactions 
$V_{\text{eff}}(q)=(1-e^{-dq}) V_q$
 \cite{Sun.2020_BaSh,5note}.
At  zero temperature, the coefficients of \equa{eqn:L_bilayer} have simple $\Delta$-independent forms: $D=2$ is the space dimension, $\nu$ is the density of states in the normal state at the band crossing energy and the bare phase mode velocity is $v_g=v_{\text{F}}/\sqrt{2}$.

If $t_k$ is zero, \equa{eqn:hamiltonian} conserves the charge in each plane  and gives a continuous family of excitonic phases parametrized by $\theta$, as manifested by the $U(1)$ symmetry under transformation $\theta\rightarrow \theta+\theta_0$ of the first two terms of \equa{eqn:L_bilayer}. 
A non-zero $t_k$ gives rise to the third term which reduces the $U(1)$ invariance to $\hat{P}$, a $Z_2$ symmetry and  implies that there are two degenerate excitonic phases characterized by $\theta=0,\pi$ (\fig{fig:bilayer}(b)). The excitonic order spontaneously breaks $\hat{P}$, giving a non-vanishing in-plane electrical polarization \cite{Portengen1996,sun2020topological} which in the BCS case is
$
P= P_{\text{2D}} \left[1- \tan\left(\frac{1}{2}\mathrm{ArcTan} |\frac{\Delta_{\text{p}}}{\Delta}| \right) 
\right] \text{Sign}[\Delta]/4
$.
Since its sign is opposite for $\theta=0,\pi$, measuring it by an electrical circuit can distinguish the two  ground states. In the BEC case \cite{8note} the polarization has a more transparent physical picture. The normal state preceding the EI phase  is a semiconductor which supports excitonic modes. $t_k$ means that these  modes have oscillating in-plane electrical dipoles.  In the EI phase, a mode softens and freezes as the static in-plane electrical polarization.

In spinful systems both singlet and triplet excitonic condensates may be defined. The triplet case exhibits spin instead of charge polarization. In the pure electronic system the two phase are degenerate at the Hartree-Fock level, but electron-lattice coupling favors the singlet state  \cite{Halperin.1968,Halperin.1968_2, Kaneko.2015}
(see Ref.~\cite{SI} Sec.~\ref{SI:spin}).  We focus on the more commonly studied singlet phase here. 

\emph{Second order Josephson effect and order parameter steering---}The  interplane current
\begin{align}
-J_z=\delta L/\delta (d A_z)= \frac{\nu}{D} \Delta_{\text{p}}^2 \sin(2\theta)\equiv
J_c \sin(2\theta)
\,
\end{align}
is periodic under $\theta\rightarrow \theta+\pi$ in contrast to the usual Josephson effect where it is periodic only under $\theta\rightarrow \theta+2\pi$; the former is thus referred to as a second order Josephson effect.  Assuming a quadratic band with effective mass $0.1 m_e$ and  $\Delta_{\text{p}}=10 \unit{meV}$, the critical current is estimated as $J_c \approx 4 \unit{mA/\mu m^2}$. 
To observe the DC Josephson effect, one can source a current at one layer and drain it on the other layer, both on the left side of the device where the in-plane counter flow current $J_a=\nu v_g^2 \partial_x \theta$ is fixed as the boundary condition \cite{Spielman2000}.
From the static limit of the Euler-Lagrange equation (charge continuity equation) implied by \equa{eqn:L_bilayer}, $\nu v_g^2 \partial_x^2 \theta = J_c \sin(2\theta)$, the phase decays to the right with a decay length $l_d=\sqrt{\nu v_g^2/J_c} \sim \sqrt{D} v_g/\Delta_{\text{p}}$ \cite{7note}. Thus in a  long junction, only the region within a distance $l_d$ to the contact contributes to the Josephson current \cite{Lozovik.1997}. 
The current phase relation can be verified by applying an in-plane magnetic field to a short junction  and measuring the critical Josephson current as a function of the magnetic flux $\Phi$ through it \cite{Tinkham}. The Fraunhofer pattern $J_c(\Phi)/J_c(0)=|\frac{\sin \left( N \pi \Phi/(2\Phi_0) \right)}{N \pi \Phi/(2\Phi_0)}|$ is expected where $\Phi_0$ is the flux quantum and the  frequency $N=2$ reveals the order of the Josephson effect (see Ref.~\cite{SI} Sec.~\ref{SI:Fraunhofer}). 

To treat the order parameter steering, we focus on spatially uniform dynamics which applies to a device with gates covering the whole sample such that $\phi_a$ is uniform, or a short EHB with side contacts. \equa{eqn:L_bilayer} in the gauge $A=0$ implies
\begin{align}
\frac{1}{C}\partial_t(\partial_t \theta +  \phi_a)+\gamma \partial_t \theta + \frac{1}{D}\Delta_{\text{p}}^2 \sin 2\theta =0
\label{eqn:EOM}
\,
\end{align}
where a  $C\neq 1$ expresses the effect of dipole-dipole interactions (charging energy) and we have added a phenomenological damping $\gamma$. 
Thus the time derivative of an interlayer voltage $\phi_a$ provides a force that pushes the phase  to increase, meaning that a suitable voltage pulse can switch the system between ground states as in \fig{fig:bilayer}(b)(c).
If $\phi_a$ is applied by side contacts or by gates immediately adjacent to the bilayer,  the external electrical circuit controls $\phi_a$ which is already the total voltage across the layers, and one has $C=1$ in \equa{eqn:EOM}.
To climb the potential hill at $\theta=\pi/2$ with energy $\nu \Delta_{\text{p}}^2/4$, the threshold voltage required for a typical pulse $\phi_a=\phi_0 e^{-(t-t_0)^2/T_0^2}$ is  $\phi_c \sim T_0  \Delta_{\text{p}}^2 C/D$, giving $\phi_c \sim 25 \unit{mV}$ for $T_0=1 \unit{ps}$, $\Delta_{\text{p}}=10 \unit{meV}$ and $C=1$. In the limit of strong drive ($\phi_a \gg \phi_c$), the equation of motion becomes $\partial_t \theta= -\phi_a$, recovering the familiar AC Josephson effect. 
Note that the switching frequency scale $1/T_0$ is upper bounded by the gap $\Delta$.

We have assumed that lattice distortions, if present, can dynamically follow the order parameter. In the opposite limit of slow lattice dynamics, one should fix the lattice distortion. For weak electron lattice coupling (ELC), the only change is that the $Z_2$ symmetry remains broken and the second minimum is at higher energy \cite{Murakami.2017}. For larger ELC the second minimum no longer exists. Thus fast phase steering can reveal the strength of ELC.

\emph{Beyond bilayers---}The second order Josephson effect generalizes to the 3D/2D systems by stacking the electron-hole bilayers/chains as in Figs.~\ref{fig:stacks}(a),(b). The stacking is along $z$ and the conjugate wavevector is $k_z \in (-\pi,\pi]/(2d)$. 
The model is invariant under translations by the z-direction lattice constant $2d$ and reflection $z \leftrightarrow -z$ with respect to a plane containing either the electron or holes.  We specialize to short ranged density-density interaction $g$ such that excitonic order 
$\Delta_{i1/2}$ only links adjacent layers as  in  \fig{fig:stacks}, 
and consider mean field solutions where the amplitude $\Delta$ is spatially uniform but  allow for the phases $\theta_{1,2}$ on the two bonds to be different. We define the symmetric and antisymmetric phase combinations $\theta_{s,a}=\left(\theta_1\pm\theta_2\right)/2$ whose domain is $\theta_s \in (-\pi,\pi],\, \theta_a \in [0,\pi)$.  In the momentum basis of field operators $\psi_{k}^\dagger=\left(\psi_{1k}^\dagger,\, \psi_{2k}^\dagger\right)=\int dr \sum_j e^{i(k_\perp r +k_z j 2d)} \left(\psi_{j1}(r),\, e^{ik_z d} \psi_{j2}(r) \right)$ where $k_\perp$ is the momentum along the planes/chains, the Lagrangian reads $L=\sum_k \psi_{k}^\dagger (\partial_\tau + H_k) \psi_{k} + \frac{2}{g} |\Delta|^2$ with the mean field Hamiltonian 
\begin{align}
H_k=
\begin{pmatrix}
\xi_1 (k_\perp)  & \Delta(k) - i\Delta_{\text{p}} f_{k} \cos d k_z \\
\Delta(k)^\ast + i\Delta_{\text{p}} f_{k} \cos d k_z & \xi_{2}(k_\perp)
\end{pmatrix}
\,
\label{eqn:H_3D}
\end{align}
where the $\Delta_{\text{p}}$ term is the intrinsic interlayer tunneling $t_k$ and 
the order parameter is
\begin{align}
	& \Delta(k) =e^{i \theta_a} \Delta \cos(d k_z + \theta_s)
	\,.
	\label{eqn:Delta_k_Ek}
\end{align}  
Our gauge choice here is that a spatially uniform electric field enters through  $k\rightarrow k+A$, including the  $\Delta(k)$ term.

At $\Delta_{\text{p}}=0$, the energy is independent of $\theta_{1}$ and $\theta_{2}$.
Nonzero $\Delta_{\text{p}}$ reduces the  symmetry to $\hat{T}$ and $\hat{P}$, and the excitonic ground state turns out to spontaneously break $\hat{T}$ instead of $\hat{P}$, corresponding to $(\theta_a,\theta_s)=(0, \pm \pi/2)$, i.e., $\theta_{i1}=\theta_{i2}=\pm\pi/2$. This is verified by expanding the Lagrangian  to quadratic order in  $\Delta_{\text{p}}$ (see Ref.~\cite{SI} Sec.~\ref{SI:stack}). Fixing $\theta_a=0$ and in the gauge $\phi=0$, one finds:
\begin{align}
L = & K[\dot{\theta}_s+d \dot{A}_z, A_x]  + c_\nu \Delta_{\text{p}}^2
\cos2\theta_s + F_0 
\label{eqn:L_stack_expansion}
\,
\end{align}
where $K$ is the kinetic term that vanishes in the static limit, $F_0(|\Delta|)$ is the ground state free energy without interlayer tunneling, and we have neglected constant $O(\Delta_{\text{p}}^2)$ terms.
The $\cos 2\theta_s$ term means a `second order Josephson' current $j_z=j_c \sin 2\theta_s$ where $j_c=2 d c_\nu \Delta_{\text{p}}^2$
and $c_\nu \sim \nu$.  In the equilibrium state, the total electrical polarization is zero but there are circulating currents $j_{\text{inter},a}=\langle \sum_k (\partial_k t_k) \sin(d k_z) \sigma_1 \rangle $ 
due to broken $\hat{T}$, as shown in Fig.~\ref{fig:stacks}(b).  Note that this state is linearly stable to lattice distortions.

Around each of the two equilibrium configurations, expanding  \equa{eqn:L_stack_expansion} to quadratic order in $\theta \equiv \theta_s\pm\pi/2$ and the EM fields $A_{x/z}$, one obtains the Gaussian action for $\theta_s$ fluctuations.
In the low energy regime $\omega\ll \Delta_{\text{p}}$ and long wavelength limit $q=0$, it reads
\begin{align}
S_s = & - \sum_{\omega} c_0(\omega) \big(\theta+ dA_z \big)_{-\omega} 
\big(\theta+ dA_z \big)_{\omega} + 
\notag \\
& 
\int dt dr \left[  c_1 \theta^2 
+\sigma_{\text{h}}(\theta+ dA_z) \partial_t A_x /d \right]  
+ S_{A_x^2}   
\label{eqn:L_s_p}
\,
\end{align}
neglecting terms subleading in $\Delta_{\text{p}}$. 
The first two terms are the kinetic and potential energies of phase fluctuations where $c_0(\omega)$ is the kinetic kernel that vanishes in the static limit and $c_1 =2c_\nu \Delta_{\text{p}}^2$ for $\Delta_{\text{p}} \ll \Delta$. The third term gives rise to an anomalous hall conductivity $\sigma_{\text{h}}$ for electric fields in the x-z plane which can also be written into an `Axion' form \cite{Armitage.2018_weyl_review}. The last term is the bare optical response in $x$ direction.

The excitonic order leads to topologically nontrivial ground states in the BCS regime ($G >0$). Setting $\xi_1(k)=-\xi_2(k)=\xi_k$ for simplicity, the quasiparticle dispersion is $E_k =\pm \sqrt{\xi_k^2 + |\Delta(k)|^2 + \Delta_{\text{p}}^2 f_k^2 \cos^2(d k_z)}$.
In the 2D stack of electron-hole chains,  
the quasiparticle is gapped with massive Dirac points at $(k_x,k_z)=(\pm k_{\text{F}},0)$ with mass $\pm \Delta_{\text{p}}$, as shown in Fig.~\ref{fig:stacks}(c). 
The Chern number of the valence band is $ \text{Sign}[\theta_s]$ so that the system is a quantum anomalous Hall insulator \cite{Liu.2016} with quantized Hall conductivity $\sigma_{\text{h}}=\text{Sign}[\theta_s] e^2/h$ and chiral edge states.  The kinetic kernel $c_0=\frac{\nu}{3} \omega^2  \Delta/\Delta_{\text{p}}$ renders the bare phase mode gap $\omega_0 \sim \Delta_{\text{p}} \sqrt{\Delta_{\text{p}}/\Delta}$.
The 3D stack of electron-hole layers is a Weyl semimetal \cite{Armitage.2018_weyl_review} with Weyl nodes at $k=(0,\pm k_{\text{F}},0)$ and anomalous hall conductivity $\sigma_{\text{h}}=\text{Sign}[\theta_s] \frac{k_{\text{F}}}{\pi}e^2/h$  (see Ref.~\cite{SI} Sec.~\ref{SI:stack}, and Sec.~\ref{SI:spin} for the effect of spins).
Note that the BEC regime ($G <0$) is topologically trivial with $\sigma_{\text{h}}$ vanishing and the minimal gap being $\sqrt{G^2+4\Delta^2}$, although there is nonzero AC hall response $\sim \Delta_{\text{p}}$ which can be measured by Kerr rotation (neglected in \equa{eqn:L_s_p}).

\emph{Order parameter steering by light---}In all these systems, the order parameter can be steered by electric fields perpendicular to the layers/chains, e.g., from ground state $|g \rangle$ to $\hat{P} |g \rangle$ for the bilayer and to $\hat{T} |g \rangle$ for the 3D/2D stacks. This can be easily verified in `pump-probe' experiments since the ground states have opposite in-plane polarization in the bilayer and opposite hall response in the stacks.
The order parameter steering follows the spirit of AC Josephson effect: the phase $\theta_s$ enters the kinetic term in \equa{eqn:L_stack_expansion} together with the vector potential as $\theta_s+d A_z$.
This term has different forms in different regimes. For example in the 2D stacks, it behaves as $K \sim \nu |\frac{\Delta}{\Delta_{\text{p}}}| \dot{\theta}_s^2$ in the slow limit of $\dot{\theta}_s \ll \Delta_{\text{p}}$ and as $K \sim \nu |\Delta| \theta_s \dot{\theta}_s$ in the moderately fast case of $ \Delta_{\text{p}} \ll \dot{\theta}_s \ll \Delta$ where we have suppressed $A_z$ for notational simplicity. Nevertheless, upon strong electric field $E_z$ such that the free energy potential $\cos 2\theta_s$ can be neglected, the equation of motion all reduces to $\dot{\theta}_s=d \dot{A}_z = -d E_z$, i.e., the electric field provides a force to rotate the phase $\theta_s$  so as to switch the system between the two ground states $\theta_s=\pm \pi/2$ (\fig{fig:stacks}(d)). The pulse that exactly delivers such a switch is  $d \int E_z(t) dt= \pi$. For a pulse duration of $1 \unit{ps}$ and $d=1 \unit{nm}$, the field needed is $E_z \sim 2 \times 10^4 \unit{V/cm}$.
For weaker fields such that the free energy potential matters, the dynamics depends on the time scale. In the  case of $ \Delta_{\text{p}} \ll \dot{\theta}_s \ll \Delta$, the equation of motion implied by \equa{eqn:L_stack_expansion} is simply
\begin{align}
\dot{\theta}_s = \frac{\Delta_{\text{p}}^2}{4 |\Delta|}   \sin (2\theta_s) - d E_z
\label{eqn:steering}
\,.
\end{align}
The threshold field to climb over the potential barrier and switch the ground states is about $E_c \sim \frac{\Delta_{\text{p}}^2}{|\Delta| d}$ which reads $E_c \sim 10^4 \unit{V/cm}$ for $\Delta_{\text{p}}=10 \unit{meV},\, |\Delta|=100 \unit{meV}$ and $d=1 \unit{nm}$.

\emph{Discussion}---The bilayer could be realized by, e.g., gating suitably stacked phosphorene bilayer \cite{Kim.2015_phorsphorene, Li.2014_phosphorene,Carvalho.2016_phosphorene} or  transition metal dichalcogenide bilayers \cite{Wang.2019, Ma.2021strongly}  to bring the conduction band of one layer and valence band (different in symmetry under $C_2$ or $C_3$ rotations around z, respectively) of the other layer closer in energy, entering the EI phase (see Ref.~\cite{SI} Sec.~\ref{SI:device}). The 3D/2D stacks may be either natural crystals such as monolayer WTe$_2$ (a 2D stack of chains) \cite{jia.2020,kwan2020theory, Varsano.2020} or artificial structures. Realizations of these topological excitonic insulators \cite{Zhu.2019, Wang2019a, Varsano.2020,Hu2019, Perfetto2020, liu2021topological} is an important research direction.

The order parameter steering also applies to the EI candidate Ta$_2$NiSe$_5$ \cite{Kaneko2013, Lu2017, Werdehausen2018, Sugimoto2018, Mazza2020,Ning.2020}.
Its basic structural unit is the Ta-Ni-Ta chain, with Ta-derived conduction band states even under reflection $\sigma_{\perp}:x\rightarrow -x$  while the Ni-derived valence band states are  odd \cite{Kaneko2013, Mazza2020}. The EI state breaks $\sigma_\perp$ and while the detailed electronic structure complicates the discussion of the Josephson effect, the phase dynamics is still described by Eqs.~\eqref{eqn:EOM} and \eqref{eqn:steering}  and  a photon pulse perpendicular to the chains can still switch the system between its two ground states (see Ref.~\cite{SI} Sec.~\ref{SI:TNS}). This may have already been observed \cite{Ning.2020}.

Fluctuations will not destroy our qualitative conclusions. Without the $U(1)$ breaking Josephson term $\cos 2\theta$, the exciton condensate in \equa{eqn:L_bilayer} has quasi long range order at temperatures $T$ below the Berezinskii-Kosterlitz-Thouless temperature $T_{\text{BKT}}$ \cite{Berezinsky.1971, Kosterlitz.1973}.  According to renormalization group analysis \cite{Jose.1977}, the Josephson coupling is a relevant  one at $T<T_{\text{BKT}}$ which renders the EI state strictly long range ordered. However, the coupling  (and the Josephson current) is renormalized by fluctuations to a  power $1/(1- \frac{T}{4T_{\text{BKT}}})$ of its bare value (see Ref.~\cite{SI} Sec.~\ref{SI:beyond_mean_field}).

\begin{acknowledgements}
Z. S. and A. J. M. acknowledge support from the
Energy Frontier Research Center on Programmable Quantum Materials funded
by the US Department of Energy (DOE), Office of Science, Basic Energy Sciences (BES), under award No. DE-SC0019443.
T. K. is supported by Grants-in-Aid for Scientific Research from JSPS (Grant Nos JP18K13509) and by the JSPS Overseas Research Fellowship. D. G. is supported by Slovenian Research Agency (ARRS) under Program J1-2455 and P1-0044. The Flatiron Institute is a division of the Simons Foundation. We thank M. M. Fogler, S. Zhang, Y. Murakami, H. Ning and Z. Meng for helpful discussions.
\end{acknowledgements}

\bibliographystyle{apsrev4-1}
\bibliography{Excitonic_Insulator.bib}

\begin{thebibliography}{87}%
\makeatletter
\providecommand \@ifxundefined [1]{%
 \@ifx{#1\undefined}
}%
\providecommand \@ifnum [1]{%
 \ifnum #1\expandafter \@firstoftwo
 \else \expandafter \@secondoftwo
 \fi
}%
\providecommand \@ifx [1]{%
 \ifx #1\expandafter \@firstoftwo
 \else \expandafter \@secondoftwo
 \fi
}%
\providecommand \natexlab [1]{#1}%
\providecommand \enquote  [1]{``#1''}%
\providecommand \bibnamefont  [1]{#1}%
\providecommand \bibfnamefont [1]{#1}%
\providecommand \citenamefont [1]{#1}%
\providecommand \href@noop [0]{\@secondoftwo}%
\providecommand \href [0]{\begingroup \@sanitize@url \@href}%
\providecommand \@href[1]{\@@startlink{#1}\@@href}%
\providecommand \@@href[1]{\endgroup#1\@@endlink}%
\providecommand \@sanitize@url [0]{\catcode `\\12\catcode `\$12\catcode
  `\&12\catcode `\#12\catcode `\^12\catcode `\_12\catcode `\%12\relax}%
\providecommand \@@startlink[1]{}%
\providecommand \@@endlink[0]{}%
\providecommand \url  [0]{\begingroup\@sanitize@url \@url }%
\providecommand \@url [1]{\endgroup\@href {#1}{\urlprefix }}%
\providecommand \urlprefix  [0]{URL }%
\providecommand \Eprint [0]{\href }%
\providecommand \doibase [0]{http://dx.doi.org/}%
\providecommand \selectlanguage [0]{\@gobble}%
\providecommand \bibinfo  [0]{\@secondoftwo}%
\providecommand \bibfield  [0]{\@secondoftwo}%
\providecommand \translation [1]{[#1]}%
\providecommand \BibitemOpen [0]{}%
\providecommand \bibitemStop [0]{}%
\providecommand \bibitemNoStop [0]{.\EOS\space}%
\providecommand \EOS [0]{\spacefactor3000\relax}%
\providecommand \BibitemShut  [1]{\csname bibitem#1\endcsname}%
\let\auto@bib@innerbib\@empty
\bibitem [{\citenamefont {Mott}(1961)}]{Mott1961}%
  \BibitemOpen
  \bibfield  {author} {\bibinfo {author} {\bibfnamefont {N.~F.}\ \bibnamefont
  {Mott}},\ }\href {\doibase 10.1080/14786436108243318} {\bibfield  {journal}
  {\bibinfo  {journal} {Philos. Mag.}\ }\textbf {\bibinfo {volume} {6}},\
  \bibinfo {pages} {287} (\bibinfo {year} {1961})}\BibitemShut {NoStop}%
\bibitem [{\citenamefont {Keldysh}\ and\ \citenamefont
  {Kopaev}(1965)}]{Keldysh1965}%
  \BibitemOpen
  \bibfield  {author} {\bibinfo {author} {\bibfnamefont {L.~V.}\ \bibnamefont
  {Keldysh}}\ and\ \bibinfo {author} {\bibfnamefont {Y.~V.}\ \bibnamefont
  {Kopaev}},\ }\href@noop {} {\bibfield  {journal} {\bibinfo  {journal} {Soviet
  Phys. Solid State}\ }\textbf {\bibinfo {volume} {6}},\ \bibinfo {pages}
  {2219} (\bibinfo {year} {1965})}\BibitemShut {NoStop}%
\bibitem [{\citenamefont {J{\'{e}}rome}\ \emph {et~al.}(1967)\citenamefont
  {J{\'{e}}rome}, \citenamefont {Rice},\ and\ \citenamefont
  {Kohn}}]{Jerome1967}%
  \BibitemOpen
  \bibfield  {author} {\bibinfo {author} {\bibfnamefont {D.}~\bibnamefont
  {J{\'{e}}rome}}, \bibinfo {author} {\bibfnamefont {T.~M.}\ \bibnamefont
  {Rice}}, \ and\ \bibinfo {author} {\bibfnamefont {W.}~\bibnamefont {Kohn}},\
  }\href {\doibase 10.1103/PhysRev.158.462} {\bibfield  {journal} {\bibinfo
  {journal} {Phys. Rev.}\ }\textbf {\bibinfo {volume} {158}},\ \bibinfo {pages}
  {462} (\bibinfo {year} {1967})}\BibitemShut {NoStop}%
\bibitem [{\citenamefont {Halperin}\ and\ \citenamefont
  {Rice}(1968{\natexlab{a}})}]{Halperin.1968}%
  \BibitemOpen
  \bibfield  {author} {\bibinfo {author} {\bibfnamefont {B.~I.}\ \bibnamefont
  {Halperin}}\ and\ \bibinfo {author} {\bibfnamefont {T.~M.}\ \bibnamefont
  {Rice}},\ }\href {\doibase 10.1103/RevModPhys.40.755} {\bibfield  {journal}
  {\bibinfo  {journal} {Rev. Mod. Phys.}\ }\textbf {\bibinfo {volume} {40}},\
  \bibinfo {pages} {755} (\bibinfo {year} {1968}{\natexlab{a}})}\BibitemShut
  {NoStop}%
\bibitem [{\citenamefont {Keldysh}\ and\ \citenamefont
  {Kozlov}(1968)}]{Keldysh1968}%
  \BibitemOpen
  \bibfield  {author} {\bibinfo {author} {\bibfnamefont {L.}~\bibnamefont
  {Keldysh}}\ and\ \bibinfo {author} {\bibfnamefont {A.}~\bibnamefont
  {Kozlov}},\ }\href@noop {} {\bibfield  {journal} {\bibinfo  {journal} {Sov.
  J. Exp. Theor. Phys.}\ }\textbf {\bibinfo {volume} {27}},\ \bibinfo {pages}
  {521} (\bibinfo {year} {1968})}\BibitemShut {NoStop}%
\bibitem [{\citenamefont {Butov}\ \emph {et~al.}(1994)\citenamefont {Butov},
  \citenamefont {Zrenner}, \citenamefont {Abstreiter}, \citenamefont {B\"ohm},\
  and\ \citenamefont {Weimann}}]{Butov.1994}%
  \BibitemOpen
  \bibfield  {author} {\bibinfo {author} {\bibfnamefont {L.~V.}\ \bibnamefont
  {Butov}}, \bibinfo {author} {\bibfnamefont {A.}~\bibnamefont {Zrenner}},
  \bibinfo {author} {\bibfnamefont {G.}~\bibnamefont {Abstreiter}}, \bibinfo
  {author} {\bibfnamefont {G.}~\bibnamefont {B\"ohm}}, \ and\ \bibinfo {author}
  {\bibfnamefont {G.}~\bibnamefont {Weimann}},\ }\href {\doibase
  10.1103/PhysRevLett.73.304} {\bibfield  {journal} {\bibinfo  {journal} {Phys.
  Rev. Lett.}\ }\textbf {\bibinfo {volume} {73}},\ \bibinfo {pages} {304}
  (\bibinfo {year} {1994})}\BibitemShut {NoStop}%
\bibitem [{\citenamefont {Butov}\ \emph {et~al.}(2002)\citenamefont {Butov},
  \citenamefont {Gossard},\ and\ \citenamefont {Chemla}}]{Butov2002}%
  \BibitemOpen
  \bibfield  {author} {\bibinfo {author} {\bibfnamefont {L.~V.}\ \bibnamefont
  {Butov}}, \bibinfo {author} {\bibfnamefont {A.~C.}\ \bibnamefont {Gossard}},
  \ and\ \bibinfo {author} {\bibfnamefont {D.~S.}\ \bibnamefont {Chemla}},\
  }\href {https://doi.org/10.1038/nature00943} {\bibfield  {journal} {\bibinfo
  {journal} {Nature}\ }\textbf {\bibinfo {volume} {418}},\ \bibinfo {pages}
  {751} (\bibinfo {year} {2002})}\BibitemShut {NoStop}%
\bibitem [{\citenamefont {Du}\ \emph {et~al.}(2017)\citenamefont {Du},
  \citenamefont {Li}, \citenamefont {Lou}, \citenamefont {Sullivan},
  \citenamefont {Chang}, \citenamefont {Kono},\ and\ \citenamefont
  {Du}}]{Du2017}%
  \BibitemOpen
  \bibfield  {author} {\bibinfo {author} {\bibfnamefont {L.}~\bibnamefont
  {Du}}, \bibinfo {author} {\bibfnamefont {X.}~\bibnamefont {Li}}, \bibinfo
  {author} {\bibfnamefont {W.}~\bibnamefont {Lou}}, \bibinfo {author}
  {\bibfnamefont {G.}~\bibnamefont {Sullivan}}, \bibinfo {author}
  {\bibfnamefont {K.}~\bibnamefont {Chang}}, \bibinfo {author} {\bibfnamefont
  {J.}~\bibnamefont {Kono}}, \ and\ \bibinfo {author} {\bibfnamefont {R.~R.}\
  \bibnamefont {Du}},\ }\href {\doibase 10.1038/s41467-017-01988-1} {\bibfield
  {journal} {\bibinfo  {journal} {Nat. Commun.}\ }\textbf {\bibinfo {volume}
  {8}},\ \bibinfo {pages} {1971} (\bibinfo {year} {2017})}\BibitemShut
  {NoStop}%
\bibitem [{\citenamefont {Li}\ \emph {et~al.}(2017)\citenamefont {Li},
  \citenamefont {Taniguchi}, \citenamefont {Watanabe}, \citenamefont {Hone},\
  and\ \citenamefont {Dean}}]{Li2017}%
  \BibitemOpen
  \bibfield  {author} {\bibinfo {author} {\bibfnamefont {J.~I.}\ \bibnamefont
  {Li}}, \bibinfo {author} {\bibfnamefont {T.}~\bibnamefont {Taniguchi}},
  \bibinfo {author} {\bibfnamefont {K.}~\bibnamefont {Watanabe}}, \bibinfo
  {author} {\bibfnamefont {J.}~\bibnamefont {Hone}}, \ and\ \bibinfo {author}
  {\bibfnamefont {C.~R.}\ \bibnamefont {Dean}},\ }\href {\doibase
  10.1038/NPHYS4140} {\bibfield  {journal} {\bibinfo  {journal} {Nat. Phys.}\
  }\textbf {\bibinfo {volume} {13}},\ \bibinfo {pages} {751} (\bibinfo {year}
  {2017})}\BibitemShut {NoStop}%
\bibitem [{\citenamefont {Burg}\ \emph {et~al.}(2018)\citenamefont {Burg},
  \citenamefont {Prasad}, \citenamefont {Kim}, \citenamefont {Taniguchi},
  \citenamefont {Watanabe}, \citenamefont {MacDonald}, \citenamefont
  {Register},\ and\ \citenamefont {Tutuc}}]{Burg.2018}%
  \BibitemOpen
  \bibfield  {author} {\bibinfo {author} {\bibfnamefont {G.~W.}\ \bibnamefont
  {Burg}}, \bibinfo {author} {\bibfnamefont {N.}~\bibnamefont {Prasad}},
  \bibinfo {author} {\bibfnamefont {K.}~\bibnamefont {Kim}}, \bibinfo {author}
  {\bibfnamefont {T.}~\bibnamefont {Taniguchi}}, \bibinfo {author}
  {\bibfnamefont {K.}~\bibnamefont {Watanabe}}, \bibinfo {author}
  {\bibfnamefont {A.~H.}\ \bibnamefont {MacDonald}}, \bibinfo {author}
  {\bibfnamefont {L.~F.}\ \bibnamefont {Register}}, \ and\ \bibinfo {author}
  {\bibfnamefont {E.}~\bibnamefont {Tutuc}},\ }\href {\doibase
  10.1103/PhysRevLett.120.177702} {\bibfield  {journal} {\bibinfo  {journal}
  {Phys. Rev. Lett.}\ }\textbf {\bibinfo {volume} {120}},\ \bibinfo {pages}
  {177702} (\bibinfo {year} {2018})}\BibitemShut {NoStop}%
\bibitem [{\citenamefont {Wang}\ \emph
  {et~al.}(2019{\natexlab{a}})\citenamefont {Wang}, \citenamefont {Rhodes},
  \citenamefont {Watanabe}, \citenamefont {Taniguchi}, \citenamefont {Hone},
  \citenamefont {Shan},\ and\ \citenamefont {Mak}}]{Wang.2019}%
  \BibitemOpen
  \bibfield  {author} {\bibinfo {author} {\bibfnamefont {Z.}~\bibnamefont
  {Wang}}, \bibinfo {author} {\bibfnamefont {D.~A.}\ \bibnamefont {Rhodes}},
  \bibinfo {author} {\bibfnamefont {K.}~\bibnamefont {Watanabe}}, \bibinfo
  {author} {\bibfnamefont {T.}~\bibnamefont {Taniguchi}}, \bibinfo {author}
  {\bibfnamefont {J.~C.}\ \bibnamefont {Hone}}, \bibinfo {author}
  {\bibfnamefont {J.}~\bibnamefont {Shan}}, \ and\ \bibinfo {author}
  {\bibfnamefont {K.~F.}\ \bibnamefont {Mak}},\ }\href
  {https://doi.org/10.1038/s41586-019-1591-7} {\bibfield  {journal} {\bibinfo
  {journal} {Nature}\ }\textbf {\bibinfo {volume} {574}},\ \bibinfo {pages}
  {76} (\bibinfo {year} {2019}{\natexlab{a}})}\BibitemShut {NoStop}%
\bibitem [{\citenamefont {Ma}\ \emph {et~al.}(2021)\citenamefont {Ma},
  \citenamefont {Nguyen}, \citenamefont {Wang}, \citenamefont {Zeng},
  \citenamefont {Watanabe}, \citenamefont {Taniguchi}, \citenamefont
  {MacDonald}, \citenamefont {Mak},\ and\ \citenamefont
  {Shan}}]{Ma.2021strongly}%
  \BibitemOpen
  \bibfield  {author} {\bibinfo {author} {\bibfnamefont {L.}~\bibnamefont
  {Ma}}, \bibinfo {author} {\bibfnamefont {P.~X.}\ \bibnamefont {Nguyen}},
  \bibinfo {author} {\bibfnamefont {Z.}~\bibnamefont {Wang}}, \bibinfo {author}
  {\bibfnamefont {Y.}~\bibnamefont {Zeng}}, \bibinfo {author} {\bibfnamefont
  {K.}~\bibnamefont {Watanabe}}, \bibinfo {author} {\bibfnamefont
  {T.}~\bibnamefont {Taniguchi}}, \bibinfo {author} {\bibfnamefont {A.~H.}\
  \bibnamefont {MacDonald}}, \bibinfo {author} {\bibfnamefont {K.~F.}\
  \bibnamefont {Mak}}, \ and\ \bibinfo {author} {\bibfnamefont
  {J.}~\bibnamefont {Shan}},\ }\href@noop {} {\enquote {\bibinfo {title}
  {Strongly correlated excitonic insulator in atomic double layers},}\ }
  (\bibinfo {year} {2021}),\ \Eprint {http://arxiv.org/abs/2104.05066}
  {arXiv:2104.05066 [cond-mat.mtrl-sci]} \BibitemShut {NoStop}%
\bibitem [{\citenamefont {Eisenstein}(2014)}]{Eisenstein2014}%
  \BibitemOpen
  \bibfield  {author} {\bibinfo {author} {\bibfnamefont {J.}~\bibnamefont
  {Eisenstein}},\ }\href {\doibase 10.1146/annurev-conmatphys-031113-133832}
  {\bibfield  {journal} {\bibinfo  {journal} {Annu. Rev. Condens. Matter
  Phys.}\ }\textbf {\bibinfo {volume} {5}},\ \bibinfo {pages} {159} (\bibinfo
  {year} {2014})}\BibitemShut {NoStop}%
\bibitem [{\citenamefont {Fogler}\ \emph {et~al.}(2014)\citenamefont {Fogler},
  \citenamefont {Butov},\ and\ \citenamefont {Novoselov}}]{Fogler2014a}%
  \BibitemOpen
  \bibfield  {author} {\bibinfo {author} {\bibfnamefont {M.~M.}\ \bibnamefont
  {Fogler}}, \bibinfo {author} {\bibfnamefont {L.~V.}\ \bibnamefont {Butov}}, \
  and\ \bibinfo {author} {\bibfnamefont {K.~S.}\ \bibnamefont {Novoselov}},\
  }\href {\doibase 10.1038/ncomms5555} {\bibfield  {journal} {\bibinfo
  {journal} {Nat. Commun.}\ }\textbf {\bibinfo {volume} {5}},\ \bibinfo {pages}
  {4555} (\bibinfo {year} {2014})}\BibitemShut {NoStop}%
\bibitem [{\citenamefont {Liu}\ \emph {et~al.}(2017)\citenamefont {Liu},
  \citenamefont {Watanabe}, \citenamefont {Taniguchi}, \citenamefont
  {Halperin},\ and\ \citenamefont {Kim}}]{Liu.2017}%
  \BibitemOpen
  \bibfield  {author} {\bibinfo {author} {\bibfnamefont {X.}~\bibnamefont
  {Liu}}, \bibinfo {author} {\bibfnamefont {K.}~\bibnamefont {Watanabe}},
  \bibinfo {author} {\bibfnamefont {T.}~\bibnamefont {Taniguchi}}, \bibinfo
  {author} {\bibfnamefont {B.~I.}\ \bibnamefont {Halperin}}, \ and\ \bibinfo
  {author} {\bibfnamefont {P.}~\bibnamefont {Kim}},\ }\href {\doibase
  10.1038/nphys4116} {\bibfield  {journal} {\bibinfo  {journal} {Nature
  Physics}\ }\textbf {\bibinfo {volume} {13}},\ \bibinfo {pages} {746}
  (\bibinfo {year} {2017})}\BibitemShut {NoStop}%
\bibitem [{\citenamefont {Kulik}\ and\ \citenamefont
  {Shevchenko}(1976)}]{Kulik.1976}%
  \BibitemOpen
  \bibfield  {author} {\bibinfo {author} {\bibfnamefont {I.~O.}\ \bibnamefont
  {Kulik}}\ and\ \bibinfo {author} {\bibfnamefont {S.~I.}\ \bibnamefont
  {Shevchenko}},\ }\href@noop {} {\bibfield  {journal} {\bibinfo  {journal}
  {Fiz. Nizk. Temp.}\ }\textbf {\bibinfo {volume} {2}},\ \bibinfo {pages}
  {1405} (\bibinfo {year} {1976})},\ \bibinfo {note} {[Sov. J. Low Temp. Phys.
  2, 687 (1976)]}\BibitemShut {NoStop}%
\bibitem [{\citenamefont {Shevchenko}(1977)}]{Shevchenko1977}%
  \BibitemOpen
  \bibfield  {author} {\bibinfo {author} {\bibfnamefont {S.~I.}\ \bibnamefont
  {Shevchenko}},\ }\href@noop {} {\bibfield  {journal} {\bibinfo  {journal}
  {Fiz. Nizk. Temp.}\ }\textbf {\bibinfo {volume} {3}},\ \bibinfo {pages} {605}
  (\bibinfo {year} {1977})},\ \bibinfo {note} {[Sov. J. Low Temp. Phys. 3, 293
  (1977)]}\BibitemShut {NoStop}%
\bibitem [{\citenamefont {Lozovik}\ and\ \citenamefont
  {Poushnov}(1997)}]{Lozovik.1997}%
  \BibitemOpen
  \bibfield  {author} {\bibinfo {author} {\bibfnamefont {Y.~E.}\ \bibnamefont
  {Lozovik}}\ and\ \bibinfo {author} {\bibfnamefont {A.~V.}\ \bibnamefont
  {Poushnov}},\ }\href {\doibase 10.1016/S0375-9601(97)00133-3} {\bibfield
  {journal} {\bibinfo  {journal} {Physics Letters A}\ }\textbf {\bibinfo
  {volume} {228}},\ \bibinfo {pages} {399} (\bibinfo {year}
  {1997})}\BibitemShut {NoStop}%
\bibitem [{\citenamefont {Wen}\ and\ \citenamefont {Zee}(1992)}]{Wen.1992}%
  \BibitemOpen
  \bibfield  {author} {\bibinfo {author} {\bibfnamefont {X.-G.}\ \bibnamefont
  {Wen}}\ and\ \bibinfo {author} {\bibfnamefont {A.}~\bibnamefont {Zee}},\
  }\href {\doibase 10.1103/PhysRevLett.69.1811} {\bibfield  {journal} {\bibinfo
   {journal} {Phys. Rev. Lett.}\ }\textbf {\bibinfo {volume} {69}},\ \bibinfo
  {pages} {1811} (\bibinfo {year} {1992})}\BibitemShut {NoStop}%
\bibitem [{\citenamefont {Shevchenko}(1994)}]{Shevchenko1994}%
  \BibitemOpen
  \bibfield  {author} {\bibinfo {author} {\bibfnamefont {S.~I.}\ \bibnamefont
  {Shevchenko}},\ }\href {\doibase 10.1103/PhysRevLett.72.3242} {\bibfield
  {journal} {\bibinfo  {journal} {Phys. Rev. Lett.}\ }\textbf {\bibinfo
  {volume} {72}},\ \bibinfo {pages} {3242} (\bibinfo {year}
  {1994})}\BibitemShut {NoStop}%
\bibitem [{\citenamefont {Spielman}\ \emph {et~al.}(2000)\citenamefont
  {Spielman}, \citenamefont {Eisenstein}, \citenamefont {Pfeiffer},\ and\
  \citenamefont {West}}]{Spielman2000}%
  \BibitemOpen
  \bibfield  {author} {\bibinfo {author} {\bibfnamefont {I.~B.}\ \bibnamefont
  {Spielman}}, \bibinfo {author} {\bibfnamefont {J.~P.}\ \bibnamefont
  {Eisenstein}}, \bibinfo {author} {\bibfnamefont {L.~N.}\ \bibnamefont
  {Pfeiffer}}, \ and\ \bibinfo {author} {\bibfnamefont {K.~W.}\ \bibnamefont
  {West}},\ }\href {\doibase 10.1103/PhysRevLett.84.5808} {\bibfield  {journal}
  {\bibinfo  {journal} {Phys. Rev. Lett.}\ }\textbf {\bibinfo {volume} {84}},\
  \bibinfo {pages} {5808} (\bibinfo {year} {2000})}\BibitemShut {NoStop}%
\bibitem [{\citenamefont {Spielman}\ \emph {et~al.}(2001)\citenamefont
  {Spielman}, \citenamefont {Eisenstein}, \citenamefont {Pfeiffer},\ and\
  \citenamefont {West}}]{Spielman2001}%
  \BibitemOpen
  \bibfield  {author} {\bibinfo {author} {\bibfnamefont {I.~B.}\ \bibnamefont
  {Spielman}}, \bibinfo {author} {\bibfnamefont {J.~P.}\ \bibnamefont
  {Eisenstein}}, \bibinfo {author} {\bibfnamefont {L.~N.}\ \bibnamefont
  {Pfeiffer}}, \ and\ \bibinfo {author} {\bibfnamefont {K.~W.}\ \bibnamefont
  {West}},\ }\href {\doibase 10.1103/PhysRevLett.87.036803} {\bibfield
  {journal} {\bibinfo  {journal} {Phys. Rev. Lett.}\ }\textbf {\bibinfo
  {volume} {87}},\ \bibinfo {pages} {036803} (\bibinfo {year}
  {2001})}\BibitemShut {NoStop}%
\bibitem [{\citenamefont {Fogler}\ and\ \citenamefont
  {Wilczek}(2001)}]{Fogler2001a}%
  \BibitemOpen
  \bibfield  {author} {\bibinfo {author} {\bibfnamefont {M.~M.}\ \bibnamefont
  {Fogler}}\ and\ \bibinfo {author} {\bibfnamefont {F.}~\bibnamefont
  {Wilczek}},\ }\href {\doibase 10.1103/PhysRevLett.86.1833} {\bibfield
  {journal} {\bibinfo  {journal} {Phys. Rev. Lett.}\ }\textbf {\bibinfo
  {volume} {86}},\ \bibinfo {pages} {1833} (\bibinfo {year}
  {2001})}\BibitemShut {NoStop}%
\bibitem [{\citenamefont {Stern}\ \emph {et~al.}(2001)\citenamefont {Stern},
  \citenamefont {Girvin}, \citenamefont {MacDonald},\ and\ \citenamefont
  {Ma}}]{Stern2001}%
  \BibitemOpen
  \bibfield  {author} {\bibinfo {author} {\bibfnamefont {A.}~\bibnamefont
  {Stern}}, \bibinfo {author} {\bibfnamefont {S.~M.}\ \bibnamefont {Girvin}},
  \bibinfo {author} {\bibfnamefont {A.~H.}\ \bibnamefont {MacDonald}}, \ and\
  \bibinfo {author} {\bibfnamefont {N.}~\bibnamefont {Ma}},\ }\href {\doibase
  10.1103/PhysRevLett.86.1829} {\bibfield  {journal} {\bibinfo  {journal}
  {Phys. Rev. Lett.}\ }\textbf {\bibinfo {volume} {86}},\ \bibinfo {pages}
  {1829} (\bibinfo {year} {2001})}\BibitemShut {NoStop}%
\bibitem [{\citenamefont {Joglekar}\ and\ \citenamefont
  {MacDonald}(2001)}]{Joglekar.2001}%
  \BibitemOpen
  \bibfield  {author} {\bibinfo {author} {\bibfnamefont {Y.~N.}\ \bibnamefont
  {Joglekar}}\ and\ \bibinfo {author} {\bibfnamefont {A.~H.}\ \bibnamefont
  {MacDonald}},\ }\href {\doibase 10.1103/PhysRevLett.87.196802} {\bibfield
  {journal} {\bibinfo  {journal} {Phys. Rev. Lett.}\ }\textbf {\bibinfo
  {volume} {87}},\ \bibinfo {pages} {196802} (\bibinfo {year}
  {2001})}\BibitemShut {NoStop}%
\bibitem [{\citenamefont {Balents}\ and\ \citenamefont
  {Radzihovsky}(2001)}]{Balents.2001}%
  \BibitemOpen
  \bibfield  {author} {\bibinfo {author} {\bibfnamefont {L.}~\bibnamefont
  {Balents}}\ and\ \bibinfo {author} {\bibfnamefont {L.}~\bibnamefont
  {Radzihovsky}},\ }\href {\doibase 10.1103/PhysRevLett.86.1825} {\bibfield
  {journal} {\bibinfo  {journal} {Phys. Rev. Lett.}\ }\textbf {\bibinfo
  {volume} {86}},\ \bibinfo {pages} {1825} (\bibinfo {year}
  {2001})}\BibitemShut {NoStop}%
\bibitem [{\citenamefont {Halperin}\ and\ \citenamefont
  {Rice}(1968{\natexlab{b}})}]{Halperin.1968_2}%
  \BibitemOpen
  \bibfield  {author} {\bibinfo {author} {\bibfnamefont {B.}~\bibnamefont
  {Halperin}}\ and\ \bibinfo {author} {\bibfnamefont {T.}~\bibnamefont
  {Rice}},\ }\href {\doibase https://doi.org/10.1016/S0081-1947(08)60740-7}
  {\emph {\bibinfo {title} {Solid State Physics}}},\ edited by\ \bibinfo
  {editor} {\bibfnamefont {F.}~\bibnamefont {Seitz}}, \bibinfo {editor}
  {\bibfnamefont {D.}~\bibnamefont {Turnbull}}, \ and\ \bibinfo {editor}
  {\bibfnamefont {H.}~\bibnamefont {Ehrenreich}},\ Vol.~\bibinfo {volume} {21}\
  (\bibinfo  {publisher} {Academic Press},\ \bibinfo {year} {1968})\ pp.\
  \bibinfo {pages} {115 -- 192}\BibitemShut {NoStop}%
\bibitem [{\citenamefont {Portengen}\ \emph {et~al.}(1996)\citenamefont
  {Portengen}, \citenamefont {{\"{O}}streich},\ and\ \citenamefont
  {Sham}}]{Portengen1996}%
  \BibitemOpen
  \bibfield  {author} {\bibinfo {author} {\bibfnamefont {T.}~\bibnamefont
  {Portengen}}, \bibinfo {author} {\bibfnamefont {T.}~\bibnamefont
  {{\"{O}}streich}}, \ and\ \bibinfo {author} {\bibfnamefont {L.~J.}\
  \bibnamefont {Sham}},\ }\href {\doibase 10.1103/PhysRevB.54.17452} {\bibfield
   {journal} {\bibinfo  {journal} {Phys. Rev. B}\ }\textbf {\bibinfo {volume}
  {54}},\ \bibinfo {pages} {17452} (\bibinfo {year} {1996})}\BibitemShut
  {NoStop}%
\bibitem [{\citenamefont {Mazza}\ \emph {et~al.}(2020)\citenamefont {Mazza},
  \citenamefont {R\"osner}, \citenamefont {Windg\"atter}, \citenamefont
  {Latini}, \citenamefont {H\"ubener}, \citenamefont {Millis}, \citenamefont
  {Rubio},\ and\ \citenamefont {Georges}}]{Mazza2020}%
  \BibitemOpen
  \bibfield  {author} {\bibinfo {author} {\bibfnamefont {G.}~\bibnamefont
  {Mazza}}, \bibinfo {author} {\bibfnamefont {M.}~\bibnamefont {R\"osner}},
  \bibinfo {author} {\bibfnamefont {L.}~\bibnamefont {Windg\"atter}}, \bibinfo
  {author} {\bibfnamefont {S.}~\bibnamefont {Latini}}, \bibinfo {author}
  {\bibfnamefont {H.}~\bibnamefont {H\"ubener}}, \bibinfo {author}
  {\bibfnamefont {A.~J.}\ \bibnamefont {Millis}}, \bibinfo {author}
  {\bibfnamefont {A.}~\bibnamefont {Rubio}}, \ and\ \bibinfo {author}
  {\bibfnamefont {A.}~\bibnamefont {Georges}},\ }\href {\doibase
  10.1103/PhysRevLett.124.197601} {\bibfield  {journal} {\bibinfo  {journal}
  {Phys. Rev. Lett.}\ }\textbf {\bibinfo {volume} {124}},\ \bibinfo {pages}
  {197601} (\bibinfo {year} {2020})}\BibitemShut {NoStop}%
\bibitem [{\citenamefont {Kaneko}\ \emph {et~al.}(2021)\citenamefont {Kaneko},
  \citenamefont {Sun}, \citenamefont {Murakami}, \citenamefont
  {Gole\ifmmode~\check{z}\else \v{z}\fi{}},\ and\ \citenamefont
  {Millis}}]{kaneko2020bulk}%
  \BibitemOpen
  \bibfield  {author} {\bibinfo {author} {\bibfnamefont {T.}~\bibnamefont
  {Kaneko}}, \bibinfo {author} {\bibfnamefont {Z.}~\bibnamefont {Sun}},
  \bibinfo {author} {\bibfnamefont {Y.}~\bibnamefont {Murakami}}, \bibinfo
  {author} {\bibfnamefont {D.}~\bibnamefont {Gole\ifmmode~\check{z}\else
  \v{z}\fi{}}}, \ and\ \bibinfo {author} {\bibfnamefont {A.~J.}\ \bibnamefont
  {Millis}},\ }\href {\doibase 10.1103/PhysRevLett.127.127402} {\bibfield
  {journal} {\bibinfo  {journal} {Phys. Rev. Lett.}\ }\textbf {\bibinfo
  {volume} {127}},\ \bibinfo {pages} {127402} (\bibinfo {year}
  {2021})}\BibitemShut {NoStop}%
\bibitem [{\citenamefont {Lenk}\ and\ \citenamefont
  {Eckstein}(2020)}]{Lenk.2020}%
  \BibitemOpen
  \bibfield  {author} {\bibinfo {author} {\bibfnamefont {K.}~\bibnamefont
  {Lenk}}\ and\ \bibinfo {author} {\bibfnamefont {M.}~\bibnamefont
  {Eckstein}},\ }\href {\doibase 10.1103/PhysRevB.102.205129} {\bibfield
  {journal} {\bibinfo  {journal} {Phys. Rev. B}\ }\textbf {\bibinfo {volume}
  {102}},\ \bibinfo {pages} {205129} (\bibinfo {year} {2020})}\BibitemShut
  {NoStop}%
\bibitem [{\citenamefont {Golubov}\ \emph {et~al.}(2004)\citenamefont
  {Golubov}, \citenamefont {Kupriyanov},\ and\ \citenamefont
  {Il'ichev}}]{Golubov.2004}%
  \BibitemOpen
  \bibfield  {author} {\bibinfo {author} {\bibfnamefont {A.~A.}\ \bibnamefont
  {Golubov}}, \bibinfo {author} {\bibfnamefont {M.~Y.}\ \bibnamefont
  {Kupriyanov}}, \ and\ \bibinfo {author} {\bibfnamefont {E.}~\bibnamefont
  {Il'ichev}},\ }\href {\doibase 10.1103/RevModPhys.76.411} {\bibfield
  {journal} {\bibinfo  {journal} {Rev. Mod. Phys.}\ }\textbf {\bibinfo {volume}
  {76}},\ \bibinfo {pages} {411} (\bibinfo {year} {2004})}\BibitemShut
  {NoStop}%
\bibitem [{\citenamefont {Tanaka}(1994)}]{Tanaka.1994}%
  \BibitemOpen
  \bibfield  {author} {\bibinfo {author} {\bibfnamefont {Y.}~\bibnamefont
  {Tanaka}},\ }\href {\doibase 10.1103/PhysRevLett.72.3871} {\bibfield
  {journal} {\bibinfo  {journal} {Phys. Rev. Lett.}\ }\textbf {\bibinfo
  {volume} {72}},\ \bibinfo {pages} {3871} (\bibinfo {year}
  {1994})}\BibitemShut {NoStop}%
\bibitem [{\citenamefont {Yip}(1995)}]{Yip.1995}%
  \BibitemOpen
  \bibfield  {author} {\bibinfo {author} {\bibfnamefont {S.}~\bibnamefont
  {Yip}},\ }\href {\doibase 10.1103/PhysRevB.52.3087} {\bibfield  {journal}
  {\bibinfo  {journal} {Phys. Rev. B}\ }\textbf {\bibinfo {volume} {52}},\
  \bibinfo {pages} {3087} (\bibinfo {year} {1995})}\BibitemShut {NoStop}%
\bibitem [{\citenamefont {Huck}\ \emph {et~al.}(1997)\citenamefont {Huck},
  \citenamefont {van Otterlo},\ and\ \citenamefont {Sigrist}}]{Huck.1997}%
  \BibitemOpen
  \bibfield  {author} {\bibinfo {author} {\bibfnamefont {A.}~\bibnamefont
  {Huck}}, \bibinfo {author} {\bibfnamefont {A.}~\bibnamefont {van Otterlo}}, \
  and\ \bibinfo {author} {\bibfnamefont {M.}~\bibnamefont {Sigrist}},\ }\href
  {\doibase 10.1103/PhysRevB.56.14163} {\bibfield  {journal} {\bibinfo
  {journal} {Phys. Rev. B}\ }\textbf {\bibinfo {volume} {56}},\ \bibinfo
  {pages} {14163} (\bibinfo {year} {1997})}\BibitemShut {NoStop}%
\bibitem [{\citenamefont {Zagoskin}(1997)}]{Zagoskin_1997}%
  \BibitemOpen
  \bibfield  {author} {\bibinfo {author} {\bibfnamefont {A.~M.}\ \bibnamefont
  {Zagoskin}},\ }\href {\doibase 10.1088/0953-8984/9/31/001} {\bibfield
  {journal} {\bibinfo  {journal} {Journal of Physics: Condensed Matter}\
  }\textbf {\bibinfo {volume} {9}},\ \bibinfo {pages} {L419} (\bibinfo {year}
  {1997})}\BibitemShut {NoStop}%
\bibitem [{\citenamefont {Il'ichev}\ \emph {et~al.}(1999)\citenamefont
  {Il'ichev}, \citenamefont {Zakosarenko}, \citenamefont {IJsselsteijn},
  \citenamefont {Hoenig}, \citenamefont {Schultze}, \citenamefont {Meyer},
  \citenamefont {Grajcar},\ and\ \citenamefont {Hlubina}}]{Ilichev.1999}%
  \BibitemOpen
  \bibfield  {author} {\bibinfo {author} {\bibfnamefont {E.}~\bibnamefont
  {Il'ichev}}, \bibinfo {author} {\bibfnamefont {V.}~\bibnamefont
  {Zakosarenko}}, \bibinfo {author} {\bibfnamefont {R.~P.~J.}\ \bibnamefont
  {IJsselsteijn}}, \bibinfo {author} {\bibfnamefont {H.~E.}\ \bibnamefont
  {Hoenig}}, \bibinfo {author} {\bibfnamefont {V.}~\bibnamefont {Schultze}},
  \bibinfo {author} {\bibfnamefont {H.-G.}\ \bibnamefont {Meyer}}, \bibinfo
  {author} {\bibfnamefont {M.}~\bibnamefont {Grajcar}}, \ and\ \bibinfo
  {author} {\bibfnamefont {R.}~\bibnamefont {Hlubina}},\ }\href {\doibase
  10.1103/PhysRevB.60.3096} {\bibfield  {journal} {\bibinfo  {journal} {Phys.
  Rev. B}\ }\textbf {\bibinfo {volume} {60}},\ \bibinfo {pages} {3096}
  (\bibinfo {year} {1999})}\BibitemShut {NoStop}%
\bibitem [{\citenamefont {Il'ichev}\ \emph {et~al.}(2001)\citenamefont
  {Il'ichev}, \citenamefont {Grajcar}, \citenamefont {Hlubina}, \citenamefont
  {IJsselsteijn}, \citenamefont {Hoenig}, \citenamefont {Meyer}, \citenamefont
  {Golubov}, \citenamefont {Amin}, \citenamefont {Zagoskin}, \citenamefont
  {Omelyanchouk},\ and\ \citenamefont {Kupriyanov}}]{Ilichev.2001}%
  \BibitemOpen
  \bibfield  {author} {\bibinfo {author} {\bibfnamefont {E.}~\bibnamefont
  {Il'ichev}}, \bibinfo {author} {\bibfnamefont {M.}~\bibnamefont {Grajcar}},
  \bibinfo {author} {\bibfnamefont {R.}~\bibnamefont {Hlubina}}, \bibinfo
  {author} {\bibfnamefont {R.~P.~J.}\ \bibnamefont {IJsselsteijn}}, \bibinfo
  {author} {\bibfnamefont {H.~E.}\ \bibnamefont {Hoenig}}, \bibinfo {author}
  {\bibfnamefont {H.-G.}\ \bibnamefont {Meyer}}, \bibinfo {author}
  {\bibfnamefont {A.}~\bibnamefont {Golubov}}, \bibinfo {author} {\bibfnamefont
  {M.~H.~S.}\ \bibnamefont {Amin}}, \bibinfo {author} {\bibfnamefont {A.~M.}\
  \bibnamefont {Zagoskin}}, \bibinfo {author} {\bibfnamefont {A.~N.}\
  \bibnamefont {Omelyanchouk}}, \ and\ \bibinfo {author} {\bibfnamefont
  {M.~Y.}\ \bibnamefont {Kupriyanov}},\ }\href {\doibase
  10.1103/PhysRevLett.86.5369} {\bibfield  {journal} {\bibinfo  {journal}
  {Phys. Rev. Lett.}\ }\textbf {\bibinfo {volume} {86}},\ \bibinfo {pages}
  {5369} (\bibinfo {year} {2001})}\BibitemShut {NoStop}%
\bibitem [{\citenamefont {Asano}(2001)}]{Asano.2001}%
  \BibitemOpen
  \bibfield  {author} {\bibinfo {author} {\bibfnamefont {Y.}~\bibnamefont
  {Asano}},\ }\href {\doibase 10.1103/PhysRevB.64.014511} {\bibfield  {journal}
  {\bibinfo  {journal} {Phys. Rev. B}\ }\textbf {\bibinfo {volume} {64}},\
  \bibinfo {pages} {014511} (\bibinfo {year} {2001})}\BibitemShut {NoStop}%
\bibitem [{\citenamefont {Zeng}\ \emph {et~al.}(2021)\citenamefont {Zeng},
  \citenamefont {Hu}, \citenamefont {Hu}, \citenamefont {You},\ and\
  \citenamefont {Wu}}]{zeng2021phasefluctuation}%
  \BibitemOpen
  \bibfield  {author} {\bibinfo {author} {\bibfnamefont {M.}~\bibnamefont
  {Zeng}}, \bibinfo {author} {\bibfnamefont {L.-H.}\ \bibnamefont {Hu}},
  \bibinfo {author} {\bibfnamefont {H.-Y.}\ \bibnamefont {Hu}}, \bibinfo
  {author} {\bibfnamefont {Y.-Z.}\ \bibnamefont {You}}, \ and\ \bibinfo
  {author} {\bibfnamefont {C.}~\bibnamefont {Wu}},\ }\href@noop {} {\enquote
  {\bibinfo {title} {Phase-fluctuation induced time-reversal symmetry breaking
  normal state},}\ } (\bibinfo {year} {2021}),\ \Eprint
  {http://arxiv.org/abs/2102.06158} {arXiv:2102.06158 [cond-mat.supr-con]}
  \BibitemShut {NoStop}%
\bibitem [{\citenamefont {Sun}\ \emph {et~al.}(2009)\citenamefont {Sun},
  \citenamefont {Yao}, \citenamefont {Fradkin},\ and\ \citenamefont
  {Kivelson}}]{Sun.2009_quadratic_touching}%
  \BibitemOpen
  \bibfield  {author} {\bibinfo {author} {\bibfnamefont {K.}~\bibnamefont
  {Sun}}, \bibinfo {author} {\bibfnamefont {H.}~\bibnamefont {Yao}}, \bibinfo
  {author} {\bibfnamefont {E.}~\bibnamefont {Fradkin}}, \ and\ \bibinfo
  {author} {\bibfnamefont {S.~A.}\ \bibnamefont {Kivelson}},\ }\href {\doibase
  10.1103/PhysRevLett.103.046811} {\bibfield  {journal} {\bibinfo  {journal}
  {Phys. Rev. Lett.}\ }\textbf {\bibinfo {volume} {103}},\ \bibinfo {pages}
  {046811} (\bibinfo {year} {2009})}\BibitemShut {NoStop}%
\bibitem [{\citenamefont {Ren}\ \emph {et~al.}(2021)\citenamefont {Ren},
  \citenamefont {Jiang}, \citenamefont {Qiao},\ and\ \citenamefont
  {Sheng}}]{Ren.2021}%
  \BibitemOpen
  \bibfield  {author} {\bibinfo {author} {\bibfnamefont {Y.}~\bibnamefont
  {Ren}}, \bibinfo {author} {\bibfnamefont {H.-C.}\ \bibnamefont {Jiang}},
  \bibinfo {author} {\bibfnamefont {Z.}~\bibnamefont {Qiao}}, \ and\ \bibinfo
  {author} {\bibfnamefont {D.~N.}\ \bibnamefont {Sheng}},\ }\href {\doibase
  10.1103/PhysRevLett.126.117602} {\bibfield  {journal} {\bibinfo  {journal}
  {Phys. Rev. Lett.}\ }\textbf {\bibinfo {volume} {126}},\ \bibinfo {pages}
  {117602} (\bibinfo {year} {2021})}\BibitemShut {NoStop}%
\bibitem [{\citenamefont {Kaneko}\ \emph {et~al.}(2013)\citenamefont {Kaneko},
  \citenamefont {Toriyama}, \citenamefont {Konishi},\ and\ \citenamefont
  {Ohta}}]{Kaneko2013}%
  \BibitemOpen
  \bibfield  {author} {\bibinfo {author} {\bibfnamefont {T.}~\bibnamefont
  {Kaneko}}, \bibinfo {author} {\bibfnamefont {T.}~\bibnamefont {Toriyama}},
  \bibinfo {author} {\bibfnamefont {T.}~\bibnamefont {Konishi}}, \ and\
  \bibinfo {author} {\bibfnamefont {Y.}~\bibnamefont {Ohta}},\ }\href {\doibase
  10.1103/PhysRevB.87.035121} {\bibfield  {journal} {\bibinfo  {journal} {Phys.
  Rev. B}\ }\textbf {\bibinfo {volume} {87}},\ \bibinfo {pages} {035121}
  (\bibinfo {year} {2013})}\BibitemShut {NoStop}%
\bibitem [{\citenamefont {Lu}\ \emph {et~al.}(2017)\citenamefont {Lu},
  \citenamefont {Kono}, \citenamefont {Larkin}, \citenamefont {Rost},
  \citenamefont {Takayama}, \citenamefont {Boris}, \citenamefont {Keimer},\
  and\ \citenamefont {Takagi}}]{Lu2017}%
  \BibitemOpen
  \bibfield  {author} {\bibinfo {author} {\bibfnamefont {Y.~F.}\ \bibnamefont
  {Lu}}, \bibinfo {author} {\bibfnamefont {H.}~\bibnamefont {Kono}}, \bibinfo
  {author} {\bibfnamefont {T.~I.}\ \bibnamefont {Larkin}}, \bibinfo {author}
  {\bibfnamefont {A.~W.}\ \bibnamefont {Rost}}, \bibinfo {author}
  {\bibfnamefont {T.}~\bibnamefont {Takayama}}, \bibinfo {author}
  {\bibfnamefont {A.~V.}\ \bibnamefont {Boris}}, \bibinfo {author}
  {\bibfnamefont {B.}~\bibnamefont {Keimer}}, \ and\ \bibinfo {author}
  {\bibfnamefont {H.}~\bibnamefont {Takagi}},\ }\href {\doibase
  10.1038/ncomms14408} {\bibfield  {journal} {\bibinfo  {journal} {Nat.
  Commun.}\ }\textbf {\bibinfo {volume} {8}},\ \bibinfo {pages} {1} (\bibinfo
  {year} {2017})}\BibitemShut {NoStop}%
\bibitem [{\citenamefont {Werdehausen}\ \emph {et~al.}(2018)\citenamefont
  {Werdehausen}, \citenamefont {Takayama}, \citenamefont {H{\"{o}}ppner},
  \citenamefont {Albrecht}, \citenamefont {Rost}, \citenamefont {Lu},
  \citenamefont {Manske}, \citenamefont {Takagi},\ and\ \citenamefont
  {Kaiser}}]{Werdehausen2018}%
  \BibitemOpen
  \bibfield  {author} {\bibinfo {author} {\bibfnamefont {D.}~\bibnamefont
  {Werdehausen}}, \bibinfo {author} {\bibfnamefont {T.}~\bibnamefont
  {Takayama}}, \bibinfo {author} {\bibfnamefont {M.}~\bibnamefont
  {H{\"{o}}ppner}}, \bibinfo {author} {\bibfnamefont {G.}~\bibnamefont
  {Albrecht}}, \bibinfo {author} {\bibfnamefont {A.~W.}\ \bibnamefont {Rost}},
  \bibinfo {author} {\bibfnamefont {Y.}~\bibnamefont {Lu}}, \bibinfo {author}
  {\bibfnamefont {D.}~\bibnamefont {Manske}}, \bibinfo {author} {\bibfnamefont
  {H.}~\bibnamefont {Takagi}}, \ and\ \bibinfo {author} {\bibfnamefont
  {S.}~\bibnamefont {Kaiser}},\ }\href {\doibase 10.1126/sciadv.aap8652}
  {\bibfield  {journal} {\bibinfo  {journal} {Sci. Adv.}\ }\textbf {\bibinfo
  {volume} {4}},\ \bibinfo {pages} {eaap8652} (\bibinfo {year}
  {2018})}\BibitemShut {NoStop}%
\bibitem [{\citenamefont {Sugimoto}\ \emph {et~al.}(2018)\citenamefont
  {Sugimoto}, \citenamefont {Nishimoto}, \citenamefont {Kaneko},\ and\
  \citenamefont {Ohta}}]{Sugimoto2018}%
  \BibitemOpen
  \bibfield  {author} {\bibinfo {author} {\bibfnamefont {K.}~\bibnamefont
  {Sugimoto}}, \bibinfo {author} {\bibfnamefont {S.}~\bibnamefont {Nishimoto}},
  \bibinfo {author} {\bibfnamefont {T.}~\bibnamefont {Kaneko}}, \ and\ \bibinfo
  {author} {\bibfnamefont {Y.}~\bibnamefont {Ohta}},\ }\href {\doibase
  10.1103/PhysRevLett.120.247602} {\bibfield  {journal} {\bibinfo  {journal}
  {Phys. Rev. Lett.}\ }\textbf {\bibinfo {volume} {120}},\ \bibinfo {pages}
  {247602} (\bibinfo {year} {2018})}\BibitemShut {NoStop}%
\bibitem [{\citenamefont {Ning}\ \emph {et~al.}(2020)\citenamefont {Ning},
  \citenamefont {Mehio}, \citenamefont {Buchhold}, \citenamefont {Kurumaji},
  \citenamefont {Refael}, \citenamefont {Checkelsky},\ and\ \citenamefont
  {Hsieh}}]{Ning.2020}%
  \BibitemOpen
  \bibfield  {author} {\bibinfo {author} {\bibfnamefont {H.}~\bibnamefont
  {Ning}}, \bibinfo {author} {\bibfnamefont {O.}~\bibnamefont {Mehio}},
  \bibinfo {author} {\bibfnamefont {M.}~\bibnamefont {Buchhold}}, \bibinfo
  {author} {\bibfnamefont {T.}~\bibnamefont {Kurumaji}}, \bibinfo {author}
  {\bibfnamefont {G.}~\bibnamefont {Refael}}, \bibinfo {author} {\bibfnamefont
  {J.~G.}\ \bibnamefont {Checkelsky}}, \ and\ \bibinfo {author} {\bibfnamefont
  {D.}~\bibnamefont {Hsieh}},\ }\href {\doibase 10.1103/PhysRevLett.125.267602}
  {\bibfield  {journal} {\bibinfo  {journal} {Phys. Rev. Lett.}\ }\textbf
  {\bibinfo {volume} {125}},\ \bibinfo {pages} {267602} (\bibinfo {year}
  {2020})}\BibitemShut {NoStop}%
\bibitem [{\citenamefont {Kim}\ \emph {et~al.}(2021)\citenamefont {Kim},
  \citenamefont {Kim}, \citenamefont {Kim}, \citenamefont {Kwon}, \citenamefont
  {Kim},\ and\ \citenamefont {Kim}}]{Kim:2021wf}%
  \BibitemOpen
  \bibfield  {author} {\bibinfo {author} {\bibfnamefont {K.}~\bibnamefont
  {Kim}}, \bibinfo {author} {\bibfnamefont {H.}~\bibnamefont {Kim}}, \bibinfo
  {author} {\bibfnamefont {J.}~\bibnamefont {Kim}}, \bibinfo {author}
  {\bibfnamefont {C.}~\bibnamefont {Kwon}}, \bibinfo {author} {\bibfnamefont
  {J.~S.}\ \bibnamefont {Kim}}, \ and\ \bibinfo {author} {\bibfnamefont
  {B.~J.}\ \bibnamefont {Kim}},\ }\href {\doibase 10.1038/s41467-021-22133-z}
  {\bibfield  {journal} {\bibinfo  {journal} {Nature Communications}\ }\textbf
  {\bibinfo {volume} {12}},\ \bibinfo {pages} {1969} (\bibinfo {year}
  {2021})}\BibitemShut {NoStop}%
\bibitem [{\citenamefont {Volkov}\ \emph {et~al.}(2021)\citenamefont {Volkov},
  \citenamefont {Ye}, \citenamefont {Lohani}, \citenamefont {Feldman},
  \citenamefont {Kanigel},\ and\ \citenamefont {Blumberg}}]{volkov2021failed}%
  \BibitemOpen
  \bibfield  {author} {\bibinfo {author} {\bibfnamefont {P.~A.}\ \bibnamefont
  {Volkov}}, \bibinfo {author} {\bibfnamefont {M.}~\bibnamefont {Ye}}, \bibinfo
  {author} {\bibfnamefont {H.}~\bibnamefont {Lohani}}, \bibinfo {author}
  {\bibfnamefont {I.}~\bibnamefont {Feldman}}, \bibinfo {author} {\bibfnamefont
  {A.}~\bibnamefont {Kanigel}}, \ and\ \bibinfo {author} {\bibfnamefont
  {G.}~\bibnamefont {Blumberg}},\ }\href@noop {} {\enquote {\bibinfo {title}
  {Failed excitonic quantum phase transition in
  {Ta$_2$Ni(Se$_{1-x}$S$_x$)$_5$}},}\ } (\bibinfo {year} {2021}),\ \Eprint
  {http://arxiv.org/abs/2104.07032} {arXiv:2104.07032 [cond-mat.str-el]}
  \BibitemShut {NoStop}%
\bibitem [{\citenamefont {Andrich}\ \emph {et~al.}(2021)\citenamefont
  {Andrich}, \citenamefont {Bretscher}, \citenamefont {Murakami}, \citenamefont
  {Gole{\v z}}, \citenamefont {Remez}, \citenamefont {Telang}, \citenamefont
  {Singh}, \citenamefont {Harnagea}, \citenamefont {Cooper}, \citenamefont
  {Millis}, \citenamefont {Werner}, \citenamefont {Sood},\ and\ \citenamefont
  {Rao}}]{andrich2020imaging}%
  \BibitemOpen
  \bibfield  {author} {\bibinfo {author} {\bibfnamefont {P.}~\bibnamefont
  {Andrich}}, \bibinfo {author} {\bibfnamefont {H.~M.}\ \bibnamefont
  {Bretscher}}, \bibinfo {author} {\bibfnamefont {Y.}~\bibnamefont {Murakami}},
  \bibinfo {author} {\bibfnamefont {D.}~\bibnamefont {Gole{\v z}}}, \bibinfo
  {author} {\bibfnamefont {B.}~\bibnamefont {Remez}}, \bibinfo {author}
  {\bibfnamefont {P.}~\bibnamefont {Telang}}, \bibinfo {author} {\bibfnamefont
  {A.}~\bibnamefont {Singh}}, \bibinfo {author} {\bibfnamefont
  {L.}~\bibnamefont {Harnagea}}, \bibinfo {author} {\bibfnamefont {N.~R.}\
  \bibnamefont {Cooper}}, \bibinfo {author} {\bibfnamefont {A.~J.}\
  \bibnamefont {Millis}}, \bibinfo {author} {\bibfnamefont {P.}~\bibnamefont
  {Werner}}, \bibinfo {author} {\bibfnamefont {A.~K.}\ \bibnamefont {Sood}}, \
  and\ \bibinfo {author} {\bibfnamefont {A.}~\bibnamefont {Rao}},\ }\href
  {\doibase 10.1126/sciadv.abd6147} {\bibfield  {journal} {\bibinfo  {journal}
  {Sci. Adv.}\ }\textbf {\bibinfo {volume} {7}},\ \bibinfo {pages} {eabd6147}
  (\bibinfo {year} {2021})}\BibitemShut {NoStop}%
\bibitem [{\citenamefont {Sun}\ and\ \citenamefont
  {Millis}(2021)}]{sun2020topological}%
  \BibitemOpen
  \bibfield  {author} {\bibinfo {author} {\bibfnamefont {Z.}~\bibnamefont
  {Sun}}\ and\ \bibinfo {author} {\bibfnamefont {A.~J.}\ \bibnamefont
  {Millis}},\ }\href {\doibase 10.1103/PhysRevLett.126.027601} {\bibfield
  {journal} {\bibinfo  {journal} {Phys. Rev. Lett.}\ }\textbf {\bibinfo
  {volume} {126}},\ \bibinfo {pages} {027601} (\bibinfo {year}
  {2021})}\BibitemShut {NoStop}%
\bibitem [{2no()}]{2note}%
  \BibitemOpen
  \href@noop {} {}\bibinfo {note} {This $t_k$ should be understood as the
  intrinsic hybridization together with any weak $p$-wave mean field
  \cite{sun2020topological} that it induces by linear coupling.}\BibitemShut
  {Stop}%
\bibitem [{\citenamefont {Kaneko}\ \emph {et~al.}(2015)\citenamefont {Kaneko},
  \citenamefont {Zenker}, \citenamefont {Fehske},\ and\ \citenamefont
  {Ohta}}]{Kaneko.2015}%
  \BibitemOpen
  \bibfield  {author} {\bibinfo {author} {\bibfnamefont {T.}~\bibnamefont
  {Kaneko}}, \bibinfo {author} {\bibfnamefont {B.}~\bibnamefont {Zenker}},
  \bibinfo {author} {\bibfnamefont {H.}~\bibnamefont {Fehske}}, \ and\ \bibinfo
  {author} {\bibfnamefont {Y.}~\bibnamefont {Ohta}},\ }\href {\doibase
  10.1103/PhysRevB.92.115106} {\bibfield  {journal} {\bibinfo  {journal} {Phys.
  Rev. B}\ }\textbf {\bibinfo {volume} {92}},\ \bibinfo {pages} {115106}
  (\bibinfo {year} {2015})}\BibitemShut {NoStop}%
\bibitem [{\citenamefont {Gole\ifmmode~\check{z}\else \v{z}\fi{}}\ \emph
  {et~al.}(2020)\citenamefont {Gole\ifmmode~\check{z}\else \v{z}\fi{}},
  \citenamefont {Sun}, \citenamefont {Murakami}, \citenamefont {Georges},\ and\
  \citenamefont {Millis}}]{Golez.2020}%
  \BibitemOpen
  \bibfield  {author} {\bibinfo {author} {\bibfnamefont {D.}~\bibnamefont
  {Gole\ifmmode~\check{z}\else \v{z}\fi{}}}, \bibinfo {author} {\bibfnamefont
  {Z.}~\bibnamefont {Sun}}, \bibinfo {author} {\bibfnamefont {Y.}~\bibnamefont
  {Murakami}}, \bibinfo {author} {\bibfnamefont {A.}~\bibnamefont {Georges}}, \
  and\ \bibinfo {author} {\bibfnamefont {A.~J.}\ \bibnamefont {Millis}},\
  }\href {\doibase 10.1103/PhysRevLett.125.257601} {\bibfield  {journal}
  {\bibinfo  {journal} {Phys. Rev. Lett.}\ }\textbf {\bibinfo {volume} {125}},\
  \bibinfo {pages} {257601} (\bibinfo {year} {2020})}\BibitemShut {NoStop}%
\bibitem [{\citenamefont {Murakami}\ \emph {et~al.}(2020)\citenamefont
  {Murakami}, \citenamefont {Gole\ifmmode~\check{z}\else \v{z}\fi{}},
  \citenamefont {Kaneko}, \citenamefont {Koga}, \citenamefont {Millis},\ and\
  \citenamefont {Werner}}]{Murakami.2020}%
  \BibitemOpen
  \bibfield  {author} {\bibinfo {author} {\bibfnamefont {Y.}~\bibnamefont
  {Murakami}}, \bibinfo {author} {\bibfnamefont {D.}~\bibnamefont
  {Gole\ifmmode~\check{z}\else \v{z}\fi{}}}, \bibinfo {author} {\bibfnamefont
  {T.}~\bibnamefont {Kaneko}}, \bibinfo {author} {\bibfnamefont
  {A.}~\bibnamefont {Koga}}, \bibinfo {author} {\bibfnamefont {A.~J.}\
  \bibnamefont {Millis}}, \ and\ \bibinfo {author} {\bibfnamefont
  {P.}~\bibnamefont {Werner}},\ }\href {\doibase 10.1103/PhysRevB.101.195118}
  {\bibfield  {journal} {\bibinfo  {journal} {Phys. Rev. B}\ }\textbf {\bibinfo
  {volume} {101}},\ \bibinfo {pages} {195118} (\bibinfo {year}
  {2020})}\BibitemShut {NoStop}%
\bibitem [{\citenamefont {Sun}\ and\ \citenamefont
  {Millis}(2020)}]{Sun.2020_BaSh}%
  \BibitemOpen
  \bibfield  {author} {\bibinfo {author} {\bibfnamefont {Z.}~\bibnamefont
  {Sun}}\ and\ \bibinfo {author} {\bibfnamefont {A.~J.}\ \bibnamefont
  {Millis}},\ }\href {\doibase 10.1103/PhysRevB.102.041110} {\bibfield
  {journal} {\bibinfo  {journal} {Phys. Rev. B}\ }\textbf {\bibinfo {volume}
  {102}},\ \bibinfo {pages} {041110(R)} (\bibinfo {year} {2020})}\BibitemShut
  {NoStop}%
\bibitem [{SI()}]{SI}%
  \BibitemOpen
  \href@noop {} {}\bibinfo {note} {See Supplemental Material for details, which
  includes Refs.~\cite{Altland.2010, Sun.2020_collective_modes,
  Herbut.2007,Tinkham,Li.2014,Dai:2014,Rodin.2014_phosphorene,Tong:2017uh,
  Basov2016,Sun.2015}}\BibitemShut {NoStop}%
\bibitem [{\citenamefont {Altland}\ and\ \citenamefont
  {Simons}(2010)}]{Altland.2010}%
  \BibitemOpen
  \bibfield  {author} {\bibinfo {author} {\bibfnamefont {A.}~\bibnamefont
  {Altland}}\ and\ \bibinfo {author} {\bibfnamefont {B.~D.}\ \bibnamefont
  {Simons}},\ }\href {\doibase 10.1017/CBO9780511789984} {\emph {\bibinfo
  {title} {{Condensed Matter Field Theory}}}}\ (\bibinfo  {publisher}
  {Cambridge University Press},\ \bibinfo {address} {Cambridge},\ \bibinfo
  {year} {2010})\BibitemShut {NoStop}%
\bibitem [{\citenamefont {Sun}\ \emph {et~al.}(2020)\citenamefont {Sun},
  \citenamefont {Fogler}, \citenamefont {Basov},\ and\ \citenamefont
  {Millis}}]{Sun.2020_collective_modes}%
  \BibitemOpen
  \bibfield  {author} {\bibinfo {author} {\bibfnamefont {Z.}~\bibnamefont
  {Sun}}, \bibinfo {author} {\bibfnamefont {M.~M.}\ \bibnamefont {Fogler}},
  \bibinfo {author} {\bibfnamefont {D.~N.}\ \bibnamefont {Basov}}, \ and\
  \bibinfo {author} {\bibfnamefont {A.~J.}\ \bibnamefont {Millis}},\ }\href
  {\doibase 10.1103/PhysRevResearch.2.023413} {\bibfield  {journal} {\bibinfo
  {journal} {Phys. Rev. Research}\ }\textbf {\bibinfo {volume} {2}},\ \bibinfo
  {pages} {023413} (\bibinfo {year} {2020})}\BibitemShut {NoStop}%
\bibitem [{\citenamefont {Herbut}(2007)}]{Herbut.2007}%
  \BibitemOpen
  \bibfield  {author} {\bibinfo {author} {\bibfnamefont {I.}~\bibnamefont
  {Herbut}},\ }\href@noop {} {\emph {\bibinfo {title} {{A Modern Approach to
  Critical Phenomena}}}}\ (\bibinfo  {publisher} {Cambridge University Press},\
  \bibinfo {address} {Cambridge},\ \bibinfo {year} {2007})\BibitemShut
  {NoStop}%
\bibitem [{\citenamefont {Tinkham}(2004)}]{Tinkham}%
  \BibitemOpen
  \bibfield  {author} {\bibinfo {author} {\bibfnamefont {M.}~\bibnamefont
  {Tinkham}},\ }\href@noop {} {\emph {\bibinfo {title} {{Introduction to
  Superconductivity}}}}\ (\bibinfo  {publisher} {Dover Publications},\ \bibinfo
  {address} {Mineola, New York},\ \bibinfo {year} {2004})\BibitemShut {NoStop}%
\bibitem [{\citenamefont {Li}\ and\ \citenamefont {Appelbaum}(2014)}]{Li.2014}%
  \BibitemOpen
  \bibfield  {author} {\bibinfo {author} {\bibfnamefont {P.}~\bibnamefont
  {Li}}\ and\ \bibinfo {author} {\bibfnamefont {I.}~\bibnamefont {Appelbaum}},\
  }\href {\doibase 10.1103/PhysRevB.90.115439} {\bibfield  {journal} {\bibinfo
  {journal} {Phys. Rev. B}\ }\textbf {\bibinfo {volume} {90}},\ \bibinfo
  {pages} {115439} (\bibinfo {year} {2014})}\BibitemShut {NoStop}%
\bibitem [{\citenamefont {Dai}\ and\ \citenamefont {Zeng}(2014)}]{Dai:2014}%
  \BibitemOpen
  \bibfield  {author} {\bibinfo {author} {\bibfnamefont {J.}~\bibnamefont
  {Dai}}\ and\ \bibinfo {author} {\bibfnamefont {X.~C.}\ \bibnamefont {Zeng}},\
  }\bibfield  {booktitle} {\emph {\bibinfo {booktitle} {The Journal of Physical
  Chemistry Letters}},\ }\href {\doibase 10.1021/jz500409m} {\bibfield
  {journal} {\bibinfo  {journal} {The Journal of Physical Chemistry Letters}\
  }\textbf {\bibinfo {volume} {5}},\ \bibinfo {pages} {1289} (\bibinfo {year}
  {2014})}\BibitemShut {NoStop}%
\bibitem [{\citenamefont {Rodin}\ \emph {et~al.}(2014)\citenamefont {Rodin},
  \citenamefont {Carvalho},\ and\ \citenamefont
  {Castro~Neto}}]{Rodin.2014_phosphorene}%
  \BibitemOpen
  \bibfield  {author} {\bibinfo {author} {\bibfnamefont {A.~S.}\ \bibnamefont
  {Rodin}}, \bibinfo {author} {\bibfnamefont {A.}~\bibnamefont {Carvalho}}, \
  and\ \bibinfo {author} {\bibfnamefont {A.~H.}\ \bibnamefont {Castro~Neto}},\
  }\href {\doibase 10.1103/PhysRevLett.112.176801} {\bibfield  {journal}
  {\bibinfo  {journal} {Phys. Rev. Lett.}\ }\textbf {\bibinfo {volume} {112}},\
  \bibinfo {pages} {176801} (\bibinfo {year} {2014})}\BibitemShut {NoStop}%
\bibitem [{\citenamefont {Tong}\ \emph {et~al.}(2017)\citenamefont {Tong},
  \citenamefont {Yu}, \citenamefont {Zhu}, \citenamefont {Wang}, \citenamefont
  {Xu},\ and\ \citenamefont {Yao}}]{Tong:2017uh}%
  \BibitemOpen
  \bibfield  {author} {\bibinfo {author} {\bibfnamefont {Q.}~\bibnamefont
  {Tong}}, \bibinfo {author} {\bibfnamefont {H.}~\bibnamefont {Yu}}, \bibinfo
  {author} {\bibfnamefont {Q.}~\bibnamefont {Zhu}}, \bibinfo {author}
  {\bibfnamefont {Y.}~\bibnamefont {Wang}}, \bibinfo {author} {\bibfnamefont
  {X.}~\bibnamefont {Xu}}, \ and\ \bibinfo {author} {\bibfnamefont
  {W.}~\bibnamefont {Yao}},\ }\href {\doibase 10.1038/nphys3968} {\bibfield
  {journal} {\bibinfo  {journal} {Nature Physics}\ }\textbf {\bibinfo {volume}
  {13}},\ \bibinfo {pages} {356} (\bibinfo {year} {2017})}\BibitemShut
  {NoStop}%
\bibitem [{\citenamefont {Basov}\ \emph {et~al.}(2016)\citenamefont {Basov},
  \citenamefont {Fogler},\ and\ \citenamefont {{Garc{\'{i}}a de
  Abajo}}}]{Basov2016}%
  \BibitemOpen
  \bibfield  {author} {\bibinfo {author} {\bibfnamefont {D.~N.}\ \bibnamefont
  {Basov}}, \bibinfo {author} {\bibfnamefont {M.~M.}\ \bibnamefont {Fogler}}, \
  and\ \bibinfo {author} {\bibfnamefont {F.~J.}\ \bibnamefont {{Garc{\'{i}}a de
  Abajo}}},\ }\href
  {http://science.sciencemag.org/content/354/6309/aag1992.abstract} {\bibfield
  {journal} {\bibinfo  {journal} {Science}\ }\textbf {\bibinfo {volume}
  {354}},\ \bibinfo {pages} {aag1992} (\bibinfo {year} {2016})}\BibitemShut
  {NoStop}%
\bibitem [{\citenamefont {Sun}\ \emph {et~al.}(2015)\citenamefont {Sun},
  \citenamefont {Guti{\'{e}}rrez-Rubio}, \citenamefont {Basov},\ and\
  \citenamefont {Fogler}}]{Sun.2015}%
  \BibitemOpen
  \bibfield  {author} {\bibinfo {author} {\bibfnamefont {Z.}~\bibnamefont
  {Sun}}, \bibinfo {author} {\bibfnamefont {{\'{A}}.}~\bibnamefont
  {Guti{\'{e}}rrez-Rubio}}, \bibinfo {author} {\bibfnamefont {D.~N.}\
  \bibnamefont {Basov}}, \ and\ \bibinfo {author} {\bibfnamefont {M.~M.}\
  \bibnamefont {Fogler}},\ }\href {\doibase 10.1021/acs.nanolett.5b00814}
  {\bibfield  {journal} {\bibinfo  {journal} {Nano Lett.}\ }\textbf {\bibinfo
  {volume} {15}},\ \bibinfo {pages} {4455} (\bibinfo {year}
  {2015})}\BibitemShut {NoStop}%
\bibitem [{5no()}]{5note}%
  \BibitemOpen
  \href@noop {} {}\bibinfo {note} {With this term and neglecting screening from
  the gates, the phase mode (exciton density fluctuation) has the dispersion
  $\omega_q= \sqrt{(1+2\pi \nu d) \left(\Delta_p^2/D + v_g^2 q^2 \right)}$ with
  $\omega_{q=0}$ being the `Josephson plasma frequency'.}\BibitemShut {Stop}%
\bibitem [{8no()}]{8note}%
  \BibitemOpen
  \href@noop {} {}\bibinfo {note} {In the BEC regime, the coefficients of
  \equa{eqn:L_bilayer} should be changed: $1/D \sim \Delta^2/(|G|W)$ for local
  interactions with $c_{\text{f}}$ redefined as $1$ and $W$ being the typical
  band width.}\BibitemShut {Stop}%
\bibitem [{7no()}]{7note}%
  \BibitemOpen
  \href@noop {} {}\bibinfo {note} {For $\Delta_{\text{p}}=10 \unit{meV}$ and
  $v_g=10^6 \unit{m/s}$, the length scale is $l_d \sim 0.4 \unit{\mu
  m}$.}\BibitemShut {Stop}%
\bibitem [{\citenamefont {Murakami}\ \emph {et~al.}(2017)\citenamefont
  {Murakami}, \citenamefont {Gole\ifmmode~\check{z}\else \v{z}\fi{}},
  \citenamefont {Eckstein},\ and\ \citenamefont {Werner}}]{Murakami.2017}%
  \BibitemOpen
  \bibfield  {author} {\bibinfo {author} {\bibfnamefont {Y.}~\bibnamefont
  {Murakami}}, \bibinfo {author} {\bibfnamefont {D.}~\bibnamefont
  {Gole\ifmmode~\check{z}\else \v{z}\fi{}}}, \bibinfo {author} {\bibfnamefont
  {M.}~\bibnamefont {Eckstein}}, \ and\ \bibinfo {author} {\bibfnamefont
  {P.}~\bibnamefont {Werner}},\ }\href {\doibase
  10.1103/PhysRevLett.119.247601} {\bibfield  {journal} {\bibinfo  {journal}
  {Phys. Rev. Lett.}\ }\textbf {\bibinfo {volume} {119}},\ \bibinfo {pages}
  {247601} (\bibinfo {year} {2017})}\BibitemShut {NoStop}%
\bibitem [{\citenamefont {Armitage}\ \emph {et~al.}(2018)\citenamefont
  {Armitage}, \citenamefont {Mele},\ and\ \citenamefont
  {Vishwanath}}]{Armitage.2018_weyl_review}%
  \BibitemOpen
  \bibfield  {author} {\bibinfo {author} {\bibfnamefont {N.~P.}\ \bibnamefont
  {Armitage}}, \bibinfo {author} {\bibfnamefont {E.~J.}\ \bibnamefont {Mele}},
  \ and\ \bibinfo {author} {\bibfnamefont {A.}~\bibnamefont {Vishwanath}},\
  }\href {\doibase 10.1103/RevModPhys.90.015001} {\bibfield  {journal}
  {\bibinfo  {journal} {Rev. Mod. Phys.}\ }\textbf {\bibinfo {volume} {90}},\
  \bibinfo {pages} {015001} (\bibinfo {year} {2018})}\BibitemShut {NoStop}%
\bibitem [{\citenamefont {Liu}\ \emph {et~al.}(2016)\citenamefont {Liu},
  \citenamefont {Zhang},\ and\ \citenamefont {Qi}}]{Liu.2016}%
  \BibitemOpen
  \bibfield  {author} {\bibinfo {author} {\bibfnamefont {C.-X.}\ \bibnamefont
  {Liu}}, \bibinfo {author} {\bibfnamefont {S.-C.}\ \bibnamefont {Zhang}}, \
  and\ \bibinfo {author} {\bibfnamefont {X.-L.}\ \bibnamefont {Qi}},\ }\href
  {\doibase 10.1146/annurev-conmatphys-031115-011417} {\bibfield  {journal}
  {\bibinfo  {journal} {Annual Review of Condensed Matter Physics}\ }\textbf
  {\bibinfo {volume} {7}},\ \bibinfo {pages} {301} (\bibinfo {year}
  {2016})}\BibitemShut {NoStop}%
\bibitem [{\citenamefont {Kim}\ \emph {et~al.}(2015)\citenamefont {Kim},
  \citenamefont {Baik}, \citenamefont {Ryu}, \citenamefont {Sohn},
  \citenamefont {Park}, \citenamefont {Park}, \citenamefont {Denlinger},
  \citenamefont {Yi}, \citenamefont {Choi},\ and\ \citenamefont
  {Kim}}]{Kim.2015_phorsphorene}%
  \BibitemOpen
  \bibfield  {author} {\bibinfo {author} {\bibfnamefont {J.}~\bibnamefont
  {Kim}}, \bibinfo {author} {\bibfnamefont {S.~S.}\ \bibnamefont {Baik}},
  \bibinfo {author} {\bibfnamefont {S.~H.}\ \bibnamefont {Ryu}}, \bibinfo
  {author} {\bibfnamefont {Y.}~\bibnamefont {Sohn}}, \bibinfo {author}
  {\bibfnamefont {S.}~\bibnamefont {Park}}, \bibinfo {author} {\bibfnamefont
  {B.-G.}\ \bibnamefont {Park}}, \bibinfo {author} {\bibfnamefont
  {J.}~\bibnamefont {Denlinger}}, \bibinfo {author} {\bibfnamefont
  {Y.}~\bibnamefont {Yi}}, \bibinfo {author} {\bibfnamefont {H.~J.}\
  \bibnamefont {Choi}}, \ and\ \bibinfo {author} {\bibfnamefont {K.~S.}\
  \bibnamefont {Kim}},\ }\href {\doibase 10.1126/science.aaa6486} {\bibfield
  {journal} {\bibinfo  {journal} {Science}\ }\textbf {\bibinfo {volume}
  {349}},\ \bibinfo {pages} {723} (\bibinfo {year} {2015})}\BibitemShut
  {NoStop}%
\bibitem [{\citenamefont {Li}\ \emph {et~al.}(2014)\citenamefont {Li},
  \citenamefont {Yu}, \citenamefont {Ye}, \citenamefont {Ge}, \citenamefont
  {Ou}, \citenamefont {Wu}, \citenamefont {Feng}, \citenamefont {Chen},\ and\
  \citenamefont {Zhang}}]{Li.2014_phosphorene}%
  \BibitemOpen
  \bibfield  {author} {\bibinfo {author} {\bibfnamefont {L.}~\bibnamefont
  {Li}}, \bibinfo {author} {\bibfnamefont {Y.}~\bibnamefont {Yu}}, \bibinfo
  {author} {\bibfnamefont {G.~J.}\ \bibnamefont {Ye}}, \bibinfo {author}
  {\bibfnamefont {Q.}~\bibnamefont {Ge}}, \bibinfo {author} {\bibfnamefont
  {X.}~\bibnamefont {Ou}}, \bibinfo {author} {\bibfnamefont {H.}~\bibnamefont
  {Wu}}, \bibinfo {author} {\bibfnamefont {D.}~\bibnamefont {Feng}}, \bibinfo
  {author} {\bibfnamefont {X.~H.}\ \bibnamefont {Chen}}, \ and\ \bibinfo
  {author} {\bibfnamefont {Y.}~\bibnamefont {Zhang}},\ }\href
  {https://doi.org/10.1038/nnano.2014.35} {\bibfield  {journal} {\bibinfo
  {journal} {Nature Nanotechnology}\ }\textbf {\bibinfo {volume} {9}},\
  \bibinfo {pages} {372} (\bibinfo {year} {2014})}\BibitemShut {NoStop}%
\bibitem [{\citenamefont {Carvalho}\ \emph {et~al.}(2016)\citenamefont
  {Carvalho}, \citenamefont {Wang}, \citenamefont {Zhu}, \citenamefont {Rodin},
  \citenamefont {Su},\ and\ \citenamefont
  {Castro~Neto}}]{Carvalho.2016_phosphorene}%
  \BibitemOpen
  \bibfield  {author} {\bibinfo {author} {\bibfnamefont {A.}~\bibnamefont
  {Carvalho}}, \bibinfo {author} {\bibfnamefont {M.}~\bibnamefont {Wang}},
  \bibinfo {author} {\bibfnamefont {X.}~\bibnamefont {Zhu}}, \bibinfo {author}
  {\bibfnamefont {A.~S.}\ \bibnamefont {Rodin}}, \bibinfo {author}
  {\bibfnamefont {H.}~\bibnamefont {Su}}, \ and\ \bibinfo {author}
  {\bibfnamefont {A.~H.}\ \bibnamefont {Castro~Neto}},\ }\href
  {https://doi.org/10.1038/natrevmats.2016.61} {\bibfield  {journal} {\bibinfo
  {journal} {Nature Reviews Materials}\ }\textbf {\bibinfo {volume} {1}},\
  \bibinfo {pages} {16061} (\bibinfo {year} {2016})}\BibitemShut {NoStop}%
\bibitem [{\citenamefont {Jia}\ \emph {et~al.}(2020)\citenamefont {Jia},
  \citenamefont {Wang}, \citenamefont {Chiu}, \citenamefont {Song},
  \citenamefont {Yu}, \citenamefont {J.}, \citenamefont {Lei}, \citenamefont
  {Klemenz}, \citenamefont {Cevallos}, \citenamefont {Onyszczak}, \citenamefont
  {Fishchenko}, \citenamefont {Liu}, \citenamefont {Farahi}, \citenamefont
  {Xie}, \citenamefont {Xu}, \citenamefont {Watanabe}, \citenamefont
  {Taniguchi}, \citenamefont {Bernevig}, \citenamefont {Cava}, \citenamefont
  {Schoop}, \citenamefont {Yazdani},\ and\ \citenamefont {Wu}}]{jia.2020}%
  \BibitemOpen
  \bibfield  {author} {\bibinfo {author} {\bibfnamefont {Y.}~\bibnamefont
  {Jia}}, \bibinfo {author} {\bibfnamefont {P.}~\bibnamefont {Wang}}, \bibinfo
  {author} {\bibfnamefont {C.-L.}\ \bibnamefont {Chiu}}, \bibinfo {author}
  {\bibfnamefont {Z.}~\bibnamefont {Song}}, \bibinfo {author} {\bibfnamefont
  {G.}~\bibnamefont {Yu}}, \bibinfo {author} {\bibfnamefont {B.}~\bibnamefont
  {J.}}, \bibinfo {author} {\bibfnamefont {S.}~\bibnamefont {Lei}}, \bibinfo
  {author} {\bibfnamefont {S.}~\bibnamefont {Klemenz}}, \bibinfo {author}
  {\bibfnamefont {F.~A.}\ \bibnamefont {Cevallos}}, \bibinfo {author}
  {\bibfnamefont {M.}~\bibnamefont {Onyszczak}}, \bibinfo {author}
  {\bibfnamefont {N.}~\bibnamefont {Fishchenko}}, \bibinfo {author}
  {\bibfnamefont {X.}~\bibnamefont {Liu}}, \bibinfo {author} {\bibfnamefont
  {G.}~\bibnamefont {Farahi}}, \bibinfo {author} {\bibfnamefont
  {F.}~\bibnamefont {Xie}}, \bibinfo {author} {\bibfnamefont {Y.}~\bibnamefont
  {Xu}}, \bibinfo {author} {\bibfnamefont {K.}~\bibnamefont {Watanabe}},
  \bibinfo {author} {\bibfnamefont {T.}~\bibnamefont {Taniguchi}}, \bibinfo
  {author} {\bibfnamefont {B.~A.}\ \bibnamefont {Bernevig}}, \bibinfo {author}
  {\bibfnamefont {R.~J.}\ \bibnamefont {Cava}}, \bibinfo {author}
  {\bibfnamefont {L.~M.}\ \bibnamefont {Schoop}}, \bibinfo {author}
  {\bibfnamefont {A.}~\bibnamefont {Yazdani}}, \ and\ \bibinfo {author}
  {\bibfnamefont {S.}~\bibnamefont {Wu}},\ }\href@noop {} {\enquote {\bibinfo
  {title} {Evidence for a monolayer excitonic insulator},}\ } (\bibinfo {year}
  {2020}),\ \Eprint {http://arxiv.org/abs/2010.05390} {arXiv:2010.05390
  [cond-mat.mes-hall]} \BibitemShut {NoStop}%
\bibitem [{\citenamefont {Kwan}\ \emph {et~al.}(2020)\citenamefont {Kwan},
  \citenamefont {Devakul}, \citenamefont {Sondhi},\ and\ \citenamefont
  {Parameswaran}}]{kwan2020theory}%
  \BibitemOpen
  \bibfield  {author} {\bibinfo {author} {\bibfnamefont {Y.~H.}\ \bibnamefont
  {Kwan}}, \bibinfo {author} {\bibfnamefont {T.}~\bibnamefont {Devakul}},
  \bibinfo {author} {\bibfnamefont {S.~L.}\ \bibnamefont {Sondhi}}, \ and\
  \bibinfo {author} {\bibfnamefont {S.~A.}\ \bibnamefont {Parameswaran}},\
  }\href@noop {} {\enquote {\bibinfo {title} {Theory of competing excitonic
  orders in insulating {WTe$_2$} monolayers},}\ } (\bibinfo {year} {2020}),\
  \Eprint {http://arxiv.org/abs/2012.05255} {arXiv:2012.05255
  [cond-mat.str-el]} \BibitemShut {NoStop}%
\bibitem [{\citenamefont {Varsano}\ \emph {et~al.}(2020)\citenamefont
  {Varsano}, \citenamefont {Palummo}, \citenamefont {Molinari},\ and\
  \citenamefont {Rontani}}]{Varsano.2020}%
  \BibitemOpen
  \bibfield  {author} {\bibinfo {author} {\bibfnamefont {D.}~\bibnamefont
  {Varsano}}, \bibinfo {author} {\bibfnamefont {M.}~\bibnamefont {Palummo}},
  \bibinfo {author} {\bibfnamefont {E.}~\bibnamefont {Molinari}}, \ and\
  \bibinfo {author} {\bibfnamefont {M.}~\bibnamefont {Rontani}},\ }\href
  {https://doi.org/10.1038/s41565-020-0650-4} {\bibfield  {journal} {\bibinfo
  {journal} {Nature Nanotechnology}\ }\textbf {\bibinfo {volume} {15}},\
  \bibinfo {pages} {367} (\bibinfo {year} {2020})}\BibitemShut {NoStop}%
\bibitem [{\citenamefont {Zhu}\ \emph {et~al.}(2019)\citenamefont {Zhu},
  \citenamefont {Tu}, \citenamefont {Tong},\ and\ \citenamefont
  {Yao}}]{Zhu.2019}%
  \BibitemOpen
  \bibfield  {author} {\bibinfo {author} {\bibfnamefont {Q.}~\bibnamefont
  {Zhu}}, \bibinfo {author} {\bibfnamefont {M.~W.-Y.}\ \bibnamefont {Tu}},
  \bibinfo {author} {\bibfnamefont {Q.}~\bibnamefont {Tong}}, \ and\ \bibinfo
  {author} {\bibfnamefont {W.}~\bibnamefont {Yao}},\ }\href {\doibase
  10.1126/sciadv.aau6120} {\bibfield  {journal} {\bibinfo  {journal} {Sci.
  Adv.}\ }\textbf {\bibinfo {volume} {5}},\ \bibinfo {pages} {eaau6120}
  (\bibinfo {year} {2019})}\BibitemShut {NoStop}%
\bibitem [{\citenamefont {Wang}\ \emph
  {et~al.}(2019{\natexlab{b}})\citenamefont {Wang}, \citenamefont {Erten},
  \citenamefont {Wang},\ and\ \citenamefont {Xing}}]{Wang2019a}%
  \BibitemOpen
  \bibfield  {author} {\bibinfo {author} {\bibfnamefont {R.}~\bibnamefont
  {Wang}}, \bibinfo {author} {\bibfnamefont {O.}~\bibnamefont {Erten}},
  \bibinfo {author} {\bibfnamefont {B.}~\bibnamefont {Wang}}, \ and\ \bibinfo
  {author} {\bibfnamefont {D.~Y.}\ \bibnamefont {Xing}},\ }\href {\doibase
  10.1038/s41467-018-08203-9} {\bibfield  {journal} {\bibinfo  {journal} {Nat.
  Commun.}\ }\textbf {\bibinfo {volume} {10}},\ \bibinfo {pages} {1} (\bibinfo
  {year} {2019}{\natexlab{b}})}\BibitemShut {NoStop}%
\bibitem [{\citenamefont {Hu}\ \emph {et~al.}(2020)\citenamefont {Hu},
  \citenamefont {Zhang}, \citenamefont {Zhang},\ and\ \citenamefont
  {Wu}}]{Hu2019}%
  \BibitemOpen
  \bibfield  {author} {\bibinfo {author} {\bibfnamefont {L.-H.}\ \bibnamefont
  {Hu}}, \bibinfo {author} {\bibfnamefont {R.-X.}\ \bibnamefont {Zhang}},
  \bibinfo {author} {\bibfnamefont {F.-C.}\ \bibnamefont {Zhang}}, \ and\
  \bibinfo {author} {\bibfnamefont {C.}~\bibnamefont {Wu}},\ }\href {\doibase
  10.1103/PhysRevB.102.235115} {\bibfield  {journal} {\bibinfo  {journal}
  {Phys. Rev. B}\ }\textbf {\bibinfo {volume} {102}},\ \bibinfo {pages}
  {235115} (\bibinfo {year} {2020})}\BibitemShut {NoStop}%
\bibitem [{\citenamefont {Perfetto}\ and\ \citenamefont
  {Stefanucci}(2020)}]{Perfetto2020}%
  \BibitemOpen
  \bibfield  {author} {\bibinfo {author} {\bibfnamefont {E.}~\bibnamefont
  {Perfetto}}\ and\ \bibinfo {author} {\bibfnamefont {G.}~\bibnamefont
  {Stefanucci}},\ }\href {\doibase 10.1103/PhysRevLett.125.106401} {\bibfield
  {journal} {\bibinfo  {journal} {Phys. Rev. Lett.}\ }\textbf {\bibinfo
  {volume} {125}},\ \bibinfo {pages} {106401} (\bibinfo {year}
  {2020})}\BibitemShut {NoStop}%
\bibitem [{\citenamefont {Liu}\ \emph {et~al.}(2021)\citenamefont {Liu},
  \citenamefont {Hu}, \citenamefont {Chen}, \citenamefont {Zhou},\ and\
  \citenamefont {Xu}}]{liu2021topological}%
  \BibitemOpen
  \bibfield  {author} {\bibinfo {author} {\bibfnamefont {Z.-R.}\ \bibnamefont
  {Liu}}, \bibinfo {author} {\bibfnamefont {L.-H.}\ \bibnamefont {Hu}},
  \bibinfo {author} {\bibfnamefont {C.-Z.}\ \bibnamefont {Chen}}, \bibinfo
  {author} {\bibfnamefont {B.}~\bibnamefont {Zhou}}, \ and\ \bibinfo {author}
  {\bibfnamefont {D.-H.}\ \bibnamefont {Xu}},\ }\href {\doibase
  10.1103/PhysRevB.103.L201115} {\bibfield  {journal} {\bibinfo  {journal}
  {Phys. Rev. B}\ }\textbf {\bibinfo {volume} {103}},\ \bibinfo {pages}
  {L201115} (\bibinfo {year} {2021})}\BibitemShut {NoStop}%
\bibitem [{\citenamefont {Berezinsky}(1971)}]{Berezinsky.1971}%
  \BibitemOpen
  \bibfield  {author} {\bibinfo {author} {\bibfnamefont {V.~L.}\ \bibnamefont
  {Berezinsky}},\ }\href@noop {} {\bibfield  {journal} {\bibinfo  {journal}
  {Sov. Phys. JETP}\ }\textbf {\bibinfo {volume} {32}},\ \bibinfo {pages} {493}
  (\bibinfo {year} {1971})}\BibitemShut {NoStop}%
\bibitem [{\citenamefont {Kosterlitz}\ and\ \citenamefont
  {Thouless}(1973)}]{Kosterlitz.1973}%
  \BibitemOpen
  \bibfield  {author} {\bibinfo {author} {\bibfnamefont {J.~M.}\ \bibnamefont
  {Kosterlitz}}\ and\ \bibinfo {author} {\bibfnamefont {D.~J.}\ \bibnamefont
  {Thouless}},\ }\href {\doibase 10.1088/0022-3719/6/7/010} {\bibfield
  {journal} {\bibinfo  {journal} {Journal of Physics C: Solid State Physics}\
  }\textbf {\bibinfo {volume} {6}},\ \bibinfo {pages} {1181} (\bibinfo {year}
  {1973})}\BibitemShut {NoStop}%
\bibitem [{\citenamefont {Jos\'e}\ \emph {et~al.}(1977)\citenamefont {Jos\'e},
  \citenamefont {Kadanoff}, \citenamefont {Kirkpatrick},\ and\ \citenamefont
  {Nelson}}]{Jose.1977}%
  \BibitemOpen
  \bibfield  {author} {\bibinfo {author} {\bibfnamefont {J.~V.}\ \bibnamefont
  {Jos\'e}}, \bibinfo {author} {\bibfnamefont {L.~P.}\ \bibnamefont
  {Kadanoff}}, \bibinfo {author} {\bibfnamefont {S.}~\bibnamefont
  {Kirkpatrick}}, \ and\ \bibinfo {author} {\bibfnamefont {D.~R.}\ \bibnamefont
  {Nelson}},\ }\href {\doibase 10.1103/PhysRevB.16.1217} {\bibfield  {journal}
  {\bibinfo  {journal} {Phys. Rev. B}\ }\textbf {\bibinfo {volume} {16}},\
  \bibinfo {pages} {1217} (\bibinfo {year} {1977})}\BibitemShut {NoStop}%
\end{thebibliography}%

\pagebreak
\widetext
\begin{center}
	\textbf{\large Supplemental Material for `Second order Josephson effect in excitonic insulators'}
\end{center}
\setcounter{equation}{0}
\setcounter{figure}{0}
\setcounter{table}{0}
\setcounter{page}{1}
\makeatletter
\renewcommand{\theequation}{S\arabic{equation}}
\renewcommand{\thefigure}{S\arabic{figure}}
\renewcommand{\bibnumfmt}[1]{[S#1]}

\tableofcontents

\section{The electron-hole bilayer}
\label{SI:bilayer}
In this section, we show the detailed derivation of the effective low energy Lagrangian  (Eq.~(2) of the main text) for the order parameter phase of the electron-hole bilayer. We reproduce the Hamiltonian in real space
\begin{align}
	H_{\text{EHB}} =& \int {dr} \psi^\dagger
	\begin{pmatrix}
		\xi_1(p+A_1)  + \phi_1  & e^{id A_z} t_{p+A} \\
		e^{-id A_z} t^\ast_{p+A} & \xi_2(p+A_2)  + \phi_2 
	\end{pmatrix}
	\psi 
	+ 
	\int {dr dr^\prime} V(r-r^\prime) \psi^\dagger(r) \psi(r) \psi^\dagger(r^\prime) \psi(r^\prime)
	\label{eqn:hamiltonian_SI}
\end{align}
here for convenience.
The Ginzburg-Landau action for order parameter fields and the EM field is obtained by integrating out the fermions  in the Hubbard-Stratonovich decoupled action  ($e^{-S[\Delta, A]}
\equiv \int  D[\bar{\psi},\psi]  e^{-S[\psi,\Delta, A]}$), resulting in
\begin{equation}
S[\Delta, A]=\mathrm{Tr\, ln} \left[\partial_\tau +H_m \right]
+
\int dr d\tau \frac{|\Delta|^2}{g}
\equiv \int dr d\tau L(\Delta, A) \,.
\label{eqn:grand_GL_si}
\end{equation}
where 
\begin{equation}
H_m=	
\begin{pmatrix}
\xi_1(p+A_1)  + \phi_1  & e^{id A_z} t_{p+A} + \Delta \\
e^{-id A_z} t^\ast_{p+A} + \Delta^\ast & \xi_2(p+A_2)  + \phi_2 
\end{pmatrix}
\,.
\end{equation}
The $\mathrm{Tr\, ln}$ means trace of logarithm of the infinite dimensional matrix where $k$ should be interpreted as the spatial derivative $-i\nabla$ acting on the fermion fields, i.e., the matrix is just the kernel $\partial_\tau +
H_m$ for all Fermion fields at all $(r, t)$ \cite{Altland.2010, sun2020topological}. Performed in Fourier basis, it involves a summation over momenta $k$, the fermion Matsubara frequencies $\omega_n= (2n+1) \pi T$ ($n\in \mathbb{Z}$, $T$ is the temperature and we have set the Boltzmann constant to be $1$) and a trace of logarithm of the $2\times 2$ matrices. The static limit of $L$ is just the static free energy function.

Using a local gauge transformation $(\psi_1(r), \psi_2(r)) \rightarrow (e^{i\theta(r)}\psi_1(r), e^{-i\theta(r)} \psi_2(r))$ that shifts the phase in Eq.~(2) of the main text to the diagonal terms, it is straightforward to see that the gradients of phase always appear together with the asymmetric EM fields as $\partial_t \theta +\phi_a$ and $\nabla \theta-A_a$,
formally analogous to superconductors \cite{Altland.2010, Sun.2020_collective_modes, Sun.2020_BaSh}.
Same as superconductors, the coefficients of the leading quadratic terms $L=K^{\mu \nu} (\partial_\mu \theta + A_{a\mu}) (\partial_\nu \theta  + A_{a\nu}) $ are just $K^{\mu \nu}=\text{diag}(-\nu, n/m)$ where $n/m=\nu v_g^2$ and $v_g=v_F/\sqrt{2}$ in the BCS weak coupling case.

The potential term for $\theta$ in Eq.~(2) of the main text is beyond $O(\theta^2)$ but comes from expanding the static free energy $F$ to second order in $\Delta_p$. In the BCS weak coupling case,
\begin{align}
	F &=\frac{1}{g}|\Delta|^2 -\frac{\nu}{2\pi}\int d\phi |\Delta+i\Delta_p f(\phi)|^2 \ln \frac{2\Lambda}{|\Delta+i\Delta_p f(\phi)|} 
	\notag\\
	&= \frac{1}{g}|\Delta|^2 -\frac{\nu}{2\pi}\int d\phi \left(|\Delta|^2 + \Delta_p^2 f^2(\phi)  + 2i \Delta_2 \Delta_p f(\phi)
	\right)
	\left(
	\ln 2\Lambda- \frac{1}{2} \ln \left(|\Delta|^2 + \Delta_p^2 f^2(\phi)  + 2i \Delta_2 \Delta_p f(\phi)
	\right)
	\right)
	\notag\\
	&\approx \frac{1}{g}|\Delta|^2-
	\nu |\Delta|^2 \ln \frac{2\Lambda}{|\Delta|}  - \Delta_p^2 \frac{\nu}{2\pi}\int d\phi f^2(\phi) \left( \ln \frac{2\Lambda}{|\Delta|}
	-\frac{1}{2} - \sin^2 \theta
	\right)
	\notag\\
	&\approx F_0(|\Delta|) - \frac{1}{4} \Delta_p^2 \nu \left(
	2\ln \frac{2\Lambda}{|\Delta|}
	-2 + \cos2\theta
	\right)
\end{align}
where $\Delta_2$ is the imaginary part of $\Delta$ and we have made use of $f(\phi)=\cos \phi$ and $\int d\phi f^2(\phi)=\pi$.

At temperatures close to $T_c$, the free energy reads \cite{Sun.2020_collective_modes} $F=-\frac{\nu}{2\pi}\int d\phi \left(-\left(\ln\frac{\Lambda}{T} \right) |\Delta+i\Delta_p f(\phi)|^2 + \frac{c_4}{T^2} |\Delta+i\Delta_p f(\phi)|^4  \right) + \frac{1}{g}|\Delta|^2$ where $c_4$ is an $O(1)$ constant, and the third term in Eq.~(3) of the main text becomes $- \nu \frac{\Delta^2}{T^2} \Delta_p^2 \cos(2(\theta+A_z d))$. In the deep BEC regime ($G \ll -|\Delta|$), this term is replaced by $-\nu_0 \frac{\Delta^2}{GW} \Delta_p^2 \cos(2(\theta+A_z d))$ where $\nu_0$ is a characteristic density of state in the normal state and $W$ is the band width.

Note that due to Josephson coupling which breaks the $U(1)$ invariance, the excitonic insulating state is not a perfect superfluid, meaning that system won't display in-plane counterflow superconductivity \cite{Wen.1992, Lozovik.1997}. This is exactly what the Josephson effect means.

\subsection{Effects beyond mean field}
\label{SI:beyond_mean_field}
In this subsection, we discuss the fluctuation effects beyond mean field. Specifically, we recall the basic theory of fluctuation effects in weakly anisotropic 2D XY systems \cite{Jose.1977} in order to establish the notations and to estimate parameters relevant to the bilayer described by Eq.~2 of the main text. At zero temperature, the quantum fluctuations may be negligibly small (suppressed by the small parameter $\frac{\Delta}{G} \ll 1$ in the BCS weak coupling regime in the same way as BCS superconductors \cite{Altland.2010}) or relatively big (away from BCS weak coupling).   However, it is well known from the map from the 2D quantum model to a 3D classical model that the system is in the long range ordered phase given strong enough stiffness, and that a $U(1)$ symmetry breaking term ($\cos N\theta$) further stabilizes the long range order \cite{Altland.2010,Herbut.2007}. 
Therefore, we focus on the effect of thermal fluctuations of the bilayer at nonzero temperature which is more relevant to experiments. 
Its thermal phase transition is described by the XY model decorated with the $U(1)$ symmetry breaking Josephson term  \cite{Jose.1977}:
\begin{align}
	S=\int d^2r \left[\frac{1}{2} K_0 (\nabla \theta)^2 + g_N \cos (N\theta)\right] ,
	\quad
	Z=\int D[\theta] e^{- S},
	\quad
	\theta \in [0,2\pi)
	\label{eqn:XY_cos}
\end{align}
where $Z$ is the partition function, $D[\theta]$ means the functional integral over the phase field, $K_0= \frac{\nu v_g^2}{k_B T} = \frac{n_s/m}{k_B T}$, $n_s$ is the bare superfluid density, $m$ is the band effective mass, $k_B$ is the Boltzmann constant, $T$ is the temperature,  $N$ is the order of the Josephson effect and $g_N$ is its coupling constant. For the second order Josephson term, one has $g_2 \sim \nu \Delta_p^2/(k_BT)$.  The short distance cutoff is $a$. Since the inter layer tunneling is usually weak, the dimensionless Josephson coupling $g_2 a^2$ is presumably a small number. For example, one has $g_2 a^2 \approx 0.02$ for $m=0.3 m_e$, $\Delta_p=1 \unit{meV}$, $T= 50 \unit{K}$ and $a=10 \unit{nm}$. Note that \equa{eqn:XY_cos} is \emph{not} the Sine-Gordon model because the domain of $\theta$ is not $(-\infty,\infty)$.
At the level of mean field theory, the system described by \equa{eqn:XY_cos} is always in the ordered (excitonic insulator) state since  \equa{eqn:XY_cos} itself exists only in the mean field EI phase. 
Phase fluctuations beyond mean field tends to destroy the long range order. 

Without the Josephson ($\cos N\theta$) terms, the system has $U(1)$ (or equivalently, $O(2)$) invariance which describes the XY model (amended by the vortex fugacity $y_0=e^{-E_{\text{core}}/(k_B T)}>0$ where $E_{\text{core}}$ is the vortex core energy). It is thermal-dynamically equivalent to the neutral 2D Coulomb plasma and the Sine-Gordon model \cite{Altland.2010,Herbut.2007}. The phase transition is of Berezinskii–Kosterlitz–Thouless (BKT) type  \cite{Berezinsky.1971, Kosterlitz.1973} with transition temperature $T_{\text{BKT}} = \frac{\pi}{2} \frac{\hbar^2 n^\ast_s}{k_B  m}$ where $n^\ast_s$ is the universal drop of superfluid density across the transition. Assuming $n_s^\ast=10^{12} \unit{cm^{-2}}$ and $m=0.3 m_e$, typical parameters for transition metal dichalcogenide bilayers \cite{Wang.2019, Ma.2021strongly},  one has $T_{\text{BKT}} \approx 50 \unit{K}$. Above $T_{\text{BKT}}$, the system is a plasma of free vortices, with exponentially decaying correlation of $\theta$. Below $T_{\text{BKT}}$, it is in an ordered (superfluid) phase of bound vortex pairs, with a `quasi long rang order' characterized by the power law decay of the phase field correlation function $\langle e^{i(\theta(0)-\theta(r))} \rangle \sim \left( \frac{a}{r} \right)^{\frac{1}{2\pi K_{\text{eff}}}}$ where $K_{\text{eff}}$ is the renormalized value of $K_0$ on the fix line $(K_0^{-1}<\pi/2,\, y_0=0)$ led by renormalization group (RG)  flow. 

With the Josephson terms, the system tends to be long range ordered since there is no spontaneously broken continuous symmetry. At the level of  Gaussian fluctuations around the mean field minimum, the correlation function is 
\begin{align}
	C(r)=\langle e^{i(\theta(0)-\theta(r))} \rangle = e^{-\langle(\theta(0)-\theta(r))^2 \rangle/2 } 
	= e^{-\frac{1}{4\pi^2 K_0}  \int d^2\mathbf{k} \frac{1-e^{i \mathbf{k} \mathbf{r}}}{k_0^2+k^2}} \approx
	\left\{
	\begin{array}{lc}
		\left( \frac{a}{r} \right)^{\frac{1}{2\pi K_0}} \,,
		& r \ll 1/k_0 \\
		(ak_0)^{\frac{1}{2\pi K_0}}\,,
		&  r \gg 1/k_0 
	\end{array}
	\right. 
	\label{eqn:correlation}
\end{align}
where $k_0 = N \sqrt{g_N/K_0}$ is from the phase mode gap. The correlation decays as a power of $r$ at short distances (starting as $1$ at the short distance cutoff $a$), and then approaches a nonzero value at large distance (long range order). Of course, a very small $g_N$ may not immediately suppress the strong fluctuations of the XY model if nonperturbative effects of fluctuations are included. This physics is contained in the RG analysis by Jos\'{e}, Kadanoff, Kirkpatrick and Nelson \cite{Jose.1977}. For readers' convenience, we reproduce the RG equation here (Eq. 5.17 of Jos\'{e} \textit{et al.} \cite{Jose.1977}, perturbative in $y_0$ and $g_N$):
\begin{align}
	\frac{d K_0^{-1}}{d l} = 2\pi^3 y_0^2  - \frac{\pi}{2}  \frac{N^2 (a^2 g_N)^2}{K_0^2} e^{-N^2/ (4K_0)} 
	,\quad
	\frac{d y_0}{d l} = \left(2-\pi K_0 \right) y_0
	,\quad
	\frac{d g_N}{d l} = \left( 2-\frac{N^2}{4\pi K_0} \right) g_N
	\label{eqn:RG_flow}
\end{align}
where $l$ is the increase of length scale and note that we have different definition of  $y_0$ from Jos\'{e} \textit{et al.} The scaling (engineering) dimensions of the coupling constants are $[K_0]=0, [y_0]=[g_N]=2$. Setting $g_N=0$, one recovers the RG flow of the XY model with the quasi long range ordered phase occurring at $K_0^{-1} < \frac{\pi}{2} - \frac{\pi^2}{\sqrt{2}} y_0$. 
Nonzero $g_N$ adds a negative  flow to the `temperature' $K_0^{-1}$ which is a manifestation of the Josephson potential reducing the fluctuations.
In the flow for $g_N$, the $2$ is from its scaling dimension while the second term is due to leading order feedback of fluctuations, which is negative since short wave length fluctuations on top of the long wave length fluctuations of $\theta$ obscures the effective $\cos (N\theta)$ potential seen by the latter. 

Now we focus on the $N<4$ situation which is not discussed by Jos\'{e} \textit{et al.} \cite{Jose.1977} but covers our case. Close to the fixed line $y_0=g_N=0$, $g_N$ is relevant at $K_0^{-1} < 8\pi/N^2$ and irrelevant at $K_0^{-1} > 8\pi/N^2$. Specifically, for second order Josephson effect (N=2), its coupling $g_2$ is relevant if the temperature is low enough: $K_0^{-1} < 2\pi$. Therefore, starting with a point in the ordered phase of the XY model, $K_0^{-1} < \frac{\pi}{2} - \frac{\pi^2}{\sqrt{2}} y_0$, the vortex fugacity $y_0$ is irrelevant while the Josephson coupling $g_N$ is relevant, and the system should flow to a fixed point for ordered phase: $y_0 \rightarrow 0  ,\, a^2 g_N \rightarrow \infty ,\, K_0^{-1} \rightarrow 0$. Since the flow in \equa{eqn:RG_flow} is perturbative  in $g_N$ and $g_0$, we terminate it at $a^2 g_N(l) \sim 1$, before which the flow can be approximated by $y_0(l) \sim y_0 e^{(2-\pi K_0)l} ,\, g_N(l) \sim  g_N  e^{(2-\frac{N^2}{4\pi K_0})l},\, K_0^{-1}(l) \sim K_0^{-1}$ to leading order in $y_0$ and $g_N$.  
Scaling back by the factor $e^{-2l}$ since $[g_N]=2$, one obtains the renormalized Josephson coupling: $a^2 g_{N \text{eff}} \sim  \left( a^2 g_N \right)^{\frac{2}{2-\frac{N^2}{4\pi K_0}}}$ in the original scale where $\frac{2}{2-\frac{N^2}{4\pi K_0}} \approx \frac{1}{1-\frac{N^2}{16} \frac{T}{T_{\text{BKT}}} }>1$ is a temperature dependent power, which ranges from the no-fluctuation value $1$ at zero temperature to about $4/3$ at $T=T_{\text{BKT}}$ for the second order Josephson effect.  This should be interpreted as the renormalized effective Josephson coupling (and the Josephson current) after integrating out the high energy fluctuations.  
At the termination point  of the flow $a^2 g_N(l) \sim 1$, one can also compute the correlation function $C^\ast(r)$ with Gaussian fluctuations, which renders \equa{eqn:correlation} with $k_0 \sim  N \sqrt{K_0^{-1}}/a$ and $C^\ast(\infty) \sim (N^2 K_0^{-1})^{\frac{1}{4\pi K_0}}$. After recalling back to the original scale, we conclude that the renormalized correlation function $C(r)$  starts from $C(0)=1$ and decays to its long range limit $C(\infty) =  C^\ast(\infty)>0$ within a length scale of $k_{0\text{eff}}^{-1}=N^{-1} \sqrt{K_0/g_{N \text{eff}} }$.
Therefore, if the parameters are such that the system without the Josephson term is below the BKT transition, it is guaranteed to enter a strictly long range ordered phase after the Josephson term is added. The actual critical temperature for the $Z_N$ symmetry breaking should be even higher than $T_{\text{BKT}}$ \cite{Jose.1977,zeng2021phasefluctuation}. This picture holds for Josephson effects of order $N < 4$.

We  note that for $N>4$ Josephson effects, being below the BKT transition temperature $K_0^{-1} < \frac{\pi}{2} - \frac{\pi^2}{\sqrt{2}} y_0$ is no longer a sufficient condition for the Josephson coupling to be relevant since the point of reversing flow direction for $g_N$, $K_0^{-1}=8\pi/N^2$, is now smaller that $\pi/2$. In this case, one has to further lower the temperature to enter a strictly long range ordered phase \cite{Jose.1977}.

\subsection{Fraunhofer interference experiment}
\label{SI:Fraunhofer}
\begin{figure}
	\includegraphics[width=0.8 \linewidth]{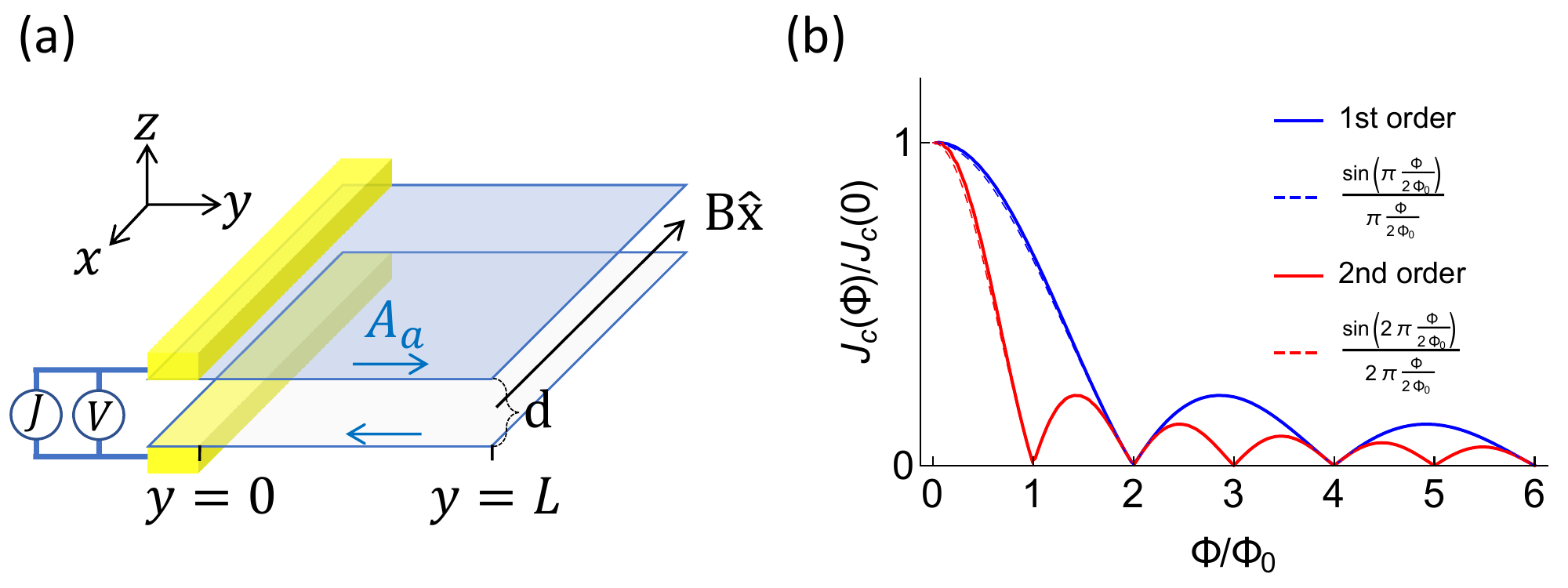}
	\caption{(a) Schematic of the quantum interference experiment where  the Josephson critical current of an electron hole bilayer is measured with in plane magnetic field applied along the $x$ direction. The blue arrows represent the vector potential of the magnetic field. The yellow objects are the metalic contacts. One needs to subtract the contact resistance (if there is any) in the current voltage measurement. (b) The Josephson critical current $J_c$ in a short junction ($L=0.5 l_d$) as functions of the magnetic flux through it. Solid blue/red line is prediction of the first/second order Josephson effect. Dashed lines are analytical approximations $J_c(\Phi)/J_c(0)=|\frac{\sin \left( N \pi \Phi/(2\Phi_0) \right)}{N \pi \Phi/(2\Phi_0)}|$ that become exact in the short junction limit.}
	\label{fig:fraunhofer}
\end{figure}
The second order current phase relation $J=J_c \sin 2\theta$ can be verified  in quantum interference experiments \cite{Tinkham, Lozovik.1997} such as the one shown in Fig.~\ref{fig:fraunhofer}. There one can apply an in plane magnetic field $B$ along $x$ direction, and measure the Josephson critical current $J_c$ of the junction as a function of magnetic field  by sourcing and draining the current on its edge at $y=0$. We represent the magnetic field with the anti-symmetric vector potential: $A_a=\frac{Bd}{2}\hat{y}$ on the two planes. A good approximation to the critical current is $J_c(\Phi)/J_c(0)=|\frac{\sin \left( N \pi \Phi/(2\Phi_0) \right)}{N \pi \Phi/(2\Phi_0)}|$ where $N$ is the order of the Josephson effect, $\Phi=dLB$ is the magnetic flux through the junction and $\Phi_0=\pi \hbar c/e$ is the flux quantum. This formula is accurate in the strong field regime such that the phase varies linearly as $\theta(y)=\int dy A_a=yBde/(2c)+\theta_0$, as shown by its agreement with  the exact result in \fig{fig:fraunhofer}(b). Note that the in plane polarization $P(\theta)$ changes with the phase ($P(\theta)$ is along $x$ because the microscopic tunneling is $t_k=i\Delta_p c_f \sin k_x$ in our Hamiltonian). Therefore, if the magnetic field is along $y$, the phase varies in the $x$ direction and $P(\theta)$ leads to in-plane charge density wave which costs potential energy and complicates the physics. Thus we choose $B$ to be along $x$ such that the phase varies in $y$ direction.

In general, the phase does not vary linearly in space. To obtain the static configuration,
one needs to minimize the Lagrangian of the junction (Eq.~2 of the main text) with appropriate boundary conditions in the magnetic field.  Neglecting the screening of the external magnetic field by the current in the junction (which will be shown to be indeed negligible below), the static saddle point equation reads 
\begin{align}
	l_d^2 \partial_y^2 \theta= \sin (N\theta)\,,\quad
	\partial_y\theta|_{y=0}-A_a=J/(\nu v_g^2) \,,\quad
	\partial_y\theta|_{y=L}-A_a=0
	\label{eqn:junction_equation}
\end{align}
where $N$ is the order of the Josephson effect, $l_d=\sqrt{\nu v_g^2/J_c}=\sqrt{n_s/(mJ_c)}$ is the decay length,  $n_s$ is the `superfluid' density and $m$ is the band effective mass. The last two equations are the boundary conditions that the current flowing in is $J$ on the left edge and $0$ on the right edge. Note that $l_d$ is the decaying length scale of the  phase and tunneling current (induced by the current source or the magnetic field)  from the edge to the bulk, and is also the size of isolated solitons \cite{Fogler2001a} in a long junction ($L\gg l_d$). It limited by the small in plane superfluid stiffness $n_s/m$ and  the requirement of charge continuity. Although its appearance in the Sine-Gordon equation resembles the Josephson penetration depth in a conventional superconducting Josephson junction \cite{Tinkham}, they have difference physical origins. In the latter case,   the finite penetration depth is due to screening of the magnetic field by the tunneling supercurrent, while the superfluid stiffness (much bigger  because it is essentially a three dimensional one) is not the bottleneck for the decaying length scale. Defining the dimensionless coordinates $y^\prime=y/l_d$, $L^\prime=L/l_d$, \equa{eqn:junction_equation} becomes
\begin{align}
	\partial_{y^\prime}^2 \theta= \sin (N\theta)\,,\quad
	\partial_{y^\prime}\theta|_{{y^\prime}=0}=\pi \frac{\Phi}{L^\prime \Phi_0}+\frac{2J}{\sqrt{N}J_{\text{cl}}} \,,\quad
	\partial_{y^\prime}\theta|_{{y^\prime}=L^\prime}=\pi \frac{\Phi}{L^\prime  \Phi_0}
	\label{eqn:sine_gordon}
\end{align}
where $J_{\text{cl}}=2 l_d J_c/\sqrt{N}$ is the Josephson critical current of a long junction in zero magnetic field. If one views $y^\prime$ as time, \equa{eqn:sine_gordon} is the dynamics of a generalized pendulum with conserved `energy' $E^\prime=\frac{1}{2}\dot{\theta}^2+\frac{1}{N}\cos(N\theta)$ and initial/final velocity fixed by the boundary conditions. 

In a long junction, without applied current ($J=0$), one can define a critical magnetic field $B_c=\frac{2}{\pi \sqrt{N}}\frac{\Phi_0}{d l_d}$ (i.e., $\Phi/(L^\prime \Phi_0)=2/(\pi\sqrt{N})$). For $B<B_c$, the phase $\theta$ decays exponentially from nonzero values on the two edges to zero in the bulk (note that for $B_{c1}<B<B_c$ where $B_{c1}=2 B_c/\pi$, this solution is only metastable and soliton excitations in the bulk can be favorable \cite{Tinkham}). The exact profile on the left edge is $\theta(y^\prime)=2\text{Arctan}[e^{-\sqrt{2} (y^\prime-y^\prime_0)}]$ where $y_0^\prime$ is chosen such that $\theta(0)$ matches its value on the left edge.  The asymmetric in-plane currents $j_a=A_a n_s/m=Bdn_s/(2m)$ are closed by tunneling currents $J_c \sin (N\theta)$ close to the edges. The screening magnetic field created by the current loop is $B_{\text{screen}}=-\frac{4\pi}{c}j_a=c_{\text{screen}} B$ where $c_{\text{screen}} =\frac{4\pi}{c^2} \frac{n_s}{m}e^2 d = 4\pi n_s d a_0 \alpha^2 \ll 1$, $a_0=\frac{\hbar^2}{me^2}$ and $\alpha=\frac{e^2}{\hbar c}$. For reasonable parameters such as $n_s \sim 10^{13} \unit{cm}^{-2}$, $m=0.1 m_e$ and $d=1 \unit{nm}$, one has $c_{\text{screen}} \sim 10^{-5}$. Therefore, the junction is weakly diamagnetic, but the screening of the magnetic field can be neglected. 
For $B>B_c$, the initial `velocity' $\partial_{y^\prime}\theta|_{{y^\prime}=0}$ means the initial `kinetic energy' exceeds the maximum possible potential energy barrier, and that the pendulum keeps rotating. Thus the junction is filled with phase solitons. 

When sourced with current ($J\neq0$), the maximum possible $J$ which has a solution will be the critical current. It can be found in the following way: start with the initial phase $\theta_L$ and initial velocity $\partial_{y^\prime}\theta|_{{y^\prime}=L^\prime}=\pi \frac{\Phi}{L^\prime  \Phi_0}$ at the right edge (zero in-plane current);  evolve the pendulum equation along `time' $y^\prime$ to the left edge, and obtain the in-plane current there $J(\theta_L, \Phi)=\frac{\sqrt{N}}{2} J_{cl} \left(\partial_{y^\prime}\theta|_{{y^\prime}=0}-\partial_{y^\prime}\theta|_{{y^\prime}=L^\prime} \right)$; the critical current is just $J_c(\Phi)=\max_{\theta_L} J(\theta_L, \Phi)$. The numerical result for $J_c(\Phi)$ is shown in \fig{fig:fraunhofer}(b) for a short junction ($L\ll l_d$). Here at nonzero flux $\Phi=M \Phi_0$, the magnetic field $B=M \Phi_0/(d L) \gg B_c$ such that the `kinetic energy' dominates over the pendulum potential, and the phase soliton profile inside the junctions is approximately linear: $\theta \approx yBde/(2c)+\theta_0$. Therefore, the conventional Fraunhofer formula $J_c(\Phi)/J_c(0)=|\frac{\sin \left( N \pi \Phi/(2\Phi_0) \right)}{N \pi \Phi/(2\Phi_0)}|$ is a good approximation to the result. It is obvious that the period in  $\Phi/\Phi_0$ is $2$ in the first order Josephson effect while it is $1$ in the second order Josephson effect.  For $n_s \sim 10^{13} \unit{cm}^{-2}$, $m=0.1 m_e$ and $J_c=0.25 \unit{nA/\mu m^2}$, one has $l_d \approx 4 \unit{\mu m}$.  In a short junction with $L= 2 \unit{\mu m}$ and $d=1 \unit{nm}$, the magnetic field required for $\Phi/\Phi_0=1$ is about $1 $ Tesla.
Note that in superconducting Josephson junctions, the  first order Josephson effect has period $1$ in the  Fraunhofer interference pattern because the cooper pair has charge $2e$.

\subsection{Circulating currents in the state $\theta=\pm \pi/2$}
\begin{figure}[b]
	\includegraphics[width=0.5 \linewidth]{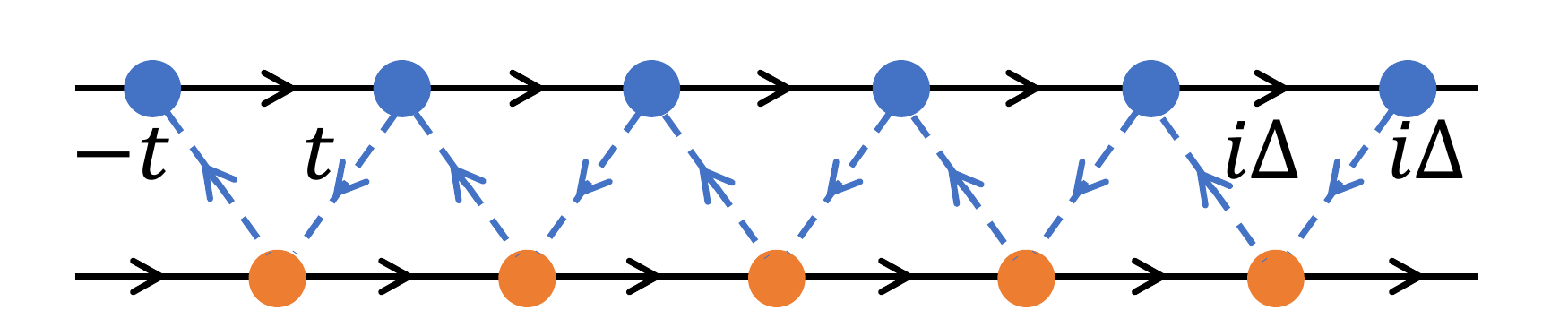}
	\caption{Circulating current pattern in the $\theta=\pi/2$ state (order parameter is $i\Delta$) drawn on the $x-z$ cross section. Shown is a tight binding example with blue (black) arrows indicating the directions of interlayer (intralayer) currents.}
	\label{fig:double_chain}
\end{figure}
During the order parameter steering described in the main text, the system would inevitably pass  $\theta=\pi/2$, the imaginary order state. In this state, the polarization is zero but there are microscopic circulating currents with zero uniform component. Assuming $\Delta$ is momentum independent for simplicity, the mean field Hamiltonian is 
\begin{align}
	H_k=\xi_k \sigma_3 - (\Delta_p f_k +\Delta)\sigma_2
	\label{eqn:imaginary_state}
\end{align}
and the in plane current operator is 
\begin{align}
	j_x=\sum_k \psi^\dagger_k \left( \partial_k H_k \right) \psi_k=\sum_k \psi^\dagger_k \left( v_p \sigma_3 -\Delta_p \partial_k f_k \sigma_2 \right) \psi_k = j_{\text{intra}} + j_{\text{inter}}
\end{align}
where $j_{\text{intra}}$
is the current within each layer and $j_{\text{inter}}$ is the current between the layers. Integrated over space, its expectation value is
\begin{align}
	j_x \propto \sum_k \langle \psi_k | \partial_k H_k | \psi_k \rangle 
	=\sum_k \partial_k \langle \psi_k |  H_k | \psi_k \rangle =0
	\label{eqn:zero_current}
	\,.
\end{align}
However, the $j_{\text{inter}}$ component is nonzero since $\partial_k f_k$ is even in $k$ and $\langle \sigma_2 \rangle_k$ has the same sign at both directions of momentum. In the BCS limit and taking $f(k)=k/k_F$, it can be directly verified that the currents are
\begin{align}
	j_{\text{inter}} &=\mathrm{Tr[j_{\text{inter}} G]}=\sum_{k, i\omega_n} \mathrm{Tr[j_{\text{inter},k} G_{k,i\omega_n}]} =-\frac{\Delta_p}{k_F}\sum_{k, i\omega_n} \mathrm{Tr[\frac{(i\omega_n+H_k) \sigma_2}{(i\omega_n)^2-E_k^2}]} 
	\notag \\
	&=2\pi\frac{\Delta_p \Delta}{k_F}\sum_{k} \frac{1}{E_k}
	\sim n v_F 
	\frac{\Delta \Delta_p}{\varepsilon_F^2} \ln \frac{\Lambda}{\Delta}
\end{align} 
where $n$ is the carrier density in the normal state, $\varepsilon_F=G/2$ is the fermi energy, $\Lambda \sim G$ is an UV cutoff and we have assumed $\Delta \gg \Delta_p$.
Therefore, there is a nonzero current $j_{\text{inter}}$ between the orbitals lying on adjacent layers, whose in plane component must be compensated by $j_{\text{intra}}=-j_{\text{inter}} $ within the layers to satisfy \equa{eqn:zero_current}. In other words, this state has microscopic circulating currents between orbitals while there is no uniform current, as shown in \fig{fig:double_chain}.

If the order parameter depends on momentum, the mean field state might appear to have a nonzero uniform current since the mean field Hamiltonian is no longer  consistent with the current operator ($j \neq \sum_k \psi^\dagger_k \left( \partial_k H_k \right) \psi_k$), i.e., local gauge invariance under $\psi(r) \rightarrow \psi(r) e^{i\phi},\, A_\mu \rightarrow A_\mu + \partial_\mu \phi$ appears broken.  \equa{eqn:imaginary_state} will be such an example if the $\Delta_p$ term comes from a $p$-wave excitonic order instead of from intrinsic hybridization: the only contribution to the current should now be $j_{\text{intra}} \neq 0$. To fix this problem, one should note that there is another phase degree of freedom $\theta_p$ in the $p$-wave order parameter ($f_k \rightarrow f_{k-\theta_p}$) \cite{sun2020topological} that couples linearly with the vector potential, and the true ground state has nonzero $\theta_p$ which finally renders a zero net current.

\subsection{Device realization}
\label{SI:device}
\begin{figure}[t]
	\includegraphics[width=0.5 \linewidth]{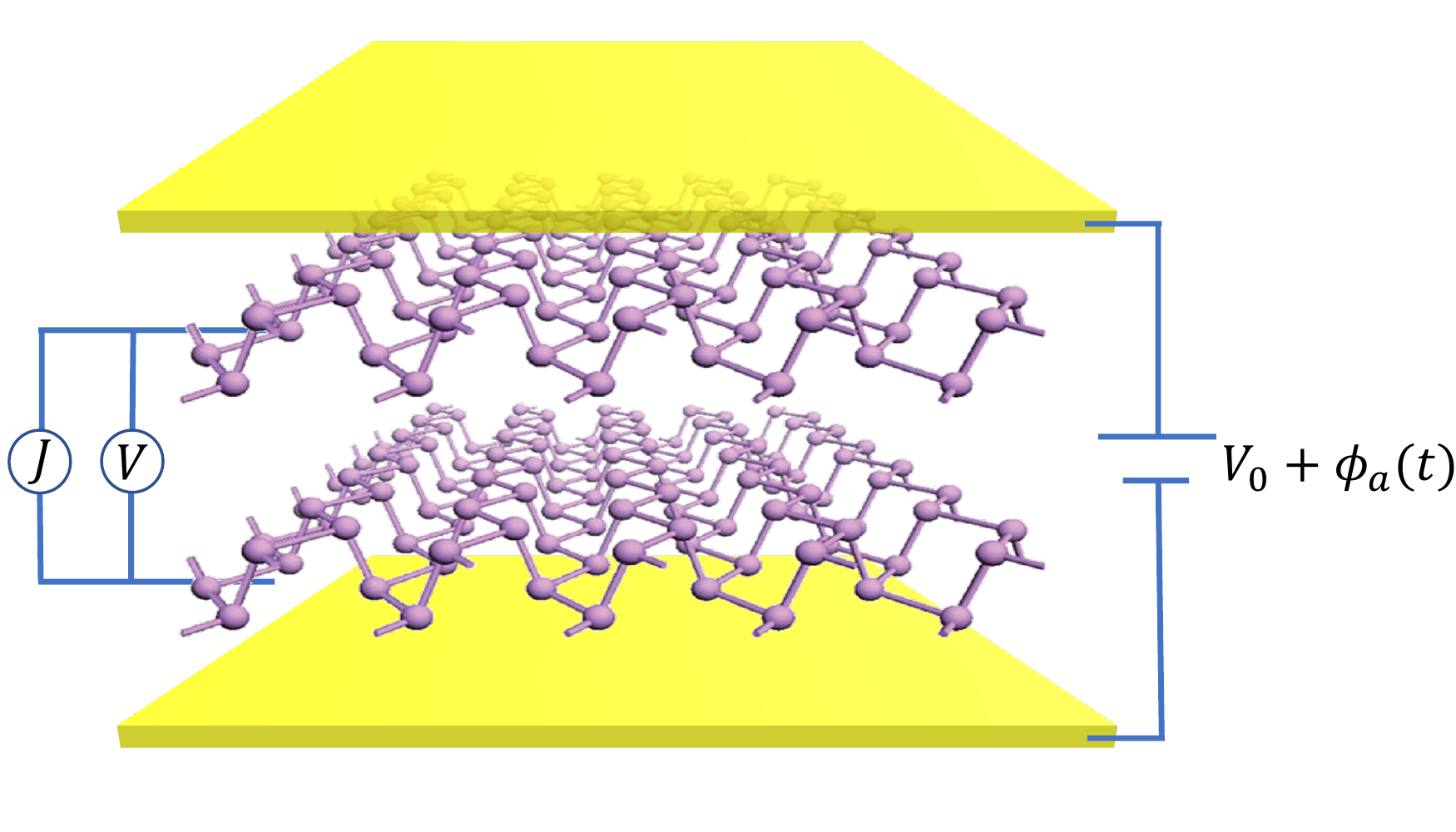}
	\caption{Illustration of the biased bilayer device for realizing the second order Josephson effect in an EI.  The bilayer system can either be phosphorene bilayer (shown in figure) \cite{Kim.2015_phorsphorene, Li.2014_phosphorene,Carvalho.2016_phosphorene} or transition metal dichalcogenide bilayers \cite{Wang.2019, Ma.2021strongly}. The yellow objects are metallic gates which provides a static z-direction electric field that shifts down/up the conduction/valence band of the top/bottom layer. The same gates can also be used to apply an additional voltage pulse $\phi_a(t)$ in an order parameter steering experiment. The `$J$' and `$V$' are ampere and voltage meters to measure the Josephson like current voltage relation.}
	\label{fig:biased_bilayer}
\end{figure}
The bilayer can be realized by the device in \fig{fig:biased_bilayer} where the z direction electric field from the bias brings the conduction band of the top layer and valence band of the bottom layer closer in energy \cite{Fogler2014a}, entering the EI phase. To realize the second order Josephson effect, the material should be such that the two relevant bands have different eigenvalues under certain point group symmetry operations (e.g., $C_2$ rotation around z, mirror operations that reverse x or y). There are two obvious candidates.
The first is phosphorene \cite{Kim.2015_phorsphorene, Li.2014_phosphorene,Carvalho.2016_phosphorene,Li.2014}. Phosphorene monolayer has a $2 \unit{eV}$ (nearly) direct band gap at gamma point, and the relevant conduction/valence band has eigenvalue $-1/1$ under a nonsymmorphic point group operation $\tau C_{2z}$ \cite{Li.2014} where $C_{2z}$ means a $180^\circ$ rotation around z axis, $\tau$ means an in-plane translation by $(a_x, a_y)/2$ where $(a_x, a_y)$ are the in-plane lattice constants. 
Similarly, the bands have eigenvalues $-1/1$ under another operation $\tau R_{x}$ where $R_{x}$ is a reflection $x \rightarrow -x$. For phosphorene bilayers with all three stable stacking orders (AA, AB, AC)\cite{Dai:2014}, the $\tau C_{2z}$ symmetry is preserved, and the top layer conduction band/bottom layer valence band has eigenvalue $-1/1$ under $\tau C_{2z}$, satisfying the requirement stated in the main text. Specifically, the leading interband tunneling will be $t_k \propto k_x$ \cite{Rodin.2014_phosphorene} due to their different $\tau R_{x}$ eigenvalues. The second candidate is the class of transition metal dichalcogenide bilayers (TMDB) where signatures of exciton condensate have been observed in biased devices \cite{Wang.2019, Ma.2021strongly}. In a monolayer, the relevant conduction and valence bands are at K points which have different eigenvalues under $C_3$ rotation around z.   In homo-bilayers, out of 6 possible types of stacking, four of them ($H_M^M$, $H_X^M$, $R_M^X$, $R_M^M$) \cite{Tong:2017uh} preserve the symmetry property such that top layer conduction band/bottom layer valence band have different eigenvalues under suitably chosen $C_3$ rotations. As a result, the interband tunneling has the chiral form \cite{Tong:2017uh} $t_k \propto k_x \pm ik_y$. In the EI state, this  falls in the class of second order Josephson effect, but with additional interesting properties. For example, in the intra valley pairing state, the direction of the in plane electrical polarization can be continuously tuned by the phase of the order parameter. Details of the second order Josephson effect in TMDBs will be in a forthcoming publication.

\section{Stack of electron-hole layers/chains}
\label{SI:stack}
\subsection{Static free energy and the ground state}
We represent $d k_z$ by $k_z$ wherever possible for notational simplicity.
Eq.~(7) of the main text is obtained by expanding the zero temperature static free energy to second order in $\Delta_p$:
\begin{align}
	F &= F_0(|\Delta|) -\frac{1}{4} \Delta_p^2 \sum_k \frac{f_k^2 \cos^2 (k_z)}{E_{0k}} \mathrm{Tr}\left[
	\sigma_2^2 - \frac{H_{0k} \sigma_2 H_{0k} \sigma_2}{E_{0k}^2}
	\right] 
	\notag \\
	&= F_0(|\Delta|) -\frac{1}{2} \Delta_p^2 \sum_k \frac{f_k^2 \cos^2 (k_z)}{E_{0k}} 
	\left[
	1+\frac{1}{E_{0k}^2} \left( \xi_k^2 +|\Delta|^2 \cos 2\theta_a \cos^2(k_z +\theta_s)   \right)
	\right] 
	\notag \\
	&= F_0(|\Delta|) - \Delta_p^2 \sum_k \frac{f_k^2 \cos^2 (k_z)}{E_{0k}^3} 
	\left[\xi_k^2 + \frac{1}{2}|\Delta(k)|^2 (\cos 2\theta_a+1)   
	\right] 
	\label{eqn:L_s_p_SI}
	\,.
\end{align}
where $\Delta(k)=e^{i \theta_a} \Delta \cos(k_z + \theta_s)$, $E_{0k}=\sqrt{\xi_k^2+|\Delta(k)|^2}$ and $H_{0k}=\xi_k \sigma_3 + \Delta \cos{\theta_a}\cos(k_z + \theta_s) \sigma_1 -\Delta \sin{\theta_a}\cos(k_z + \theta_s) \sigma_2$. 
In the BCS weak coupling case $|\Delta|, |\Delta_p| \ll G$, the momentum summation leads to
\begin{align}
	F & \xrightarrow{\text{BCS}} F_0(|\Delta|) - \Delta_p^2 
	\nu_{2D} \int \frac{d k_z}{2\pi} \cos^2 (k_z) \int \frac{d\phi_k}{2\pi}  f_{\phi_k}^2
	\left( 
	(\cos 2\theta_a+1)
	+  2 \ln \frac{2\Lambda}{|\Delta(k)|} 
	\right)
	\notag \\
	& = F_0(|\Delta|) - \Delta_p^2 
	\frac{\nu}{4} 
	\left( (\cos 2\theta_a+1) +
	\frac{4}{\pi}\int d k_z \cos^2 (k_z) 
	\ln \frac{2\Lambda}{|\Delta(k)|} 
	\right)
	\notag \\
	& = F_0(|\Delta|) - \Delta_p^2 
	\frac{\nu}{4} 
	\left( (\cos 2\theta_a+1) +
	h(\theta_s) 
	\right)
	\,.
\end{align}
Note that the coefficient  $\frac{\nu}{4}$ holds for 3D and should be replaced by $\frac{\nu}{2}$ in 2D.
Here we have defined the function $h(\theta_s)$ which is obviously odd in $\theta_s$ and periodic in $\theta_s$ with period $\pi$. It can thus be Fourier expanded as $h(\theta_s)=\sum_{n \in Z} a_n \cos(2n \theta_s)$, where
\begin{align}
	a_0&=\frac{4}{\pi^2}\int_0^\pi d\theta dk_z \cos^2 (k_z) 
	\ln \frac{2\Lambda}{|\Delta \cos (k_z+\theta)|}=  2 \ln \frac{4\Lambda}{|\Delta|}
	\,,\notag\\
	a_1&=\frac{8}{\pi^2}\int_0^\pi d\theta dk_z \cos (2\theta) \cos^2 (k_z) 
	\ln \frac{2\Lambda}{|\Delta \cos (k_z+\theta)|}
	= -1
	\,.
\end{align}
Therefore, to leading Fourier expansion the free energy reads
\begin{align}
	F = F_0(|\Delta|) + \frac{\nu}{4}\Delta_p^2  
	\left( -\cos(2\theta_a) +  \cos(2\theta_s)  -1-2 \ln \frac{4\Lambda}{|\Delta|}
	\right)
	\label{eqn:L_s_p_SI_bcs}
	\,
\end{align}
in the BCS weak coupling case. 

For systems close to the transition temperature $T_c$ such that  $T_c \gg |\Delta, \Delta_p|$, or in the BEC case $G \ll -|\Delta, \Delta_p|$, the free energy can be expanded in powers of $\Delta_k= e^{i \theta_a} \Delta \cos(k_z + \theta_s) - i\Delta_p f_{k} \cos  k_z$ as
\begin{align}
	F = \sum_k \left[ c_2 |\Delta_k|^2 +  c_4 |\Delta_k|^4 + O(|\Delta_k|^6)   \right]
	\label{eqn:F_s_p_SI_BEC}
	\,.
\end{align}
One has $c_2 \sim -1/T,\, c_4 \sim 1/T^3$ for $\xi_k \ll T$ and $c_2 \sim 1/\xi_k,\, c_4 \sim 1/\xi_k^3$ for $\xi_k \gg T$. The first term of \equa{eqn:F_s_p_SI_BEC} reads $|\Delta_k|^2=\Delta^2  \cos^2(k_z + \theta_s) -
2\Delta \Delta_p f_{k} \sin\theta_a \cos(k_z + \theta_s)   \cos k_z +  \Delta^2_p f^2_{k} \cos^2 k_z$. After being summed over $k$, it has no $\theta_a$ or $\theta_s$ dependence since the cross product term  $\sim \Delta \Delta_p$ sums to zero due to $f_k$ being odd. The second term of  \equa{eqn:F_s_p_SI_BEC} gives phase dependence:
\begin{align}
	F  &= F_0 + 2\Delta^2   \Delta^2_p \left(2 \sin^2\theta_a  + 1   
	\right)  \frac{2}{(2\pi)^D} \int dk_z dk^{D-1}_\perp c_4(k_\perp)  f^2_{k_\perp}   \cos^2(k_z + \theta_s)   \cos^2 k_z
	\notag\\
	&=
	F_0 + c_\theta \Delta^2   \Delta^2_p \left(2 \sin^2\theta_a  + 1   
	\right)  (2\cos^2 \theta_s + 1)
	\label{eqn:F_phase_SI_BEC}
	\,
\end{align}
where $c_\theta = \frac{1}{4(2\pi)^{D-1}}  \int dk^{D-1}_\perp c_4(k_\perp)  f^2_{k_\perp}$. In the BCS limit and close to $T_c$, one has $c_\theta  \sim \nu/T^2$. In the BEC regime, we redefine $f_{k_\perp}= \sin k_x$ since there is no longer a "Fermi surface" to normalize at. At zero temperature, one has $c_\theta  \sim \nu_0/(GW)$ where $\nu_0$ is the characteristic density of states of the band $\xi(k)$ in the normal phase and $W$ is the band width.

\begin{figure}
	\subfloat[][]{
		\includegraphics[width=0.3\linewidth]{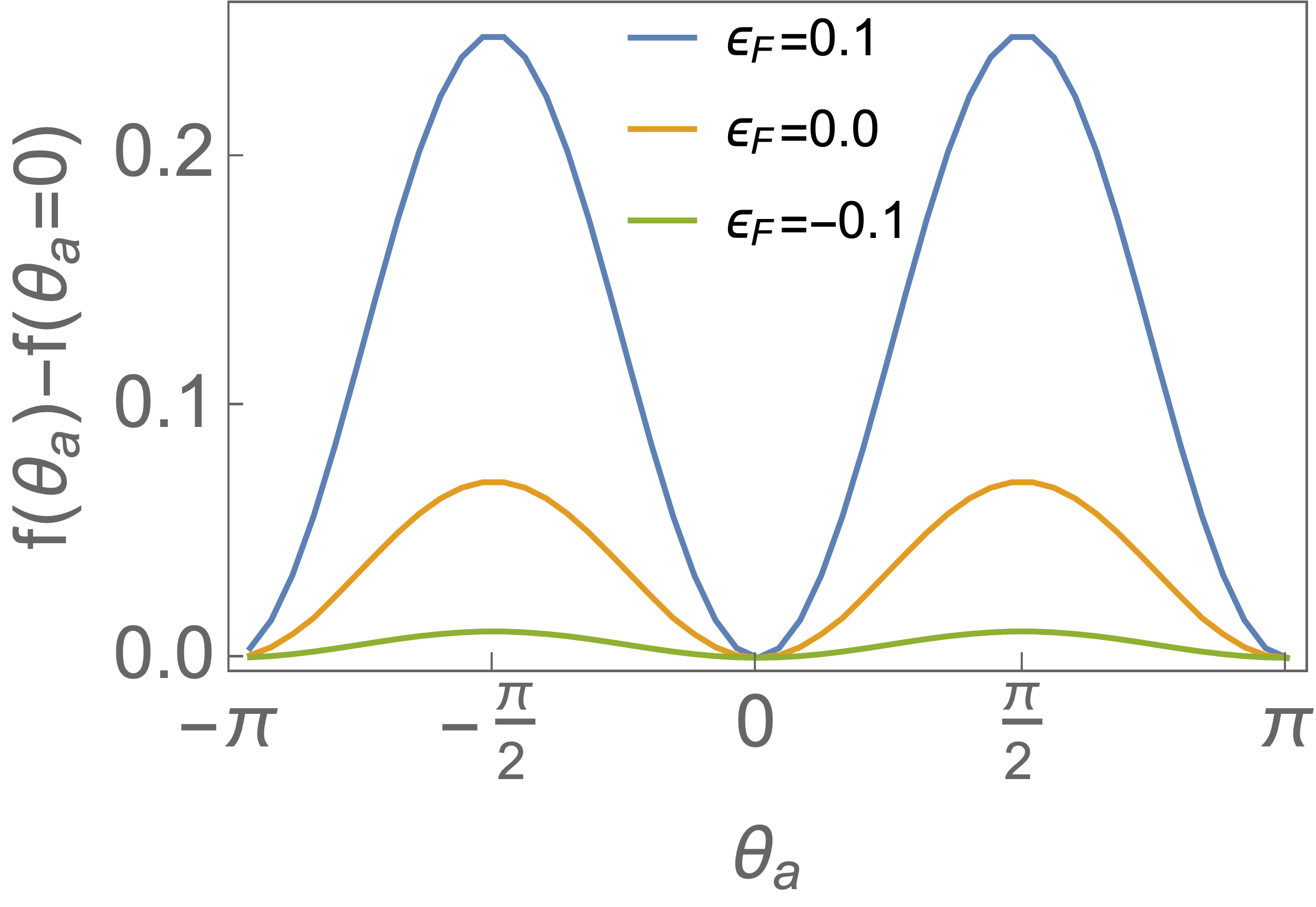}
		\label{fig:sfig1}
	}
	\subfloat[][]{
		\centering
		\includegraphics[width=0.3\linewidth]{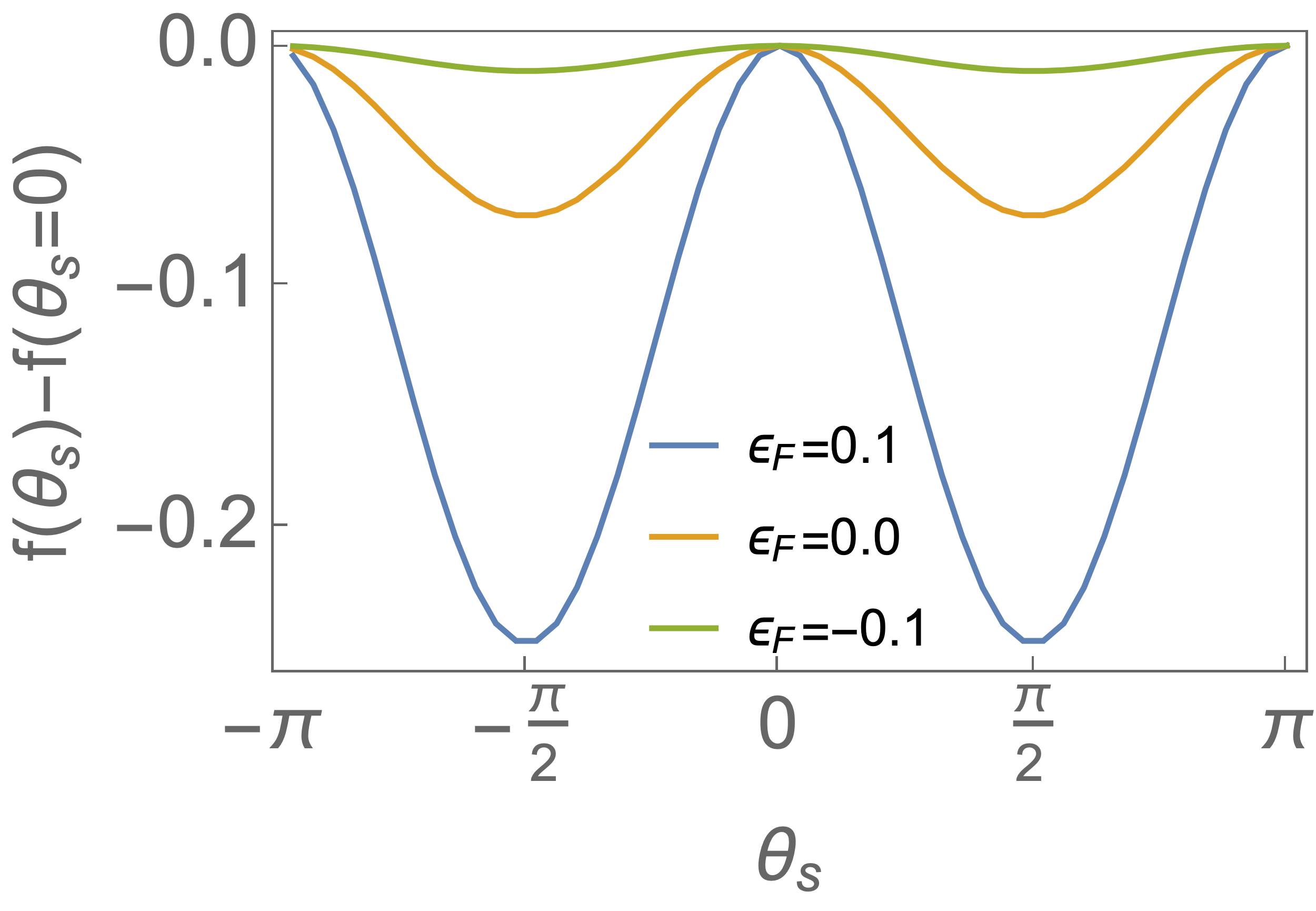}
		\label{fig:sfig2}
	}
	\subfloat[][]{
		\centering
		\includegraphics[width=.3\linewidth]{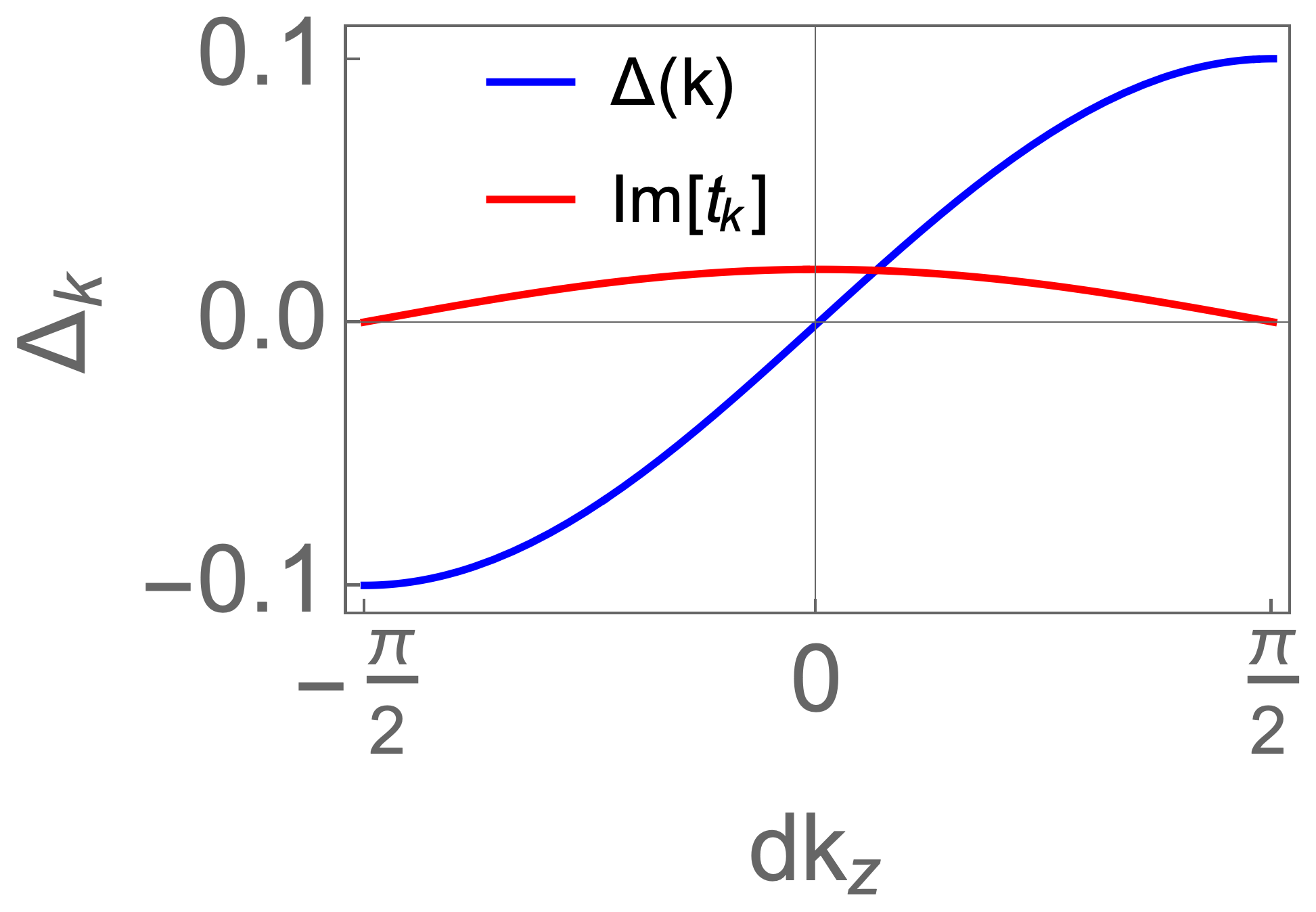}
		\label{fig:sfig2}
	}
	\caption{ The free energies (a) $f(\theta_a,\theta_s=\pi/2)$ (b)  $f(\theta_a=0,\theta_s)$ in arbitrary units for several values of  $\epsilon_F$ for the 2D stack.  (c) Blue curve is the optimized order parameter $\Delta(k_F,k_z)$ as a function of $k_z$. Red curve is the imaginary intrinsic tunneling $i\Delta_p f_{k_F} \cos k_z$. The band dispersion is $\xi_1=-\xi_2=-\cos k_x+1-\epsilon_F$. The $s$-wave gap magnitude is fixed at $\Delta=0.1$, the intrinsic hybridization is $\Delta_p=0.01$. The dependence on the phases are similar in both BCS  ($\epsilon_F > 0$) and BEC ($\epsilon_F < 0$)  regimes.}
	\label{fig:delta_theta}
\end{figure}

In any case, the intrinsic tunneling has reduced the symmetry group from $U(1)\times U(1)$ to $Z_2 \times Z_2$, i.e., inversion $\hat{P}$ that maps $(\theta_a,\theta_s)$ to $(\theta_a, -\theta_s+\pi)$ and time reversal $\hat{T}$ that maps $(\theta_a, \theta_s)$ to $(-\theta_a, -\theta_s)$. Since the free energy minima lie at $(\theta_a,\theta_s)=(0, \mp\pi/2)$), the ground state spontaneously breaks $\hat{T}$.
Numerically exact results (non perturbative in $\Delta_p$) for the 2D stack are shown in \fig{fig:delta_theta}. 

We take the BCS weak coupling limit to describe the ground state and edge states. In the 2D stack, the ground state is a quantum anamolous hall insulator with chiral edge states. For example, along the edges parallel to $z$, the edge states have dispersion $E(k_z)= \pm \Delta(k_z)$ and wave function $\psi_{\pm}(k_z)=(1,\, \pm 1) \sin (k_F x) e^{\mp \Delta_p \cos(k_z) x} e^{ik_z  z}$ with the $\pm$ sign denoting the left/right edge \cite{sun2020topological}. In the 3D stack of layers, the ground state is a Weyl semi metal with fermi arc surface states. On the $y-z$ surface, the surface states have dispersion $E(k_y, k_z)= \pm \Delta(k_z)$ and wave function $\psi_{\pm}(k_y, k_z)=(1,\, \pm 1)  \sin (k_F x) e^{\mp \Delta_p f_k \cos(k_z) x} e^{i(k_y  y+k_z z)}$. Here $f_k$ means the function $f$ evaluated on the fermi surface at $(k_x,k_y)=(\sqrt{k_F^2-k_y^2},k_y)$.

Note that in the limit of only one bilayer with periodic boundary condition, $k_z$ can only take the value $0$, and the ground state favors the order parameter $\Delta(k)=\pm \Delta \cos(k_z)$ which does not break  $\hat{T}$. As number of layers is increased beyond two bilayers, the $\hat{T}$ breaking order parameter $\Delta(k)=\pm \Delta \sin(k_z)$ starts to have lower energy. 

In the order parameter steering process, as $\theta_s$ crosses the regimes of $0,\pi$, the system has electrical polarizations in $x$ direction similar to the ground states of the bilayer. The resulting energy cost due to the surface charge is a marginal effect in 2D but may be an obstacle in 3D. However, one can use metallic gates to screen out this effect.

\subsection{Kinetic terms}
\subsubsection{$U(1)$ invariant case}
In this subsection, we discuss the $\Delta_p=0$ limit where the system has $U(1) \times U(1)$ invariance corresponding to varying $\theta_{1}$ and $\theta_{2}$, and the ground state has $\Delta_{i1}=\Delta_{i2}=\Delta$. Without the EM field and assuming $\xi_1=-\xi_2=\xi$, the quasiparticle dispersion is $E=\pm \sqrt{\xi_k^2 + \Delta^2\cos^2(k_z)}$ with Dirac nodal lines at $(k_{\perp},k_z) =(k_F, \pm \pi/2)$.
In the continuous limit, neglecting dissipative terms contributed by the nodes, the low energy Lagrangian for anti symmetric phase fluctuations is 
\begin{align}
	L_a=& \frac{1}{2} \nu \left[- (\partial_t \theta_{a} + \phi_a)^2  + v_g^2 (\nabla \theta_{a}-A_a)^2 
	\right]
	\,
\end{align}
where $\nabla$ means the in plane gradient and $(\phi_a, A_a)=(\phi_1-\phi_2, A_1-A_2)/2$ is the layer-antisymmetric EM field in a unit cell, defined in similar fashion to the single bilayer. In the gauge $A_a=0$, including the electrical field energy $L_{EM}=\frac{1}{4\pi d} \phi_a^2$ (charging energy of the bilayers viewed as capacitors) and integrating it out, one obtains the Lagrangian of Coulomb renormalized phase modes:
\begin{align}
	L_a=& - \frac{1}{2} \frac{1}{1/\nu + 2\pi d} (\partial_t \varphi_{a})^2  + \frac{1}{2} \nu v_g^2 (\nabla \varphi_{a})^2
	\,.
\end{align}
The low energy Lagrangian for  symmetric fields is
\begin{align}
	L_s \sim \frac{1}{2} \nu v_F^2 \frac{1}{\Delta^2} E_s^2 + \nu |\Delta| (\theta_s -dA_z) \partial_t (\theta_s -dA_z)
	\label{eqn:A_s}
	\,
\end{align}
where $E_s$ is the in-plane electric field. The coefficient of the first term is determined by in-plane optical conductivity neglecting dissipative terms.
As expected, there is no z-direction current response since $A_z$ drops out after integrating out $\theta_s$.

\subsubsection{With $P$-type intrinsic tunneling}
To second order in $\Delta_p$, we write the action for quadratic fluctuations of $\theta \equiv \theta_s-\theta_0$ around the ground state as
\begin{align}
	S_s = & - \sum_{\omega} c_0(\omega) \big(\theta+ dA_z \big)_{-\omega} 
	\big(\theta+ dA_z \big)_{\omega} + 
	\int dt dr \Big(  c_1 \theta^2 + 
	\chi \theta \partial_t A_x/d
	+\sigma_{\text{h}}A_z \partial_t A_x\Big)  
	+ S_{A_x^2}   
	\label{eqn:L_s_p_SI}
	\,.
\end{align}
The kinetic kernel $c_0(\omega)$ in the first term is the $\omega$ dependent part of the correlation function
\begin{align}
	c(\omega) &\equiv \chi_{\theta,\theta}=
	\Delta^2 \chi_{\cos (k_z) \sigma_1,\cos (k_z) \sigma_1}(\omega)  
	=  -4\Delta^2 
	\sum_{k} \cos^2k_z \frac{E^2- \Delta^2 \sin^2 k_z}{(\omega^2 - 4E^2)E}
	\,.
\end{align}	
For the 3D stack
\begin{align}
	c_{3D}(\omega) -c_{3D}(0) 
	&\approx  \pi i \Delta^2 
	\sum_{k} 
	\frac{E^2- \Delta^2 \sin^2(k_z)}{E^2} \left( \delta(\omega-2E)-\delta(\omega+2E) \right)
	+ \text{Real Part}
	\notag \\
	&\approx 
	\left\{
	\begin{array}{lc}
		\frac{1}{3} \nu \frac{\Delta}{\Delta_p}  \omega^2 \left(i\text{Sign}(\omega) + \frac{1}{\pi} \ln \frac{\Delta_p}{|\omega|} \right)
		& \omega \ll \Delta_p \\
		\frac{\pi \nu}{4} (i\Delta \omega+ \frac{1}{\pi}\omega^2)
		& \Delta_p \ll \omega \ll \Delta \\
		\nu \Delta^2 (i\text{Sign}(\omega) + \frac{1}{\pi}\ln \frac{\Delta}{|\omega|} )
		& \omega \gg \Delta
	\end{array}
	\right. 
\end{align}	
where the real part is obtained from the imaginary part through Kramers-Kronig relation and we have neglected subleading terms.
For the 2D stack, there is no dissipation when $\omega <\Delta_p$ and one can expand to $O(\omega^2)$:
\begin{align}
	c_{2D}(\omega)-c_{2D}(0) &
	\xrightarrow{O(\omega^2)} 
	\Delta^2 
	\sum_{k} \cos^2(k_z) \omega^2 \frac{E^2- \Delta^2 \sin^2(k_z)}{4E^5}
	\notag \\
	&\xrightarrow{\Delta \ll \varepsilon_F} 
	\Delta^2  \omega^2
	\nu \frac{d}{\pi} \int dk_z \cos^2(k_z) \left(\frac{1/2}{\Delta(k)^2} - \Delta^2 \sin^2(k_z)  \frac{1/3}{\Delta(k)^4}
	\right)
	\notag \\
	&= \Delta^2  \omega^2
	\nu \frac{1}{\pi} \int_{-1}^1 dt \sqrt{1-t^2} \left(\frac{1/2}{\Delta^2 t^2+\Delta_p^2  (1-t^2)} - \Delta^2 t^2  \frac{1/3}{\left(\Delta^2 t^2 +\Delta_p^2 (1-t^2) \right)^2}
	\right)
	\notag \\
	&= \Delta^2  \omega^2
	\nu  \frac{1}{\pi} 
	\left( \frac{\pi}{2}  \frac{-\Delta_p^2+|\Delta\Delta_p|}{\Delta_p^2(\Delta^2-\Delta_p^2)}
	- \frac{\pi}{6} \frac{-2\Delta^2\Delta_p^2+2|\Delta\Delta_p^3|+|\Delta\Delta_p|(\Delta^2-\Delta_p^2)}{\Delta^2 \Delta_p^2(\Delta^2-\Delta_p^2)}
	\right)
	\notag \\
	&\xrightarrow{\Delta_p \ll \Delta} 
	\frac{\nu}{3} \frac{|\Delta|}{|\Delta_p|} \omega^2
\end{align}
In the regime $2\Delta_p \ll \omega \ll \Delta$, the minimal gap $\Delta_p$ can be neglected such that there are Dirac nodes at $(k_x,\, k_z)=(\pm k_F,\, -\theta_s+\pi/2)$ which renders the dynamics of $\theta_s$ dissipative. The kinetic kernel for $\theta_s$ becomes 
\begin{align}
	c_{2D}(\omega) -c_{2D}(0)
	&= \Delta^2 \pi i
	\sum_{k} \frac{\xi^2}{E^2} \left( \delta(\omega-2E)- \delta(\omega+2E)\right) + \text{Real Part}
	\notag \\
	&=\frac{1}{8 v_F  d} (i \Delta \omega + \omega^2) = \frac{\pi \nu}{4} (i \Delta \omega + \frac{1}{\pi} \omega^2)
\end{align}	
where the real part is obtained through Kramer-Kronig relation by noting that the $i\omega$ behavior of the imaginary part has an UV cutoff $\sim \Delta$ beyond which it approaches a constant.

The second term in \equa{eqn:L_s_p_SI} comes from expanding \equa{eqn:L_s_p_SI_bcs}. For example for the 2D stack, the static free energy as a function of $\theta_s$ is
\begin{align}
	F(\Delta,\Delta_p,\theta_s)=-\frac{\nu}{\pi}\int^{\pi/2}_{-\pi/2} dk_z |\Delta \sin(k_z+\theta_s)+i\Delta_p \cos(k_z)|^2 \ln \frac{2\Lambda}{|\Delta \sin(k_z+\theta_s)+i\Delta_p \cos(k_z)|}
	\label{eqn:f_theta}
\end{align}
where $k_z$ means the $z$ direction momentum times the interlayer thickness.
It can be expanded to $O(\theta_s^2)$ as
\begin{align}
	c_1 \theta_s^2 &=\frac{\nu}{2\pi}\int^{\pi/2}_{-\pi/2} d k_z   
	\left[
	\left(-\ln\frac{4\Lambda^2}{|\Delta(k)|^2} +1\right)
	\left(- \Delta_p^2 \cos(2k_z) \theta_s^2
	\right)
	+ \frac{1}{2} \frac{1}{|\Delta(k)|^2} 4\Delta_p^4 \sin^2(k_z)\cos^2(k_z) \theta_s^2
	\right]
	\notag\\
	&=\frac{\nu}{4\pi} \theta_s^2 \int^{\pi/2}_{-\pi/2} dk_z   
	\left[
	\Delta_p^2 \ln \frac{4\Lambda^2}{|\Delta(k)|^2} \cos(2k_z)
	+
	\frac{\Delta_p^4}{\Delta^2 \sin^2(k_z)+\Delta_p^2 \cos^2(k_z)}  \sin^2(2k_z) 
	\right]
	\notag\\
	&=\frac{\nu}{4} \left(\frac{1-\frac{\Delta_p}{\Delta}}{1+\frac{\Delta_p}{\Delta}} \Delta_p^2 +
	\frac{ \Delta_p^4}{\Delta(\Delta+\Delta_p)} \right) \theta_s^2
	\xrightarrow{O(\Delta_p^2)} 
	\frac{\nu}{4} \Delta_p^2 \theta_s^2
\end{align}	

The third term is due to the fact that $\theta_s$ fluctuations are accompanied by  x direction polarization. In the BCS limit, each $k_x$ chain at $k_z$ in momentum space contributes a polarization density of 
$P_{k_z}= \frac{P_{1D}}{2\pi}(\pi/2- \varphi_{k_z})$ 
in the 2D stack, and each $k_x-k_y$ surface at $k_z$ contributes $
P_{k_z}=
\frac{1}{4} P_{2D} \left[1- \tan(\varphi_{k_z}/2) \right]
$
in the 3D stack where $\varphi_{k_z}=\mathrm{ArcTan} \frac{\Delta_p \cos k_z}{\Delta(k_z)}$. Noting that $\Delta(k_z)=\Delta \sin(k_z+\theta)$, the change of polarization is related to $\theta$ as 
\begin{align}
	\chi =\partial_{\theta}P_x&=\frac{P_{1D}}{2\pi} 2\int_0^{\pi/2} \frac{d k_z}{2\pi}  \frac{\Delta \cos(k_z)}{\Delta_p \cos(k_z)} \sin^2\varphi
	=\frac{P_{1D}}{2\pi^2} \frac{\Delta}{\Delta_p}  
	\int_0^{\pi/2} d k_z  \frac{1}{1+\frac{\Delta^2}{\Delta_p^2} \tan^2(k_z)}
	=\frac{P_{1D}}{4\pi} 
	\frac{\Delta/\Delta_p}{1+\Delta/\Delta_p} 
\end{align}
for the 2D stack and as
\begin{align}
	\chi =\partial_{\theta}P_x&=\frac{1}{4} P_{2D}  2  \int_0^{\pi/2} \frac{dk_z }{2\pi}  \frac{\Delta \cos (k_z)}{\Delta_p \cos(k_z)}  \frac{\sin^2 \varphi}{\cos^2 (\varphi/2)}
	= \frac{P_{2D}}{2\pi} \frac{\Delta }{\Delta_p }  \int_0^{\pi/2} dk_z   \frac{\sin^2 \varphi}{1+\cos\varphi}
	\notag\\
	&= \frac{P_{2D}}{2\pi}  \frac{\Delta  }{\Delta_p }
	\int_0^{\pi/2} dk_z  \frac{\cos^2k_z/(\frac{\Delta^2}{\Delta_p^2} \sin^2k_z+\cos^2k_z)}
	{1+\frac{\Delta}{\Delta_p} \sin k_z/\sqrt{\frac{\Delta^2}{\Delta_p^2} \sin^2k_z+ \cos^2k_z}}
	\notag\\
	&= \frac{P_{2D}}{2\pi}  f\left(\frac{\Delta_p}{\Delta} \right)
\end{align}  
for the 3D stack where $f(x)\rightarrow 1$ as $x \rightarrow 0$ and $f(x)\rightarrow \pi/(2x)$ as $x \rightarrow \infty$. Note that $\chi$ reduces to $\sigma_h$ in the limit of $\Delta_p \ll \Delta$ which we assume in Eq.~(10) of the main text for notational simplicity. In general, $\chi \neq \sigma_h$ because $\theta_s$ does not act completely in the same way as $A_z$ since it does not enter the $\Delta_p$ term in the Hamiltonian.

Finaly, the $S_{A_x^2}$ term is determined by the bare optical conductivity along $x$. In the 3D stack, the bare optical conductivity along $x$ direction in the low frequency regime is controlled by the Weyl nodes  with the Hamiltonian for one of them being $H_W=v_F k_y \sigma_3 + v_x k_x \sigma_2 + v_z k_z \sigma_1$ where $v_x=\Delta_p/k_F$ and $v_z= \Delta d$. The optical conductivity in the regime $\omega \ll \Delta_p$ reads
\begin{align}
	\sigma_{x0} &= \frac{v_x^2}{v_z^2} d^2 \frac{i}{\omega} c_{3D}(\omega) 
	= \frac{1}{3}  \frac{\nu}{k_F^2}  \frac{\Delta_p}{\Delta}   \left(|\omega| - \frac{i}{\pi} \omega \ln \frac{\Delta_p}{|\omega|} \right)
	\,.
\end{align}
In the 2D stack of chains, the optical conductivity is controlled by the massive Dirac nodes with Hamiltonian $H_D = v_F k_x \sigma_3 + v_z k_z \sigma_1 + \Delta_p \sigma_2$. The optical conductivity reads
\begin{align}
	\sigma_{x0} &= \frac{v_F^2}{v_z^2} d^2 \frac{i}{\omega} c_{2D}(\omega) 
	= 
	\frac{v_F^2}{\Delta^2} \nu
	\left\{
	\begin{array}{lc}
		-\frac{1}{3} \frac{|\Delta|}{|\Delta_p|} i\omega
		& \omega \ll \Delta_p \\ 
		\frac{\pi}{4} (\Delta - \frac{i}{\pi} \omega)
		& \Delta_p \ll \omega \ll \Delta 
	\end{array}
	\right. 
	\,.
\end{align}

\subsection{Optical conductivity and hyperbolic phase polaritons}
Integrating out the phase in \equa{eqn:L_s_p_SI}, one obtains the EM Lagrangian:
\begin{align}
	L_{EM}
	&\sim -\frac{1}{c_1-c_0} 
	\left( c_0 c_1 d^2 A_z^2  
	+ \chi^2 (\partial_t A_x)^2/d^2  
	+ 2c_0\chi A_z \partial_t A_x \right)
	+ \sigma_{\text{h}}A_x \partial_t A_z 
	+ S_{A_x^2}   
	\label{eqn:L_s_EM_SI}
	\,
\end{align}
in the gauge $\phi_s=0$.
Since the EM action is related to the optical conductivity as $S_{\text{EM}}=-\frac{i}{2} \sum_\omega \omega \sigma_{\text{ij}} A_{\text{i}} (-\omega) A_{\text{j}} (\omega)$, one obtains
\begin{align}
	\hat{\sigma}= 
	\frac{1}{c_0 - c_1}
	\begin{pmatrix}
		i \frac{\sigma_{\text{h}} ^2}{d^2} \omega +(c_0 - c_1) \sigma_{x0}  & -c_0 \sigma_{\text{h}} \\
		c_0 \sigma_{\text{h}}&  i c_0 c_1 d^2/\omega
	\end{pmatrix}
	+ \sigma_{\text{h}}\epsilon_{xz}
	\label{eqn:sigma_p}
	\,
\end{align}
where $i/j$ takes the value of $x,z$ and we have set $\chi=\sigma_h$ which holds for $\Delta_p \ll \Delta$.
In the DC limit, the stacks do not exhibit superconductivity along $z$ since $c_0(\omega)$ vanishes faster than $\omega$, but have a hall response $\sigma_{\text{h}}$ due to broken $\hat{T}$. 
Note that for the 2D stack, the conductivity simplifies to 
\begin{align}
	\sigma= 
	\frac{1}{\omega^2 - \omega_0^2}
	\begin{pmatrix}
		\frac{1}{4} \chi^2 i\omega/c_0 +(\omega^2 - \omega_0^2) \sigma_{x0} & -\omega^2  \chi  \\
		\omega^2  \chi &   i\omega c_1
	\end{pmatrix}
	+ \sigma_{\text{h}}\epsilon_{xz}
	\label{eqn:sigma_2D}
	\,
\end{align}
in the low frequency regime.
To compare, a layered superconductor with interlayer Josephson tunneling has the Lagrangian $L\sim -\nu (\partial_t \varphi_s + \phi_s)^2 + v_z^2(\partial_z \varphi_s - A_z)^2$ along $z$ direction which corresponds to superconductivity: $j_z\sim v_z^2 A_z$. 

Note that in the BEC case and in the limit of $\Delta_p \rightarrow 0$, one has $\chi \rightarrow \sigma_{\text{h}}=0$. Thus there is no Hall response even at nonzero frequency, consistent with the fact that time reversal symmetry is effectively unbroken if $\Delta_p=0$ (tunneling  between the layers is not possible). In general, $\Delta_p \neq 0$ and there is nonzero AC Hall response.

Being optically active, the bulk phase mode hybridizes with photons to form hyperbolic phase polaritons whose dispersion is determined by $1+\frac{4\pi i}{\omega} \sigma_{ij} q_i q_j/q^2=0$ for the 3D stack \cite{Basov2016,Sun.2015} and by $1+\frac{2\pi i}{\omega} \sigma_{ij} q_i q_j/|q|=0$ for the 2D stack. Due to the hall response, there are also chiral surface phase polaritons on the side surfaces/edges circulating the $x-z$ plane which we leave for future study.

\section{Stacks with $s$-type tunneling}
In this section we assume $s$-type intrinsic hybridization $t_p=t>0$ which modifies the free energy already to linear order in $t$.  The ground state order parameter is a real $\Delta$ with the same sign as $t$ which reads $\Delta(k)=\Delta \cos(k_z)$ in momentum space, meaning $(\theta_a,\theta_s)=(0, 0)$. The quasiparticle dispersion is $E=\pm \sqrt{\xi_k^2 + 4(\Delta+t)^2\cos^2(k_z)}$ with Dirac nodes/lines at $(k_{\perp},k_z) =(k_F, \pm \pi/(2d))$ in the BCS case.
We focus on the physics related to Josephson effect, meaning we neglect spatial fluctuations. In the continuous limit, the relevant degrees of freedom are the $\theta_s$ and the uniform $A_z$. 
The low energy Lagrangian valid for $\omega \ll \Delta$ is
\begin{align}
	L_s \sim \nu \Delta \left[ (\theta_s +dA_z) \partial_t (\theta_s + dA_z)
	+ \omega_0 \theta_s^2
	\right]
	\label{eqn:L_A_s_z}
	\,.
\end{align}
where $\omega_0=t \ln \frac{\Lambda}{\Delta}$.
Thus the phase mode is overdamped due to the gapless quasi particle dispersion. Integrating out the phase $\theta_s$, one obtains the optical conductivity along z 
\begin{align}
	\sigma_z=\frac{\nu \Delta  \omega_0 d^2}{-i\omega + \omega_0} 
	\,
\end{align}
which has a Drude form with width $\omega_0$. This is a surprising result because simple Dirac nodes/lines do not give such a Drude conductivity at zero temperature since there are no quasi particles. The Drude behavior  is the result of collective coupling to the collective phase mode $\theta_s$. Note that the DC conductivity $\sigma_{DC}=\nu \Delta d^2$ does not depend on $t$ although there cannot be interlayer tunneling without $t$.  The reason is that as electric field increases beyond $E_c \sim \omega_0/d \sim t/d \ln \frac{\Lambda}{\Delta}$, the system won't have a fixed $\theta_s$ but enters the AC Josephson effect regime $\dot{\theta}_s=d E_z$ and the current oscillates with frequency $d E_z$. Therefore, as $t$ increases from zero, it expands the field regime for linear DC response  while the linear response DC conductivity is a constant value.

\section{Three chain model for $\text{Ta}_2\text{NiSe}_5$}
\label{SI:TNS}
\begin{figure}
	\includegraphics[width=0.5 \linewidth]{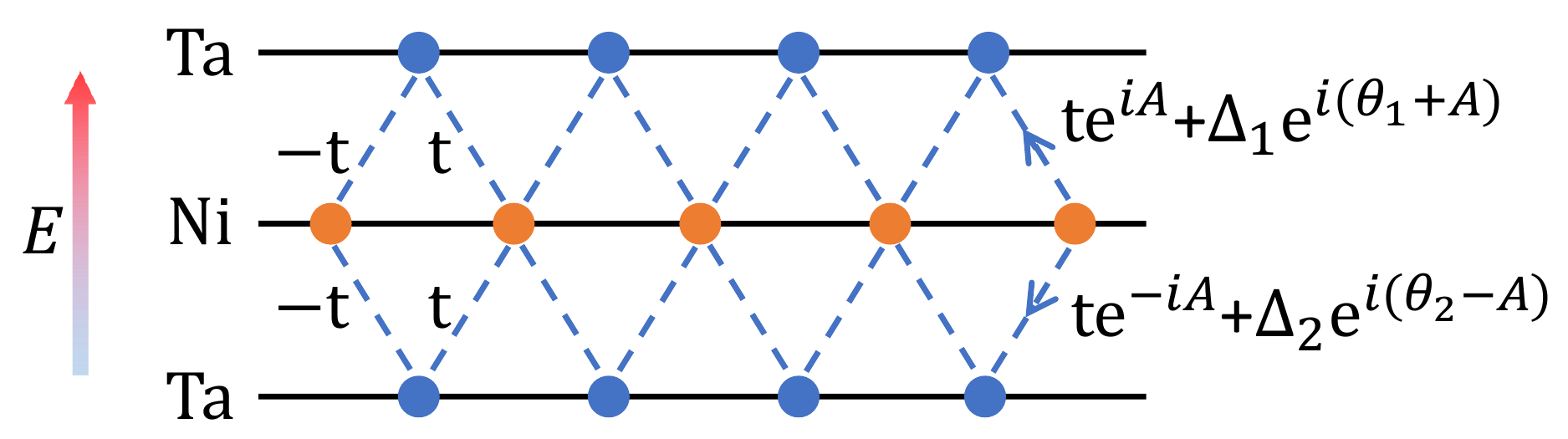}
	\caption{Schematic of the three chain model for Ta$_2$NiSe$_5$. There is a mirror symmetry $\sigma_{\perp}$ of reversing the direction of the chain. The Ni chain with active d orbitals (odd under $\sigma_{\perp}$, giving one valence band) is sandwiched by two Ta chains with active d orbitals (even under $\sigma_{\perp}$, giving two degenerate conduction bands).}
	\label{fig:TNS_chain}
\end{figure}
Ta$_2$NiSe$_5$ can be viewed as a 3D stack of its basic element, a composite Ta-Ni-Ta chain \cite{Kaneko2013, Mazza2020} shown in \fig{fig:TNS_chain}.  Its excitonic physics can be reasonably captured by this 1D chain, with the excitonic order as bond variables $\Delta_1$ and $\Delta_2$. The Lagrangian within the three band model is $L=\sum_k \psi_{k}^\dagger (\partial_\tau + H_k) \psi_{k} + \frac{1}{g} (|\Delta_1|^2 + |\Delta_2|^2)$  with the mean field Hamiltonian 
\begin{align}
	H_{k} =
	\begin{pmatrix}
		\xi_c(k) & 	\phi_1(k) & 0\\
		\phi^\ast_1(k) & 	\xi_v(k) & \phi^\ast_2(k) \\
		0 & 	\phi_2(k) & \xi_c(k)
	\end{pmatrix}
	,\,
	\begin{array}{l}
		\phi_1(k)= e^{iA}\left(it  \sin k  + \Delta_1 \cos k \right)
		\\
		\phi_2(k)=e^{-iA} \left(it  \sin k  + \Delta_2  \cos k \right)
	\end{array}
	\,.
	\label{eqn_H_three_chain}
\end{align}
Here and following we suppress the half lattice constant $a$ for notational simplicity.
We  also set $\xi_c=-\xi_v=\xi$ without changing the qualitative physics.
The resulting  dispersions for three quasi particle bands are
\begin{align}
	E_k= \left( \xi_k,\, \pm \sqrt{\xi_k^2+ |\phi_1(k)|^2+ |\phi_2(k)|^2}
	\right)
	\,.
\end{align}
At $t=0$, as a result of the mirror symmetry of interchanging the two Ta chains, \equa{eqn_H_three_chain} has an $U(2)$ symmetry corresponding to $( \Delta_1,  \Delta_2) \rightarrow ( \Delta_1,  \Delta_2)\hat{U}$ where $\hat{U}$ is an arbitrary unitary matrix. The intrinsic tunneling $t$ reduces this symmetry, rendering the low energy manifold to satisfy $|\Delta_1|=|\Delta_2|$.
Fixing $\Delta_1=\Delta e^{i\theta_1}, \Delta_2=\Delta e^{i\theta_2}$, the low energy Lagrangian in terms of the two phase angles $\theta_1$ and $\theta_2$ is 
\begin{align}
	L=K[\theta_1+A,\, \theta_2-A] + c_{\text{tns}}  \left(\sin \theta_1 + \sin \theta_2 \right)^2 + F_0
	\label{eqn:L_TNS_SI}
\end{align}
where $K$ is the kinetic energy term that vanishes in the static limit, $F_0$ is the ground state free energy at $t=0$ and we have expanded to $O(t^2)$ terms. The coefficient $c_{\text{tns}}$ is $c_{\text{tns}}= c_{\text{tns1}} t^2 \Delta^2$ in the BEC regime. This term can be computed by noting that the condensation energy is 
\begin{align}
	\delta F= \sum_k (|\xi_k|-|E_k|)
	&=\sum_k \left[ -\frac{|\phi_1(k)|^2 +|\phi_2(k)|^2}{2|\xi_k|}
	+ \frac{(|\phi_1(k)|^2 +|\phi_2(k)|^2)^2}{8|\xi_k|^3}
	+O(|\phi|^6)
	\right] 
	\,,
	\notag\\
	|\phi_i(k)|^2 &= t^2 \sin^2 k + \Delta^2 \cos^2 k + t\Delta \sin {\theta_i} \sin 2k  
	\,.
\end{align}
Considering the summation over $k$, the leading  $\theta_i$ dependence comes from the $O(\phi^4)$ terms and reads
\begin{align}
	F_2= t^2 \Delta^2 \left(\sin \theta_1 + \sin \theta_2 \right)^2
	\sum_k \frac{\sin^2 2k}{8|\xi_k|^3}
	\equiv  c_{\text{tns1}} t^2 \Delta^2 \left(\sin \theta_1 + \sin \theta_2 \right)^2
	\,.
\end{align}

Therefore, the ground states lie at the lines satisfying $\theta_1=-\theta_2$ or $\theta_1=\theta_2+\pi$ in the $(\theta_1,\theta_2)$ space.
From the equation of motion implied by \equa{eqn:L_TNS_SI}, the $z$ direction current is $j_z=-\partial_A L=(\partial_{\theta_1}-\partial_{\theta_2}) F_2 = 4 c_{\text{tns}} \sin^2 \left(\frac{\theta_1+\theta_2}{2}\right) \sin{(\theta_1-\theta_2)}$.
Mean field analysis considering effects beyond the three chain model has found that the actual symmetry is  $Z_2$, and there are only two degenerate ground states lying at $(\theta_1, \theta_2)=(0,\pi)$ and $(\theta_1, \theta_2)=(\pi, 0)$ \cite{Mazza2020}.  

With a strong electric field $E=-\partial_t A$ along $z$ (perpendicular to the chain), as shown in \fig{fig:TNS_chain}, only the kinetic term in \equa{eqn:L_TNS_SI} matters in the equation of motion, giving
\begin{align}
	\partial_t \theta_1 = E=-\partial_t \theta_2
	\,.
\end{align}
In other words, the phase $\theta_i$ has to adjust to cancel the Peierls phase $A$, and
the order parameter won't catch up with the dynamics in our choice of gauge. 
Therefore, an electric field pulse can rotate $\theta_1$ counterclockwise from $0$ to $\pi$ while rotating $\theta_2$ clockwise from $\pi$ to $0$, switching the two ground states. For a pulse duration of $T_0 \sim 0.1 \unit{ps}$ and chain width $d\approx 5 \unit{\AA}$, the field needed is $E\sim \pi/(d T_0)  \sim 4 \times 10^5 \unit{V/cm}$.
In the mean time, there is nonzero $z$ direction current oscillating between the three chains. 
The bulk inversion $\hat{P}$ is unbroken so there is no spontaneous polarization to identify the change of state, but the $z$ direction Josephson current could lead to measurable radiations. 

For weaker fields, the free energy barrier matters which imposes a threshold field to achieve the steering. According to the interband hybridization \cite{Mazza2020} $t \approx 36 \unit{meV}$  and the chain width $d\approx 5 \unit{\AA}$, from Eq.~(2) of the main text in the BCS limit, we estimate the threshold field to be $E_c \sim T_0 t^2/d \sim 10^5 \unit{V/cm}$ for $0.2 \unit{eV}$ photons.

\section{Effect of spin}
\label{SI:spin}
In this section, we show that the main conclusion of this paper is not affected by spin degrees of freedom of the electron and hole bands which can be represented by the fermion creation operators $\psi^\dagger=(\psi_{c\uparrow}^\dagger,\psi_{c\downarrow}^\dagger,\psi_{v\uparrow}^\dagger,\psi_{v\downarrow}^\dagger)$.
After decomposing the interband part of the density-density interaction $V=	\int {dr dr^\prime} V(r-r^\prime) \sum_{s_1,s_2} \psi_{c, s_1}^\dagger(r) \psi_{c, s_1}(r) \psi_{v, s_2}^\dagger(r^\prime) \psi_{v, s_2}(r^\prime)$ in the $s$-wave electron-hole pairing channel (the dominant channel), one obtains the spinful Lagrangian 
\begin{align}
	L= \psi^\dagger (\partial_\tau + H_k) \psi + \frac{1}{g} \text{Tr}[\hat{\Delta}^\dagger \hat{\Delta}]
	+ \frac{1}{g_X} \left(X^2 + \frac{1}{\omega_0^2} \dot{X}^2 \right)
	\label{eqn:lagrangian}
\end{align}
where the mean field Hamiltonian is 
\begin{align}
	H_{k} =
	\begin{pmatrix}
		\xi_c \hat{I} & 	\hat{\Delta}(k) + i\Delta_p f_k \hat{I} + X\hat{I} \\
		\hat{\Delta}^\dagger(k) - i\Delta_p f_k \hat{I} + X\hat{I}  & \xi_v \hat{I}
	\end{pmatrix}
	,\quad
	\hat{\Delta} = \Delta_0 \hat{I} + \overrightarrow{\Delta} \cdot \overrightarrow{\sigma}
	\,.
	\label{eqn_H_spin}
\end{align}
Here we have added the shear phonon $X$ which does not flip the spins. The Lagrangian \equa{eqn:lagrangian} has an $\text{SU}(2)$ symmetry of spin rotations, whose element is represented as $\hat{U}_g: (\psi_{\uparrow}^\dagger,\psi_{\downarrow}^\dagger) \rightarrow (\psi_{\uparrow}^\dagger,\psi_{\downarrow}^\dagger) \hat{U},\, \hat{\Delta} \rightarrow  \hat{U}^\dagger \hat{\Delta} \hat{U} =  \Delta_0 \hat{I} + (\hat{R} \overrightarrow{\Delta}) \cdot \overrightarrow{\sigma}$ where $\hat{U}$ is a $2\times 2$ $\text{SU}(2)$ rotation matrix and $\hat{R}$ is a $3\times 3$ $\text{SO}(3)$ rotation matrix. It is obvious that the $\Delta_0$ component of the order parameter is invariant under spin rotation, and thus corresponds to a spin singlet condensate. The $\overrightarrow{\Delta}$ transforms as a 3-vector and corresponds to a triplet condensate. The singlet and triplet form two irreducible representations of the symmetry. The phonon appears in the $\hat{I}$ channel and couples linearly only to the singlet condensate \cite{Halperin.1968,Halperin.1968_2}. 

Without the phonons, i.e., at $g_X \rightarrow 0$, the singlet and triplet states are degenerate in energy. The pure singlet state is simply two copies of  the ferroelectric states (or time reversal broken states in the stacks) discussed in the main text, both having identical orbital properties such as the electrical polarization. Therefore, all conclusions are the same as spinless models discussed in the main text. The pure triplet state is formed by two copies of the ferroelectric states with opposite electrical polarization,  corresponding to a spin polarized state. For example, the $\overrightarrow{\Delta}=(0,0, \Delta)$ state has nonzero $\langle s_z \rangle$ on one side of the sample and opposite $\langle s_z \rangle$  on the other side.

In the bilayer system in Fig.~1 of the main text, the phonon $X$ couples linearly and  cooperates with the singlet condensate, lowering its energy than the triplet state by some energy $\delta F$ which is about $\delta F \sim \Lambda^2 \left( \frac{1}{g+g_X} e^{-\frac{1}{\nu(g+g_X)}} -  \frac{1}{g} e^{-\frac{1}{\nu g}}\right)$ in the BCS weak coupling case. Therefore, as long as the shear phonon mode of the bilayer is not infinitely rigid, the ground state is the singlet state which is also ferroelectric.

In the 3D stacks of alternating electron-hole layers and 2D stacks of electron-hole chains, the   shear phonon does not couple linearly with the grounds state order parameters, and thus does not lift the degeneracy of singlet and triplet states. Among the effects not considered in this paper, direct exchange interaction favors the triplet state while kinetic exchange favors the singlet \cite{Halperin.1968_2}. The singlet state is made of two identical copies of time reversal breaking ones with spin up and down, respectively. The triplet state has opposite time reversal breaking for spin up and down, exhibiting microscopically circulating spin currents instead of charge currents \cite{Halperin.1968}. As a result, the 2D stack is in a quantum spin hall state, the 3D stack is a dirac semimetal to which spin orbit coupling might open a gap, and the anomalous hall response vanishes due to time reversal symmetry. However, the order parameter steering and Josephson current is orbital physics which works universally for both singlet and triplet states.

\section{Correlation functions}
\label{app:correlation_function}
This section defines the two point correlation functions that are coefficients of the quadratic terms in the effective action after integrating out the fermions.
The correlation function $\chi_{\sigma_i \sigma_j}$ is defined as
\begin{align}
	\chi_{\sigma_i \sigma_j}(q) = 
	\left\langle \hat{T} 
	\left(\psi^\dagger \sigma_i \psi \right)_{(r,t)}
	\left(\psi^\dagger \sigma_j \psi \right)_0  
	\right\rangle \bigg|_q 
	=
	\sum_{\omega_n, k}
	\mathrm{Tr}\left[ G(k,i\omega_n) \sigma_i G(k+q,i(\omega_n+\Omega))  \sigma_j \right]
	\,.
	\label{eqn:chi_defi}
\end{align}
where $\hat{T}$ is the time order symbol, $\mathrm{Tr}$ is over the spinor basis,  $q=(\mathbf{q},i\Omega)$ and
\begin{align}
	G(k,i\omega_n) = 
	\left\langle \hat{T} 
	\psi (x)
	\psi^\dagger(0) 
	\right\rangle \bigg|_{k,i\omega_n}
	=\frac{1}{i\omega_n - H_k }
	\,
\end{align}
is the electron Green's function. At zero temperature, rotating $i\Omega$ to $\omega$, \equa{eqn:chi_defi} reads
\begin{align}
	\chi_{\sigma_i \sigma_j}(\omega, q) 
	= \frac{1}{2}
	\sum_{k} \frac{1}{\omega^2 - (E+E^\prime)^2}
	\Bigg\{& 
	(E+E^\prime)
	\mathrm{Tr}\left[
	\sigma_i \sigma_j 
	-
	\frac{H_k \sigma_i H_{k^\prime} \sigma_j}{E E^\prime}
	\right]
	+
	\omega \mathrm{Tr}\left[
	\frac{\sigma_i H_{k^\prime} \sigma_j}{E^\prime}
	-
	\frac{H_{k} \sigma_i \sigma_j}{E}
	\right]
	\Bigg\}
	\label{eqn:chi_zero_t}
\end{align}
where $H_k$ is the mean field Hamiltonian.

\end{document}